\documentclass[12pt]{elsarticle}

\usepackage[margin=1.1in]{geometry}

\usepackage{amsmath, amssymb}
\usepackage{bm}
\usepackage{mathrsfs}
\usepackage{graphicx, fancyhdr}
\usepackage{placeins} 
\usepackage{floatrow}

\usepackage{ifpdf}
\usepackage{multirow}
\usepackage{caption}
\usepackage{subcaption}

\newdimen\figrasterwd
\figrasterwd\textwidth

\usepackage{setspace}
\usepackage{xcolor}

\newcommand{\Bxi}{\boldsymbol{\xi}}

\newcommand{\Bxirml}{\boldsymbol{\xi}_\text{rml}}
\newcommand{\Bxiuc}{\boldsymbol{\xi}_p}

\newcommand{\argmin}[1]{\underset{#1}{\text{argmin}}}

\newcommand{\Bd}{\mathbf{d}} 
\newcommand{\Bdobs}{\mathbf{d}}

\newcommand{\Cdinv}{C^{-1}_{\text{d}}}

\usepackage{hyperref}
\hypersetup{
 colorlinks=true,
 linkcolor=blue,
 citecolor=blue,
 urlcolor=blue
 }

\journal{Journal of Computational Physics}

\begin{document}

\begin{frontmatter}



\title{Deep reinforcement learning for optimal well control in subsurface systems with uncertain geology}

\author{Yusuf Nasir\corref{mycorrespondingauthor}}
\ead{nyusuf@stanford.edu}

\author{Louis J.~Durlofsky}
\ead{lou@stanford.edu}
\address{Department of Energy Resources Engineering, Stanford University, Stanford, CA 94305, USA}

\cortext[mycorrespondingauthor]{Corresponding author}

\begin{abstract}

A general control policy framework based on deep reinforcement learning (DRL) is introduced for closed-loop decision making in subsurface flow settings. Traditional closed-loop modeling workflows in this context involve the repeated application of data assimilation/history matching and robust optimization steps. Data assimilation can be particularly challenging in cases where both the geological style (scenario) and individual model realizations are uncertain. The closed-loop reservoir management (CLRM) problem is formulated here as a partially observable Markov decision process, with the associated optimization problem solved using a proximal policy optimization algorithm. This provides a control policy that instantaneously maps flow data observed at wells (as are available in practice) to optimal well pressure settings. The policy is represented by a temporal convolution and gated transformer blocks. Training is performed in a preprocessing step with an ensemble of prior geological models, which can be drawn from multiple geological scenarios. Example cases involving the production of oil via water injection, with both 2D and 3D geological models, are presented. The DRL-based methodology is shown to result in an NPV increase of 15\% (for the 2D cases) and 33\% (3D cases) relative to robust optimization over prior models, and to an average improvement of 4\% in NPV relative to traditional CLRM. The solutions from the control policy are found to be comparable to those from deterministic optimization, in which the geological model is assumed to be known, even when multiple geological scenarios are considered. The control policy approach results in a 76\% decrease in computational cost relative to traditional CLRM with the algorithms and parameter settings considered in this work. 

\end{abstract}


\end{frontmatter}


\section{Introduction}

Closed-loop modeling is utilized for decision making in a variety of domains including chemical plant operations, the control of wind farms, and the management of subsurface resources. The closed-loop modeling framework typically entails the updating of system settings (decisions) at different stages based on new information. In subsurface flow applications, e.g., the management of groundwater resources, CO$_2$ sequestration, geothermal operations and oil/gas production, closed-loop modeling traditionally involves data assimilation (for incorporating new well-based data) and robust optimization steps. This robust optimization entails the determination of well settings (flow rates or pressures) that are optimal, in an average sense, over a set of realizations that are representative of the uncertain subsurface geology. Both the data assimilation (also referred to as history matching) and robust optimization steps are computationally intensive. This is especially true if the models are generated from multiple geological scenarios, as will be the case if we wish to consider, for example, systems with sand channels of different orientation, sinuosity, thickness, etc.

Our goal in this work is to introduce a new control policy approach for efficient closed-loop decision making. The control policy is represented by a deep neural network that is trained using a reinforcement learning technique. In contrast to traditional closed-loop procedures, the training of the control policy entails only the use of prior (as opposed to history matched) geological models, thus avoiding the need for the repeated application of the computationally demanding data assimilation and robust optimization steps. In practical cases where nonlinear output constraints are present (an example of which is a maximum field water production rate), robust optimization with traditional workflows can lead to overly conservative solutions because these constraints must be satisfied for all realizations. Our control policy approach circumvents this limitation as it is able to adjust well settings, based on observations, in a model-specific manner.

A number of techniques that can reduce the computational burden of the optimization and/or data assimilation steps in traditional closed-loop workflows have been introduced. These include surrogate or proxy models involving reduced-order numerics \cite{van2006reduced, he2011enhanced, zalavadia2018parametric}, deep learning models \cite{zhu2018bayesian, tang2020deep, wang2021efficient, kim2021recurrent, wang2021theory}, and machine learning models \cite{nwachukwu2018fast, nasir2020hybrid, tang2021use}. Recently, a deep learning surrogate based on convolutional and recurrent neural networks for robust optimization and closed-loop modeling was proposed \cite{kim2022recurrent}. While the above-mentioned approaches have been shown to reduce computational demands to varying degrees, none directly addresses the conservative nature of the solutions obtained through robust optimization, nor were any of these approaches designed for the combined treatment of multiple geological scenarios.

In the context of robust production optimization for oil/gas reservoirs, which involves the determination of optimal controls for injection and production wells, the potentially conservative nature of robust optimization has motivated the development of rule-based or control policy approaches. Addiego-Guevara et al.~\cite{addiego2008insurance}, for example, proposed a policy that determines well settings/controls based on well water cuts (water cut is ratio of water production rate to total liquid production rate). The parameters of the policy in that study were obtained by optimizing a base geological model. More recently, Hanssen et al.~\cite{hanssen2017closed} introduced a control policy that distributes the total production and injection rates of fluids between wells based on a notion of priority, with this priority expressed as a function of water cut. In that work, the control policy was represented by a set of implicit algebraic equations that augment the original flow simulation equations. The use of this type of control policy thus requires access to the simulator source code. Although often effective, these (and related) control policy approaches have strong heuristic components, and they do not consider the full suite of available data, much of which may be useful in defining policies.

Advances in machine learning have led to the use of deep reinforcement learning (DRL) techniques for obtaining policies for sequential decision making. DRL has been successfully applied to train artificial intelligence agents that can play a variety of games at human or superhuman levels \cite{mnih2015human, silver2017mastering}. Due to the remarkable capabilities of DRL in generating policies for playing strategy games, these approaches have seen increasing adoption in other domains. Fan et al.~\cite{fan2020reinforcement}, for example, considered the use of DRL to obtain active control strategies for drag reduction in turbulent flow. Hachem et al.~\cite{hachem2021deep} applied DRL for open-loop control of conjugate heat transfer systems. DRL has also been used for shape optimization \cite{viquerat2021direct} in aerodynamics problems. The use of DRL for the development and management of oil reservoirs has also been the subject of a considerable amount of recent work \cite{ma2019waterflooding, miftakhov2020deep, he2021deep, nasir2021deep, zhang2022training}, as we now discuss.

Ma et al.~\cite{ma2019waterflooding} evaluated different DRL algorithms to optimize water injection and oil production well controls for multiple geological realizations. Their policy is represented by a fully connected neural network that maps quantities such as pressure at each grid block to well controls. DRL has also been applied to similar problem where the policy maps the pressure and saturation at each grid block of a particular reservoir model to the well controls \cite{miftakhov2020deep, zhang2022training}. He et al.~\cite{he2021deep} and Nasir et al.~\cite{nasir2021deep} introduced policies for optimizing the number, location and drilling sequence of production wells. The policies in these studies are represented by a convolutional neural network and trained with multiple geological models under varying economic conditions. A common assumption in studies involving the use of DRL for subsurface flow problems is that the reservoir model is known, and state quantities such as pressure and saturation are available. In practice, however, these quantities are uncertain and only production/injection data are observed. Thus the assumptions underlying many of the existing formulations are only valid in (open-loop) deterministic settings.

In this paper, we introduce a general and nonintrusive (with respect to the flow simulator) DRL-based control policy approach for closed-loop modeling. The control policy is established using a temporal convolution and recently introduced gated transformer blocks \cite{parisotto2020stabilizing}. The training of the framework is accomplished using the proximal policy optimization algorithm \cite{schulman2017proximal}, with training data generated from flow simulations across an ensemble of prior geological models. After appropriate training, the policy instantaneously maps quantities that can be observed in practice to decision variables that prescribe (optimal) settings for existing injection and production wells. We test the framework, using both 2D and 3D geological models drawn from single and multiple geological scenarios, for problems involving the production of oil via water injection. The performance of our DRL-based methodology is compared to robust optimization over prior geological models, to deterministic realization-by-realization optimization, and to traditional closed-loop reservoir management.

This paper proceeds as follows. In Section~\ref{sec:ge_trad_clrm}, we give the governing equations for oil-water subsurface flow and then describe the traditional closed-loop reservoir modeling workflow. The use of deep reinforcement learning to determine a general control policy for closed-loop decision making is presented in Section~\ref{sec:drl_cp}. The control policy representation and training procedure are also described. Detailed computational results, for 2D and 3D systems, are presented in Section~\ref{sec:comp_res}. Comparisons of DRL-based results to those from robust optimization over prior geological models, to deterministic realization-by-realization optimization, and to traditional CLRM are provided. We conclude in Section~\ref{sec:concl} with a summary and suggestions for future work.

\section{Governing equations and traditional closed-loop modeling approach}
\label{sec:ge_trad_clrm}

In this section, we present the governing equations for the oil-water system considered in this study. We then describe the components and limitations of traditional closed-loop modeling in this setting.

\subsection{Governing equations}

We consider isothermal oil-water flow. The system is immiscible, meaning the water component exists only in the water phase, and the oil component only in the oil phase. Gravitational effects are included in the 3D models. The governing equation for each phase ($l = o$ for oil and $l = w$ for water) is obtained by combining Darcy's law for multiphase flow with a statement of mass conservation for each component. This gives
\begin{equation}
    \nabla \cdot \Big[ \textbf{k}\rho_{l} \lambda_{l} \left(\nabla p - \rho_{l}g \nabla D \right) \Big] = \frac{\partial }{\partial t}  \left(\phi \rho_{l} S_{l}\right) + q_{l}, \quad l=o,~w.
    \label{eq:2p_eqn}
\end{equation}
The convective (flow) terms appear on the left-hand side. Here \textbf{k} is the absolute permeability tensor (permeability is essentially a flow conductivity), $\rho_{l}$ is the phase density, $\lambda_{l}$ is the phase mobility, given by $\lambda_{l}=k_{rl}/\mu_{l}$, with $k_{rl}$ the phase relative permeability and $\mu_{l}$ the phase viscosity, $p$ is pressure (taken to be the same for both phases, as is often done in reservoir simulation), $g$ is gravitational acceleration, and $D$ is depth. On the right-hand side we have the accumulation term, with $\phi$ denoting porosity and $S_{l}$ phase saturation (volume fraction), and the source terms, with $q_{l}$ indicating the mass source/sink term. In this work, we consider the permeability field $\textbf{k}({\bf x})$, where {\bf x} denotes spatial location within the reservoir, to be uncertain. Eq.~\ref{eq:2p_eqn} is solved numerically using a standard finite-volume procedure, with the model represented on a Cartesian grid containing a total of $N_b$ cells.

Production and injection wells in this work are controlled through the specification of bottom-hole pressures (BHPs). BHP is the pressure in the wellbore at a particular depth (this could be the top of the reservoir, or the depth of the uppermost perforation). The pressure in the wellbore at other vertical locations is computed through a $\rho g \Delta z$ adjustment of the BHP, where $\rho$ is an average fluid density in the well and $\Delta z$ is computed relative to the BHP location. The phase flow rate for well $w$ in well-block $i$ is given by the Peaceman well model~\cite{peaceman1983interpretation}:
\begin{equation}
    \left(q_{l}^{w}\right)_{i} = WI_{i} \left(\lambda_{l} \rho_{l} \right)_{i} (p_{i} - p^{w}),
    \label{eq:rate}
\end{equation}
where $WI_{i}$ is the well index, which is a prescribed function of the well radius, well-block geometry and permeability, $p^w$ is the wellbore pressure evaluated at the center of the well block, and $p_{i}$ is the well-block pressure. Note that flow is driven by the difference between the wellbore and well-block pressures.

\subsection{Traditional closed-loop modeling approach}
\label{sec:trad_clrm}

In the context of oil reservoir management, two closed-loop modeling frameworks have been introduced -- closed-loop field development (CLFD) \cite{Shirangi2015Closed-LoopValidation, jahandideh2020closed}, in which the locations of new wells are optimized, and closed-loop reservoir management (CLRM) \cite{brouwer2004improved, aitokhuehi2005optimizing, jansen2009closed}, where the
time-varying controls for existing wells are optimized. We consider the CLRM problem in this work because our formulation is based on an existing set of wells. The optimization in our case entails the determination of the BHPs of production and injection wells, at a series of control steps,  such that an economic metric of interest is maximized. Operational constraints, for example the maximum water injection rate or liquid production rate, must be satisfied. 

\begin{figure}[htp]
	\centering
	\includegraphics[width=0.7\textwidth]{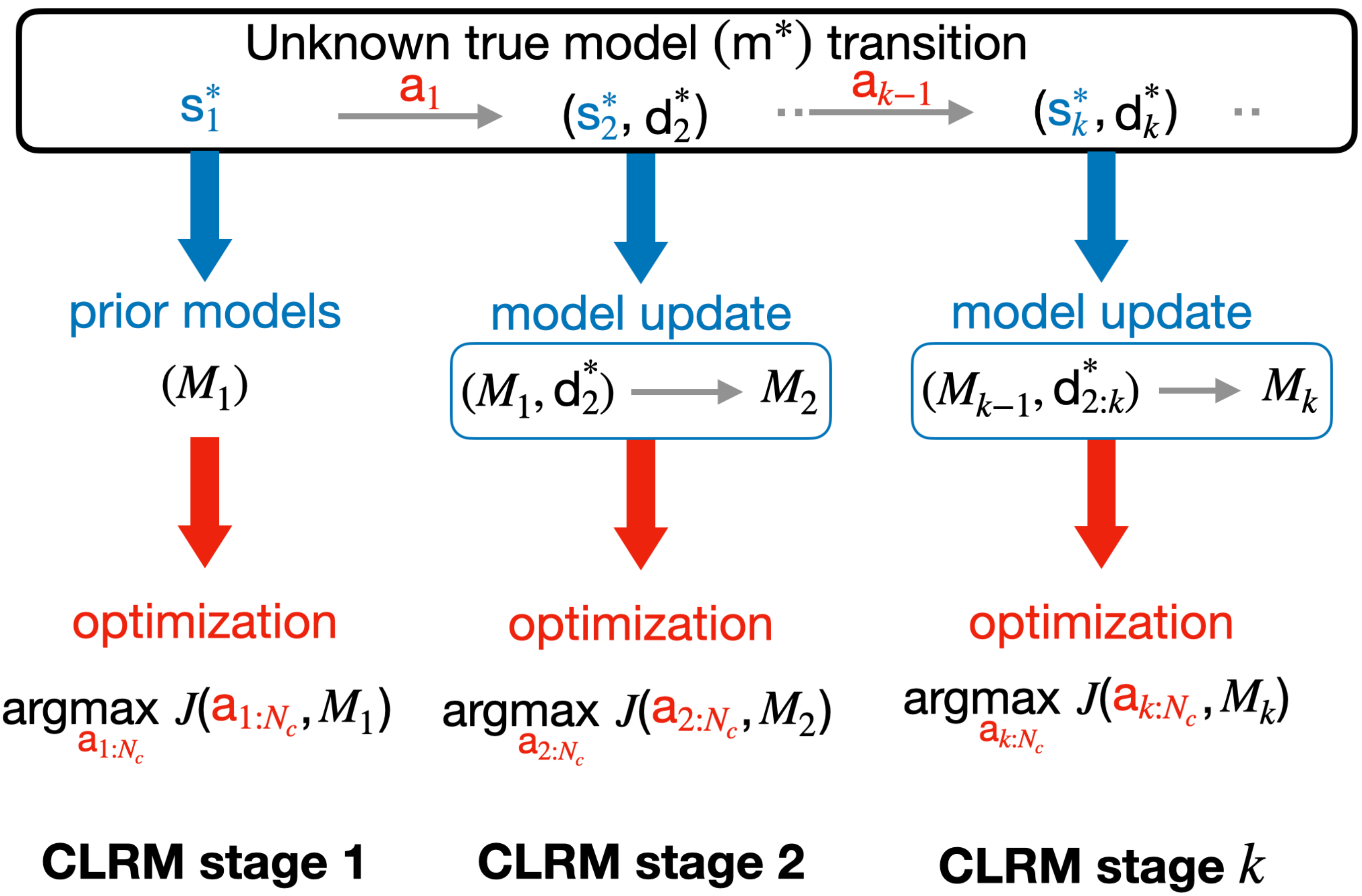}
	\caption{Traditional closed-loop reservoir management framework.}
	\label{fig:clrm_schematic}
\end{figure}

The traditional CLRM approach for this problem is shown in Fig.~\ref{fig:clrm_schematic}. The framework involves two key components, namely the update of a set of geological models based on newly observed production and injection data, followed by model-based (robust) optimization under geological uncertainty. These steps are repeated a number of times during the life of the production operation. As shown in Fig.~\ref{fig:clrm_schematic}, the time frame is divided into $N_c$ regular intervals, each of which corresponds to a control/decision-making step. The `true' (though uncertain from our perspective) reservoir model $\textbf{m}^{*}$ is initially at an unobserved state $\textbf{s}_{1}^{*}$, where $\textbf{s}^{*}$ represents the spatial distribution of the pressure and saturation. 

At the first decision-making stage, the goal is to determine the well settings $\textbf{a}_{1} \in \mathbb{R}^{N_{w}}_{+}$ (${N_{w}}$ is the total number of production and injection wells) to be prescribed in the first control step. To achieve this, we first represent our prior geological knowledge of $\textbf{m}^{*}$ through a set of $N_{r}$ model realizations, i.e., ${\{\textbf{m}^{1}_{1}, \textbf{m}^{2}_{1},\ldots, \textbf{m}^{N_{r}}_{1}\}}$. This set of prior realizations is denoted by $M_{1}$. Prior models are typically conditioned to honor `hard' data (rock properties such as permeability at well locations), and to be consistent with a particular geological scenario(s), as determined from seismic data, outcrops, analogs, etc. A robust optimization is then performed to determine the well settings $\textbf{a}_{1}$ for operating the wells in the first control step. This optimization entails the maximization of an objective function $J$, commonly defined as the expected net present value (NPV) (i.e., $J = \mathbb{E} \left[NPV\right]$), with the expectation computed over the geological models in $M_{1}$. Although the goal at this step is to determine $\textbf{a}_{1}$, a long-term production optimization that involves future well settings (i.e., $\textbf{a}_{1:N_c} = [\textbf{a}_{1}^T, \textbf{a}_{2}^T, \ldots, \textbf{a}_{N_c}^T]^T$) is performed. This is because the optimal well settings $\textbf{a}_{1}$ are dependent on the settings used in subsequent control steps.

After operating the wells until the end of the first control step, with well settings $\textbf{a}_{1}$, the reservoir $\textbf{m}^{*}$ transitions to a new (unknown) state $\textbf{s}_{2}^{*}$. However, the time-varying production and injection data $\textbf{d}_{2}^{*}$ up to the beginning of the second control step are observed. The prior models $M_{1}$ are then updated based on the observed data $\textbf{d}_{2}^{*}$. This entails a data assimilation/history matching step. The model update is typically performed using well-established approaches such as randomized maximum likelihood \cite{kitanidis1995quasi}, ensemble Kalman filtering \cite{chen2006data}, or ensemble smoothing with multiple data assimilation \cite{emerick2013ensemble}. The set of posterior models after history matching, denoted $M_2$, is given by $M_{2} = {\{\textbf{m}^{1}_{2}, \textbf{m}^{2}_{2},\ldots, \textbf{m}^{N_{r}}_{2}\}}$. With these updated models, a robust optimization is again performed to obtain the well settings $\textbf{a}_{2}$ applied in the second control step.

In general, at any decision-making stage $k \in (1, 2, \ldots, N_c)$, the history matching is performed with all existing observed data $\textbf{d}_{2:k}^{*}$. This involves updating the prior models $M_{k-1}$ to generate a set of posterior models $M_{k}$. Robust production optimization, involving all decision variables until the last stage (i.e., $\textbf{a}_{k:N_c}$), is then performed. This optimization entails the maximization of the objective function $J$, given by
\begin{gather}
\begin{array}{rrclcl}
\displaystyle \max_{\textbf{x}_{k} \in \mathbb{X}} & {J (\textbf{x}_{k}, M_k)}, \ \  \textrm{subject to } \textbf{c}(\textbf{x}_{k})\leq\textbf{0},
\end{array}
\label{eq:opt_uncertainty}
\end{gather}
\noindent 
where $\textbf{x}_{k} = \textbf{a}_{k:N_c}$ denotes the current and future well settings, and the space $\mathbb{X}$, which prescribes upper and lower bounds, defines the allowable values for the well settings. The vector \textbf{c} denotes any nonlinear output constraints. The expected NPV, which defines the objective function $J$, is computed as
\begin{gather}
	J(\textbf{x}_{k}, M_{k}) = \frac{1}{N_r} \sum\limits_{i = 1}^{N_r} \textnormal{NPV}(\textbf{x}_{k}, \textbf{m}_{k}^{i}).
    \label{eq:exp_npv}
\end{gather} 
\noindent The computation of $J(\textbf{x}_{k}, M_{k})$ requires performing $N_{r}$ flow simulations; i.e., one for each geological model  $\textbf{m}_{k}^{i} \ \forall i \in (1, 2, \ldots, N_{r})$. The time-varying well BHPs defined by $\textbf{x}_{k}$ are applied in each of these runs. 

Following \cite{nasir2021two}, we compute NPV as
\begin{align}
	\textnormal{NPV}(\textbf{x}_{k}, \textbf{m}_{k}^{i})=\sum\limits_{j = 1}^{N_t}  \left[ \sum\limits_{i = 1}^{N_p} \left(p_{o} ~q^{i}_{o,j}-c_{pw}~q^{i}_{pw,j}\right) - \sum\limits_{i = 1}^{N_i} c_{iw}~q^{i}_{iw,j} \right]\frac{\Delta t_j}{( 1+b )^{t_j/365}},
    \label{gen_field_dev_npv_eqn}
\end{align} 
\noindent
where $N_i$ and $N_p$ are the number of injectors and producers, respectively. The number of time steps in the flow simulation is denoted by $N_t$. The variables $t_j$ and $\Delta t_j$ are the time and time step size (in days) at time step $j$. The rates of oil and water production and water injection, for well $i$ at time step $j$ are, respectively, $q^{i}_{o,j}$, $q^{i}_{pw,j}$, and $q^{i}_{iw,j}$. 
The economic parameters $p_{o}$, $c_{pw}$, $c_{iw}$ and $b$ represent the oil price, cost of produced and injected water, and the annual discount rate, respectively. In contrast to \cite{nasir2021two}, we exclude the well drilling cost since, in the context of CLRM, the wells have already been drilled and do not affect the optimization solution.

A large set of geological models is typically required to capture the uncertainty in $\textbf{m}^{*}$. This can lead to substantial cost, since computational requirements for both data assimilation and robust optimization are directly proportional to $N_r$. In practical settings, where the geological scenario from which the geological realizations are drawn is (typically) also uncertain, data assimilation may have to be performed over multiple scenarios. Because of the underlying assumptions associated with the above-mentioned history matching algorithms, in many cases data assimilation is conducted one scenario at a time. This acts to further complicate the procedure and increase computational demands.

Another important limitation in the traditional CLRM workflow stems from the assumption that, at any CLRM stage, all geological models are equally probable (this assumption is implicit in Eq.~\ref{eq:exp_npv}). This treatment is required because the history matching procedures we apply only provide a set of posterior models, not their associated probabilities. Thus we seek well settings that are optimal in an average sense, even though some of the models are more likely than others. A closely related issue involves the satisfaction of nonlinear constraints. Specifically, in many CLRM procedures, all models (or in some setups the large majority of models) are required to satisfy all constraints. This tends to drive the optimization toward overly conservative solutions, even though the limiting behavior may result from a few low-probability models. The DRL-based procedures described in the next section circumvent, to a large extent, these limitations.

\section{Deep reinforcement learning control policy for closed-loop modeling}
\label{sec:drl_cp}

We now describe the deep reinforcement learning framework for closed-loop modeling. We first discuss a general control-policy-based framework for CLRM. We then introduce the policy optimization procedure used in this work.

\subsection{General control-policy-based framework for CLRM}

As discussed in the Introduction, many of the previous control policies used for reservoir management have strong heuristic components, and often base actions on a subset of the available data (e.g., on water cut). In our general control policy framework, by contrast, we consider all observable information in the decision making process.

A distribution of possible geological scenarios and associated geological models, $p(\textbf{m})$, which represents the uncertainty in the true geological model $\textbf{m}^*$, is first defined. A set of $N_r$ geological realizations, $M_{1} = {\{\textbf{m}^{1}_{1}, \textbf{m}^{2}_{1},\ldots, \textbf{m}^{N_{r}}_{1}\}}$, are sampled from $p(\textbf{m})$. We consider a parameterized class of policies $\{\pi_{\theta}: \mathbb{D} \rightarrow \mathbb{A}$; $\ \boldsymbol{\theta} \in \mathbb{R}^{N}$; s.t.~$\textbf{a}_{k} = \pi_{\theta}(\textbf{d}_{1:k}) \ \forall k \in (1, 2, \ldots, N_{c}) \}$, where $\mathbb{D}$  and $\mathbb{A}$ represent the observation and well control spaces respectively, and $\textbf{d}_1$ contains any observed data that may exist before the first decision stage. Note that $\textbf{d}_1$ is a null vector if no data are observed before the first decision-making stage. With this definition, given the policy parameters $\boldsymbol{\theta}$, the well settings for any control step $k$ can be obtained from the observed data. Instead of optimizing Eq.~\ref{eq:opt_uncertainty} at each control step, we define a single optimization problem given by 
\begin{gather}
\begin{array}{rrclcl}
\displaystyle \max_{\boldsymbol{\theta} \in \mathbb{R}^{N}} & {G (\pi_{\theta}, M_1)}, \ \  \textrm{subject to } \textbf{c}(\pi_{\theta})\leq\textbf{0}.
\end{array}
\label{eq:policy_opt}
\end{gather}
We consider each geological model in $M_1$ to be a possible true model and define $G(\pi_{\theta}, M_1)$ as
\begin{gather}
	G(\pi_{\theta}, M_1) = \frac{1}{N_r} \sum\limits_{i = 1}^{N_r} \sum\limits_{k = 1}^{N_c} \textnormal{NPV}(\textbf{a}_{k}^{i}, \ \textbf{m}_{1}^{i}).
    \label{eq:policy_exp_npv}
\end{gather} 
Here $\textbf{a}_{k}^{i} = \pi_{\theta}(\textbf{d}_{1:k}^{i})$, with $\textbf{d}_{1:k}^{i}$ the observed data up to control step ${k}$ for possible true model $\textbf{m}_{1}^{i}$. Note that the objective functions in the optimization problems defined by Eq.~\ref{eq:opt_uncertainty} and \ref{eq:policy_opt} are similar. However, the control policy optimization (Eq.~\ref{eq:policy_opt}) involves only the prior geological models $M_1$, and the well settings are defined by a policy instead of by a single solution that maximizes Eq.~\ref{eq:opt_uncertainty}. 


Once a control policy is defined based on the prior models in $M_1$, the well settings at each control step $k$, for model $i$, are obtained immediately (without any time lag) through $\textbf{d}_{1:k}^{i}$.
We reiterate that, because we are optimizing over policies that define a strategy based on the observed data for each possible model, the well settings for each model in $M_1$ will be different. Thus, less likely models have no direct influence on the well settings that are applied for the true model $\textbf{m}^{*}$ (they have some indirect influence as they are used in the determination of the policy parameters $\boldsymbol{\theta}$). This is in contrast to the traditional CLRM approach, where the likelihood of the various models does not affect the weighting.

\subsection{Deep reinforcement learning for determination of control policy}
\label{sec:drl}

\begin{figure}[htp]
	\centering
	\includegraphics[width=0.6\textwidth]{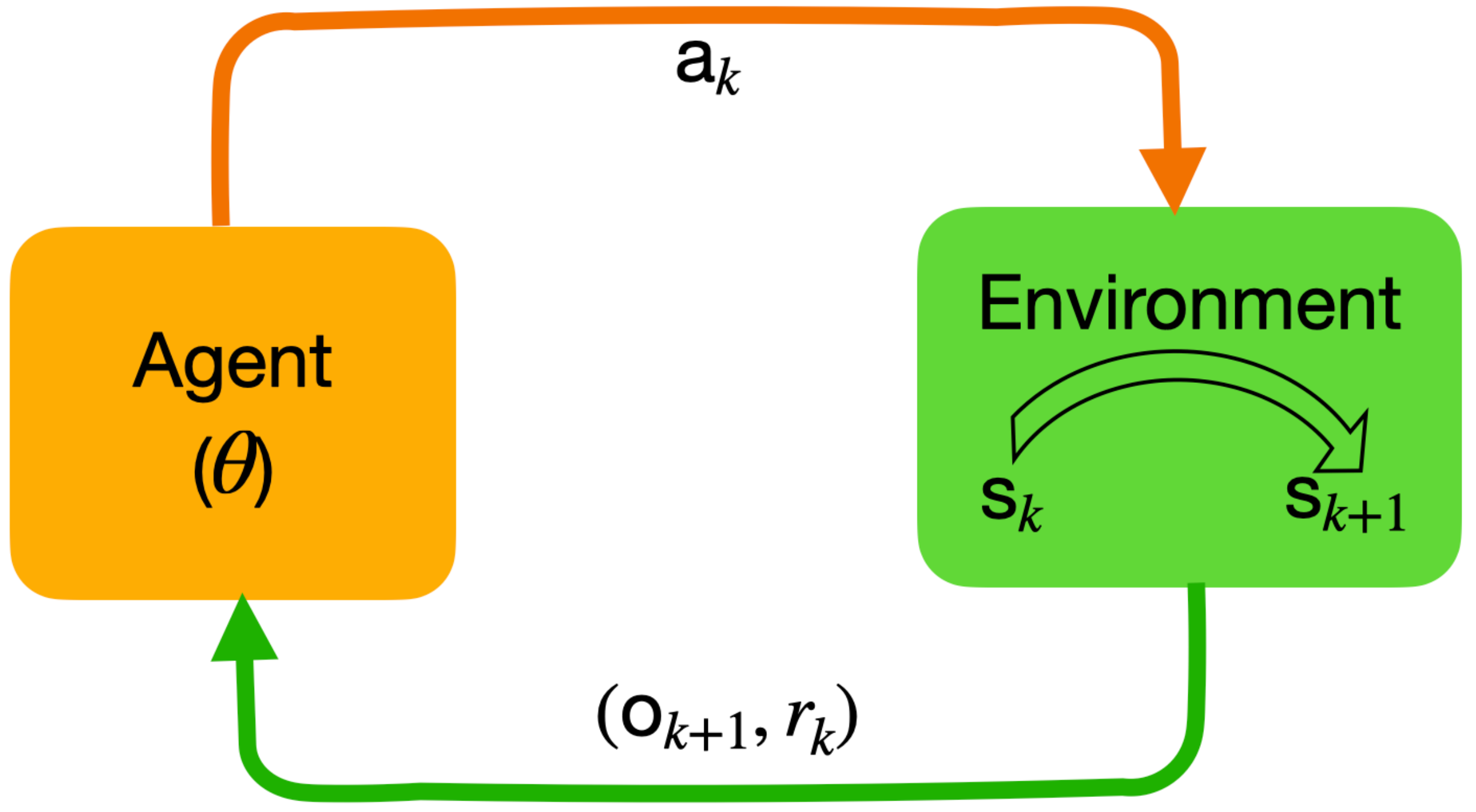}
	\caption{Schematic of the deep reinforcement learning agent-environment interface.}
	\label{fig:drl_interface}
\end{figure}

Deep reinforcement learning (DRL) is used to solve the optimization problem posed in Eq.~\ref{eq:policy_opt}, with the policy parameters $\boldsymbol{\theta}$ defined by the weights of a neural network. In reinforcement learning, an agent interacts with an environment ($\varepsilon$) through an interface, as illustrated in Fig.~\ref{fig:drl_interface}. At each stage $k$ of the decision-making process, with the environment at state $\textbf{s}_{k}$, the agent takes an action $\textbf{a}_{k}$. The environment indicates the consequences of the action to the agent through an observation, denoted $\textbf{o}_{k+1}$, and a reward, denoted $r_{k}$. The observation contains information on the change in the state of the environment, i.e., $\textbf{s}_{k+1}$ relative to $\textbf{s}_{k}$, while the reward represents the quality of the action taken in decision stage $k$. The signals exchanged between the agent and the environment provide the history, denoted $\textbf{h}_{k}$, with $\textbf{h}_{k} = (\textbf{o}_1, \textbf{a}_1, \textbf{o}_2, \ldots, \textbf{a}_{k-1}, \textbf{o}_k)$. Given this $\textbf{h}_{k}$, the agent selects an action through a policy $\pi_{\theta}$, with $\textbf{a}_{k} = \pi_{\theta}(\textbf{h}_{k})$. 

The exchange of signals ($\textbf{a}_k, \textbf{o}_{k+1}, r_k$) between the agent and the environment proceeds for an episode, defined by decision stage $k=1$ to $k=N_c$. The goal in reinforcement learning is to find a policy $\pi_{\theta}$ that maximizes the expected cumulative reward given by
\begin{equation}
    \overline{V}_{\pi_{\theta}} = \mathbb{E} \left[\sum_{k=1}^{N_c}r_{k}\mid \pi_{\theta}, \varepsilon \right],
    \label{eq:exp_return}
\end{equation}
where the expectation is with respect to the entire experience of the agent over multiple episodes. The formulation of the reinforcement learning problem can vary depending on the type of uncertainty associated with the environment. If the full state of the environment $\textbf{s}_{k}$ is observed (i.e., $\textbf{o}_{k} = \textbf{s}_{k}$), the process is referred to as a fully observable Markov decision process (MDP). In a fully observable MDP, the action $\textbf{a}_{k}$ depends on the history only through $\textbf{s}_{k}$. This implies $\textbf{a}_{k}=\pi_{\theta}(\textbf{h}_{k})=\pi_{\theta}(\textbf{s}_{k})$. Formulations of the reinforcement learning problem in such settings, in the context of oil/gas field development and management, were considered in previous studies \cite{ma2019waterflooding, miftakhov2020deep, he2021deep, nasir2021deep, zhang2022training}. Here, however, consistent with practical cases, the environment $\varepsilon$ is uncertain and the state cannot be observed. This renders our problem a partially observable Markov decision process (POMDP).

We represent the epistemic uncertainty of the environment through the prior set of geological models, with $\varepsilon = M_1$. The expectation in Eq.~\ref{eq:exp_return} is now with respect to `experiences' from the multiple possible environments or geological models. This definition of Eq.~\ref{eq:exp_return} (with $\varepsilon = M_1$) is equivalent to the policy optimization for CLRM given in Eq.~\ref{eq:policy_exp_npv}. However, as will be shown later, the maximization of Eq.~\ref{eq:exp_return}, achieved by determining optimal policy parameters $\boldsymbol{\theta}$, can be performed in an iterative manner. With this treatment, we avoid having to simulate all geological models in $M_1$ at each iteration. 

Because we consider all geological models in $M_1$ as possible true models, the true state of the environment (such as the permeability, pressure and saturation distribution) is assumed to be unknown, even though we have access to it (i.e., we set $\textbf{o}_{k}=\textbf{d}_{k}$). The maximization of Eq.~\ref{eq:exp_return} therefore also entails reduction of the epistemic uncertainty in $\varepsilon$ through the history $\textbf{h}_{k}$. This can be viewed as an implicit data assimilation.

We now define the action, observation and history for our specific application, with nonlinear output constraints $\textbf{c}(\pi_{\theta}$) included. As noted earlier, wells are operated through the specification of time-varying BHPs, subject to constraints such as maximum liquid production or water injection rate. These constraints are nonlinear because their satisfaction/violation can only be determined by solving the nonlinear flow equations defined by Eqs.~\ref{eq:2p_eqn} and \ref{eq:rate}. We define the action as $\textbf{a}_{k}=[0, 1]^{N_w}$. The resulting value of $\textbf{a}_{k}$ is then linearly mapped to be between the allowable upper and lower bounds of the well BHPs. 

The continuous action space $[0, 1]^{N_w}$ is represented by a diagonal Gaussian distribution. The actions are represented in terms of random variables to enable sampling of the action space for each history $\textbf{h}_{k}$. During policy training, exploration is achieved by sampling actions according to 
\begin{equation}
    a_k^i = sig(\mu_k^i + \sigma_k^i\epsilon_k^i),
    \label{eq:action_sampling}
\end{equation}
where $a_k^i$ is the action for well $i$ at control step $k$, $\mu_k^i$ is the action mean, $\sigma_k^i$ is the action standard deviation, $\epsilon_k^i \sim \mathcal{N}(0, 1)$ introduces stochasticity in the policy training, and $sig$ denotes the sigmoid function, which ensures the actions are between 0 and 1. After the control policy optimization, the action mean is taken as the point estimate of the optimal action, i.e., $a_k^i = sig(\mu_k^i)$.  

The output rate constraint is handled within the flow simulation using a procedure described in \cite{kourounis2014adjoint}. This approach entails switching wells from BHP control to rate control when the rate constraint is violated. The well is then operated at the specified rate constraint value until it can be switched to BHP control without violating the rate constraint. Satisfaction of the original constraints, and the feasibility of switching wells that have previously violated constraints back to BHP control, are checked at each time step of the flow simulation. 

For wells that switch from BHP control to rate control (i.e., for wells where the rate constraint is violated), the BHP used to determine the well pressure in Eq.~\ref{eq:rate} will be different from that proposed by the policy. Therefore, instead of defining the history $\textbf{h}_{k}$ as in the conventional reinforcement learning procedure, we now define it to contain only the observed data $\textbf{o}_{1:k}$. These data now include the BHPs associated with the wells operating under rate control, which can be computed from Eq.~\ref{eq:rate} with $(q_l^w)_i$ specified. Importantly, the proposed policy actions $\textbf{a}_{1:k-1}$ are now not included in the history $\textbf{h}_{k}$. Specifically, we now define the history as $\textbf{h}_{k}:=\textbf{d}_{1:k}$, with the observation at each control step $\textbf{d}_{k} \in \mathbb{R}_{+}^{N_{d}(3N_p + 2N_i)}$ defined as
\begin{equation}
   \textbf{d}_{k} = \left[\textbf{q}_{o,k}^{T}, \ \textbf{q}_{wi,k}^{T}, \ \textbf{p}_{w,k}^{T}, \ \textbf{w}_{p,k}^{T} \right]^{T}.
    \label{eq:obs}
\end{equation}
At each control step $k$, $\textbf{q}_{o} \in \mathbb{R}_{+}^{N_{d}N_{p}}$ is the oil production rate reported at $N_d$ intervals between the starting points of control step $k-1$ and $k$, $\textbf{q}_{wi} \in \mathbb{R}_{+}^{N_{d}N_{i}}$ is the water injection rate, $\textbf{p}_{w} \in \mathbb{R}_{+}^{N_{d}N_{w}}$ contains the actual BHPs (which satisfy the rate constraints), and $\textbf{w}_{p} \in \mathbb{R}_{+}^{N_{d}N_{p}}$ is the well water cut.
 
In practice, the observed data contain some amount of measurement error. We account for this by  adding random independent Gaussian noise with zero mean to the observed data. The standard deviation of the rate measurement error is taken to be 5\% of the true rate, with the minimum and maximum measurement errors specified as 1.5 and 8~m$^{3}$/day. The standard deviation of pressure measurement errors is set to 0.35~bar.

The partially observable nature of our problem leads to additional complexity in the policy representation compared to the MDP formulation, because in our case the size of the history vector $\textbf{h}_{k}$ increases with time, as more data are observed. A memory-based neural network is therefore used to represent and learn the state or memory of the agent, $\hat{\textbf{s}}_{k}$. This transforms the POMDP to an approximate MDP with decisions made through the agent state $\hat{\textbf{s}}_{k}$ instead of the environment state $\textbf{s}_{k}$. The state of the agent, which serves as a compact representation of the history, is recursively updated through application of
\begin{equation}
   \hat{\textbf{s}}_{k} = f_{\zeta}(\hat{\textbf{s}}_{k-1:k-\tau}, \textbf{d}_{k}),
    \label{eq:mem_upd}
\end{equation}
where $f_\zeta$ denotes the agent state update function, with $\boldsymbol{\zeta} \subset \boldsymbol{\theta}$, $\tau$ is the number of previous agent states/memories considered during the computation of the current agent state, and $\hat{\textbf{s}}_{k-1:k-\tau} = [\hat{\textbf{s}}_{k-1}^T, \hat{\textbf{s}}_{k-2}^T, \ldots, \hat{\textbf{s}}_{k-\tau}^T]^T$. For notational convenience, we write $\hat{S}_{k-1} = \hat{\textbf{s}}_{k-1:k-\tau}$. The action proposed by the policy is now based on the previous $\tau$ agent states ($\hat{S}_{k-1}$) and the most recent observation. This means we write $\textbf{a}_{k} = \pi_{\theta}(\hat{S}_{k-1}, \textbf{d}_{k})$ rather than $\textbf{a}_{k} = \pi_{\theta}(\textbf{h}_{k})$. 

The long short-term memory (LSTM) \cite{hochreiter1997long} neural network is one possible choice for the agent state update function $f_{\zeta}$. In LSTM, a single memory ($\hat{\textbf{s}}_{k-1}$, with $\tau=1$) composed of long and short-term components is used to evolve the state of the network. In this work, we instead use a stabilized version of the transformer network \cite{parisotto2020stabilizing} (described later), which has been shown to outperform LSTM in problems where memory of past events is important. In the prediction of the next agent state $\hat{\textbf{s}}_{k}$, transformers extract features from each previous memory through an attention mechanism \cite{vaswani2017attention}. This facilitates the use of the most important information from each previous memory for the prediction of the current agent state. 

\subsection{Control policy optimization procedure}
\label{sec:cont_pol_opt}

The proximal policy optimization (PPO) method \citep{schulman2017proximal} is used for the maximization of Eq.~\ref{eq:exp_return}. The policy parameters $\boldsymbol{\theta}$ are varied and stochastic gradient descent is applied. PPO is an `on-policy' algorithm, which means policy improvement is achieved by generating training data (over multiple episodes and geological models) from the latest policy. By following the current policy, an individual training sample contains the previous agent states $\hat{S}_{k-1}$, observation data $\textbf{d}_{k}$, action $\textbf{a}_{k}$, and reward $r_k$, at control step $k$.

We now briefly describe the PPO algorithm, adapted for our POMDP problem. Please see \cite{schulman2017proximal} for a detailed explanation of the PPO algorithm. In PPO, the maximization of Eq.~\ref{eq:exp_return} is achieved by minimizing the policy loss $L_{\pi_{\theta}}$, given by
\begin{subequations}
\begin{align}
\label{eq:policy_loss_ppo}
\begin{split}
 L_{\pi_{\theta}} = -\mathbb{E}_k[\min(p_k(\boldsymbol{\theta}) A_k,\textit{clip}(p_k(\boldsymbol{\theta}), 1-\epsilon, 1+\epsilon) A_k)],
\end{split}
\\[2ex]
\label{eq:advantage_gae}
\begin{split}
 A_k = \sum_{l=k}^{N_c}\left(\gamma\lambda\right)^{l-k}\left(r_l+\gamma V^\pi(\hat{\textbf{s}}_{l+1})-V^\pi(\hat{\textbf{s}}_{l}) \right),
\end{split}
\end{align}
\end{subequations}
where $p_k(\boldsymbol{\theta})$ is the ratio of the new policy $\pi_{\theta}(\hat{S}_{k-1}, \textbf{d}_{k})$ to the old policy $\pi_{\theta_{old}}(\hat{S}_{k-1}, \textbf{d}_{k})$ from which the training data were generated, and $A_k$ is the advantage function. Here by policy ratio we mean the ratio of parameters that define the action distribution. The policy change quantified by $p_k(\boldsymbol{\theta})$ is bounded within [$1-\epsilon, 1+\epsilon$] through the term $\textit{clip}(p_k(\boldsymbol{\theta}), 1-\epsilon, 1+\epsilon)$. Here we set $\epsilon=0.3$. This essentially introduces a trust-region into the optimization, which acts to limit changes in the policy. This prevents large updates that could lead to deterioration of the policy. 

While $p_k(\boldsymbol{\theta})$ determines the magnitude of the policy change, the advantage function $A_k$  defines the direction in which the policy is updated \cite{schulman2015high}. This function  defines the quality of an action for a specific state relative to a baseline quality. It is computed by comparing the immediate reward ($r_l$) achieved at a particular state ($\hat{\textbf{s}}_l$), plus the predicted future rewards ($V^\pi(\hat{\textbf{s}}_{l+1})$), to the predicted total baseline reward at the current agent state ($V^\pi(\hat{\textbf{s}}_{l}$)). We can thus view the minimization in Eq.~\ref{eq:policy_loss_ppo} as adjusting the policy in the direction of the state-action sequences that outperform the baseline (in which case $A_k > 0$). The hyperparameters $\gamma$ and $\lambda$ control the bias and variance introduced in the estimation of the advantage. They act to discount future rewards, thus impacting the long-term effect of an action. The values of $\gamma$ and $\lambda$ are problem dependent and are determined through numerical experimentation. Based on numerical experiments for our problem, we set $\gamma = 0.99$ and $\lambda = 1$.

The baseline in Eq.~\ref{eq:advantage_gae} is referred to as the value function $V^\pi$. This function predicts the expected cumulative discounted reward $V^\pi(\hat{\textbf{s}}_{k}) = \mathbb{E}\left[\sum^{N_c}_{l=k} \gamma^{l-k} r_l\right]$ of being in state $\hat{\textbf{s}}_{k}$ and then following the current policy $\pi$. The parameters of the value function are determined by minimizing the value function loss $L_{vf}$, given by
\begin{subequations}
\begin{align}
\label{eq:value_function_loss}
\begin{split}
 L_{\textit{vf}} = \mathbb{E}_k\left[\max\left(\left(V_\psi(\hat{\textbf{s}}_{k})-V_{target}(\textbf{h}_{k})\right)^2,\left(V_{\psi, clipped}-V_{target}(\textbf{h}_{k})\right)^2\right)\right],
\end{split}
\\[2ex]
\label{eq:v_clipped}
\begin{split}
 V_{\psi, clipped} = V_{\psi_{old}}(\hat{\textbf{s}}_{k})+ \textit{clip}(V_\psi(\hat{\textbf{s}}_{k})-V_{\psi_{old}}(\hat{\textbf{s}}_{k}),-\eta,\eta).
\end{split}
\end{align}
\end{subequations}
Here $V_{target}(\textbf{h}_{k})$ denotes the computed value function from the training samples, and $V_{\psi_{old}}$ is the value function defined by the parameters $\psi_{old}$, which are the parameters from the previous iteration. The value function loss thus quantifies the mismatch between the predicted and computed state values. The hyperparameter $\eta$ limits the magnitude of the updates of the value function parameters $\psi$.

At each control step, specifying the baseline in Eq.~\ref{eq:advantage_gae} as the cumulative discounted reward of all future control steps aids in finding policies that are robust to the epistemic uncertainty in the environment. This is because the maximization of Eq.~\ref{eq:exp_return}, and the corresponding outperformance of the baseline, requires reduction of this uncertainty. This reduction of epistemic uncertainty based on observed data can be viewed as an implicit data assimilation.

A Kullback–Leibler (KL) divergence penalty is incorporated in PPO to improve the stability of the policy updates. KL divergence measures the difference between the old and new policy, and the KL divergence penalty term ($L_{kl}$) acts to minimize this difference to avoid large policy updates. An entropy penalty $L_{ent}$ is also added to the PPO loss. This entropy measures the diversity of the action distribution of the policy. The negative-entropy penalty term improves the diversity of the action distribution, which enhances the exploration of the search space. 

The PPO loss, denoted $L_{ppo}$, with the four components described, is given by
\begin{equation}
    L_{ppo} = L_{\pi_{\theta}}+c_{vf}L_{vf}+c_{kl}L_{kl}+c_{ent}L_{ent},
    \label{eq:total_loss}
\end{equation}
where the coefficients $c_{vf}$, $c_{kl}$, and $c_{ent}$ are weighting factors for the value function, KL divergence, and entropy terms. At each PPO iteration, it would be computationally expensive to simulate all geological models in $M_1$ for the minimization of Eq.~\ref{eq:total_loss}. Instead, we select a set of realizations that capture the flow behavior of the full set of geological models. We apply the procedure described in \cite{shirangi2016general} to divide the geological models into clusters with similar flow characteristics. This entails extracting flow-based features from simulations of all geological realizations. These features are then used as input to a k-means clustering algorithm \cite{hartigan1979algorithm}. This clustering is done once before the policy optimization. At each PPO iteration, the training data are generated through simulations involving realizations sampled equally from each cluster, with well settings as proposed by the current policy. We consider $\mathcal{O}(1000)$ prior geological models, and $\mathcal{O}(100)$ are simulated in each policy training iteration. 

The number of clusters is determined through a scree plot \cite{james2013introduction}. Because the policy optimization can overfit to the geological models used in the optimization, the best policy during policy optimization might not perform well on geological models sampled from $p(\textbf{m})$ that are not part of the training. In our numerical experiments, we exclude the representative geological models (centroids of each cluster) from the training. These representative geological models, determined using a k-medoid clustering algorithm \cite{gordon1998partitions}, are used for control policy selection after training. The control policy selection entails simulating the representative models with well settings proposed by control policies at different iterations of the policy optimization. The control policy with the highest expected return $\overline{V}_{\pi_{\theta}}$ (Eq.~\ref{eq:exp_return}) is selected as the optimal policy.

\subsection{Policy and value function representation}
\label{sec:pol_val_rep}

\begin{figure}[t]
	\centering
	\includegraphics[width=1\textwidth]{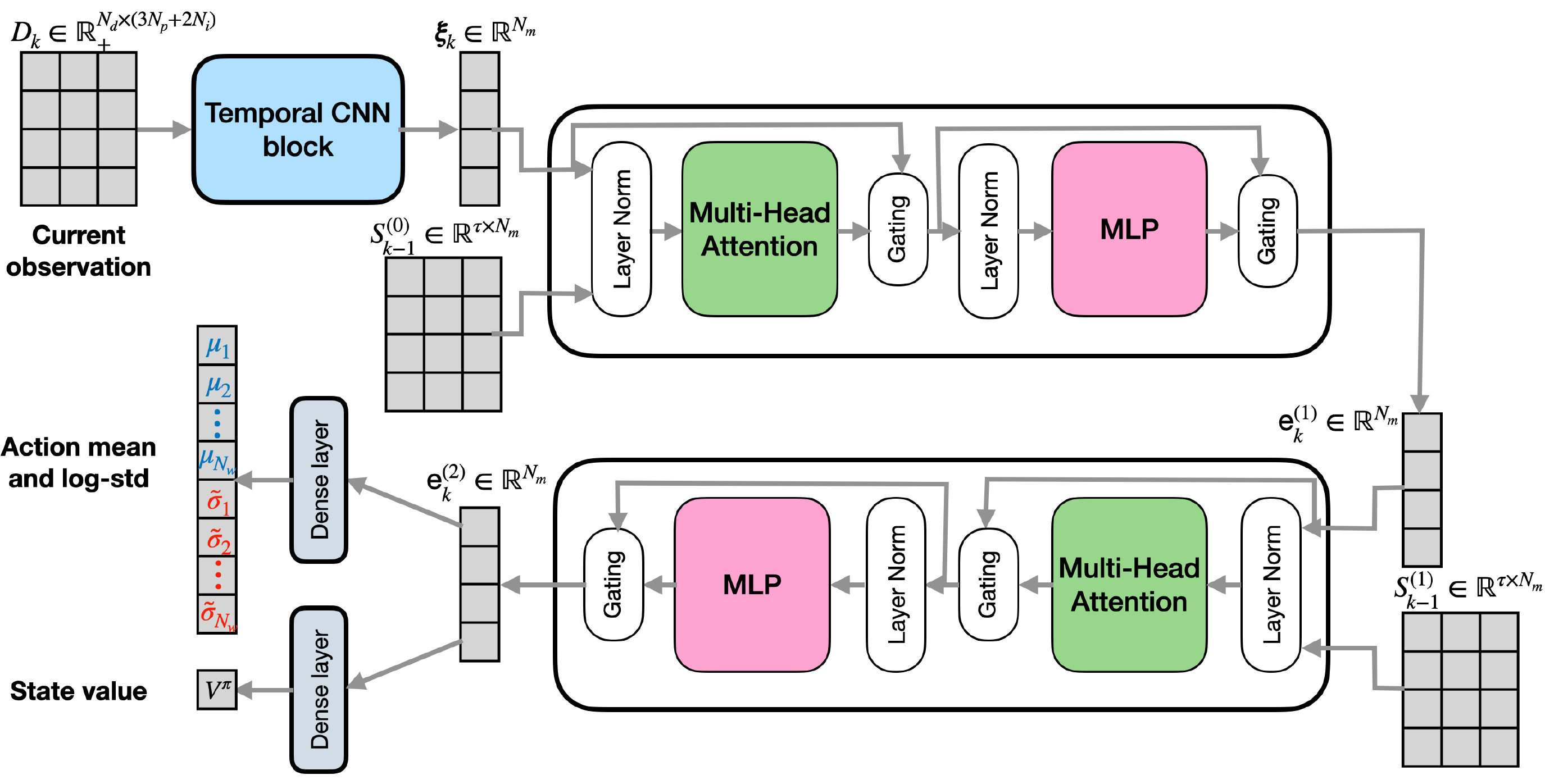}
	\caption{Policy and value functions represented by temporal block and two-layer gated transformer blocks.}
	\label{fig:network}
\end{figure}

The policy and value functions must be parameterized to utilize the PPO algorithm. We represent the policy ($\pi_\theta$) and value ($V_\psi$) functions by the neural network architecture shown in Fig.~\ref{fig:network}. This architecture is comprised of a temporal convolution block and a gated transformer block, of the type recently proposed by Parisotto et al.~\cite{parisotto2020stabilizing}.

As discussed earlier, the observations $\textbf{d}_{k} \in \mathbb{R}_{+}^{N_{d}(3N_p + 2N_i)}$ include the flow rates and BHPs at $N_d$ regular intervals between the starting points of control steps $k-1$ and $k$. This introduces a secondary time level (the control steps are the primary level) in our formulation. To maintain the temporal dependencies of the quantities in the observed data, we reshape $\textbf{d}_{k}$ to a matrix $D_{k} \in \mathbb{R}_{+}^{N_{d} \times (3N_p + 2N_i)}$. The matrix $D_{k}$ serves as input to the temporal convolution block, which is comprised of 1D convolutional neural network (CNN) layers that capture the temporal structure of the data. The temporal block outputs a latent vector $\boldsymbol{\xi}_k \in \mathbb{R}^{N_m}$, which is a compact representation of the observations. Here $N_m$ is the dimension of the extracted features.

The gated transformer block contains $L$ layers, each with relative multihead attention (RMHA) and multilayer perceptron (MLP) submodules. The state of the agent is represented by memories produced at the different layers of the gated transformer block. The previous $\tau$ agent states for the $L$ layers in the gated transformer block ($\hat{S}_{k-1} \in \mathbb{R}^{L \times \tau \times N_m}$), and the latent representation of the current observation ($\boldsymbol{\xi}_k \in \mathbb{R}^{N_m}$), serve as inputs to the gated transformer block in the current control step $k$. Note that at the first control step, the agent states $\hat{S}_{0}$ are initialized to a zero matrix. We set $\tau=N_c$, which means that, at the final control step, the agent can `see' its states in all previous control steps.

The operations in layer $l$ of the gated transformer block \cite{parisotto2020stabilizing} at control step $k$ are expressed as 
 \begin{subequations}
\begin{align}
\label{eq:rmha_ops} 
\begin{split}
 \hat{\textbf{y}}^{(l)}_k = \textnormal{RMHA}^{(l)}(\textnormal{LayerNorm([StopGrad}(\hat{S}_{k-1}^{(l-1)}),\ \textbf{e}^{(l-1)}_k])),
\end{split}
\\[2ex]
\label{eq:gate_rmha} 
\begin{split}
\textbf{y}^{(l)}_k = g_r^{(l)}(\textbf{e}^{(l-1)}_k, \ \textnormal{ReLU}(\hat{\textbf{y}}^{(l)}_k)),
\end{split}
\\[2ex]
\label{eq:pwmlp_ops}
\begin{split}
\hat{\textbf{e}}^{l}_k = \textnormal{MLP}^{(l)}(\textnormal{LayerNorm}(\textbf{y}^{(l)}_k)),
\end{split}
\\[2ex]
\label{eq:gate_pmlp} 
\begin{split}
\textbf{e}^{(l)}_k = g_p^{(l)}(\textbf{y}^{(l)}_k, \ \textnormal{ReLU}(\hat{\textbf{e}}^{(l)}_k)).
\end{split}
\end{align}
\end{subequations}
Here the input to layer $l$ is the embedding $\textbf{e}^{(l - 1)}_k \in \mathbb{R}^{N_m}$ and agent states $\hat{S}_{k-1}^{(l-1)}$ from the previous layer. The input embedding to the first layer is $\textbf{e}^{(0)}_k = \boldsymbol{\xi}_k$, and agent memory $\hat{S}_{k-1}^{(0)} = \boldsymbol{\xi}_{k-1:k-\tau}$ is the compact representation of the observations from the previous $\tau$ control steps. The agent memory $\hat{S}_{k-1}^{(l-1)}$ provides a fixed historical context to layer $l$. The StopGrad function therefore ensures $\hat{S}_{k-1}^{(l-1)}$ is treated as a constant input and is not taken into account in the computation of gradients in layer $l$. The LayerNorm function~\cite{ba2016layer} normalizes the matrix resulting from the concatenation of $\hat{S}_{k-1}^{(l-1)}$ and $\textbf{e}^{(l - 1)}_k$. This has been shown to improve the training of neural networks. 

The RMHA submodule (Eq.~\ref{eq:rmha_ops}) performs $H$ parallel attention \cite{vaswani2017attention} operations on the input, which consists of the previous set of agent states and embedding from the previous layer. The attention operation converts each previous memory and embedding to a feature vector. The output from the attention operation is a weighted sum of the feature vectors for all inputs, with higher weights assigned to more relevant features. This enables extraction of features from the previous agent states and embedding that are relevant to the computation of the current agent state. The output vectors from the $H$ attention operations are concatenated and passed through a fully connected (dense) layer to produce $\hat{\textbf{y}}^{(l)}_k \in \mathbb{R}^{N_m}$. 

ReLU denotes the rectified linear unit activation. The functions $g_r$ and $g_p$ are gating layer functions, represented by a gated recurrent unit (GRU) \cite{chung2014empirical}, used to improve the stability of the optimization. The MLP submodule consists of two fully connected layers that process the output of the RMHA submodule after gating. The agent state $\hat{\textbf{s}}^{(l)}_k$ is prescribed to be the resulting embedding from each layer, $\textbf{e}^{(l)}_k$. 
    
As shown in Fig.~\ref{fig:network}, the embedding from the final layer of the gated transformer, $\textbf{e}^{(L)}_k = \hat{\textbf{s}}^{(L)}_k$, is processed by a fully connected layer to obtain the action mean $\boldsymbol{\mu}_k \in \mathbb{R}^{N_w}$ and action log-standard deviation $\tilde{\boldsymbol{\sigma}}_k \in \mathbb{R}^{N_w}$ for the action distribution at control step $k$. The output from the neural network is prescribed to be the log-standard deviation, instead of the standard deviation, because the output quantity can be negative (after exponentiation, the resulting standard deviation will be positive, as required). Note that the components of the output from the neural network that define the action mean are in linear scale. During training, actions are sampled from the action distribution according to Eq.~\ref{eq:action_sampling} with the action mean $\boldsymbol{\mu}_k$ and standard deviation $\boldsymbol{\sigma}_k = \exp(\tilde{\boldsymbol{\sigma}}_k)$. The agent state $\hat{\textbf{s}}^{(L)}_k$ is also processed by a separate fully connected layer to determine the scalar value of the state. 

The 1D CNN layers in the temporal convolution block have 64 filters with a filter size of 2 in the first layer and 3 in the second layer. We set $H = 2$, $N_m = 128$ and $L = 2$. The first dense layer in the MLP submodule has 64 units and the second has 128 units. The full architecture involves a total of approximately 618,000 parameters.

\section{Computational results}
\label{sec:comp_res}

In this section, we apply the control policy-based CLRM procedure to two example cases. The first example involves channelized 2D models generated from a single geological scenario, while the second example involves 3D models characterized by five different (channelized) scenarios. The control policy approach is compared to robust optimization over prior geological models, to deterministic realization-by-realization optimization and, in the first case, to traditional CLRM.

\subsection{Problem setup}

The flow simulation and optimization setup is similar for the two example cases. We consider oil-water flow, with production driven by water injection. The initial reservoir pressure is set to 350~bar and the initial oil and water saturations are 0.85 and 0.15, respectively. Oil and water viscosities at the initial reservoir pressure are 1~cp and 0.3~cp, respectively. The phase relative permeabilities, which are functions of water saturation, are given as
\begin{equation}
  k_{rw}(S_w) = k_{rw}^{0} \bigg(\frac{S_w - S_{wr}}{1 - S_{wr} - S_{or}}\bigg)^{a}, \ \  k_{ro}(S_w) = k_{ro}^{0} \bigg(\frac{1 - S_w - S_{or}}{1 - S_{wr} - S_{or}}\bigg)^{b},
    \label{eq:relperm}
\end{equation}
where $k_{rw}^{0} = 0.6$, $k_{ro}^{0} = 0.9$, $S_{wr} = S_{or} = 0.15$, and $a = b = 2$. Porosity is taken to be constant and set to 0.2.

The flow simulation involves five production wells and four injection wells. For the first 200~days, producers operate at a fixed BHP of 345~bar, and injectors operate at 400~bar. Production optimization begins at day 200. The goal is to optimize the well settings at seven control steps, each of length 200~days. Thus we have a total production life of 1600~days. There are nine decision variables at each control step (63 decision variables in total). Producer BHPs are constrained to be between 280 and 345~bar, and injector BHPs between 370 and 500~bar. A maximum well liquid production rate constraint of 1526~m$^3$/day is imposed on each production well. The economic parameters are specified as $p_{o} = \$386$/m$^3$ (\$55/STB), $c_{pw} = c_{iw} = \$31$/m$^3$ (\$5/STB), and $b = 0.1$.  

The PPO hyperparameters were determined from a set of numerical experiments. The coefficients for the PPO loss are set to $c_{vf}= 1$ and $c_{kl} = 0.2$, and $c_{ent}$ is specified to decay linearly from $5 \times 10^{-4}$ to $10^{-7}$ at the last iteration. The use of a decaying entropy coefficient allows for the control of the level of exploration at different stages of the training. The PPO implementation in the open-source Ray RLlib library~\citep{liang2018rllib} is used in this work. The training of the control policy through stochastic gradient descent is achieved using the Adam optimizer \cite{kingma2014adam}. A linear learning rate decay schedule is applied, with an initial learning rate of $10^{-4}$ and a final learning rate of $5 \times 10^{-6}$. The mini-batch size and number of epochs are set to 256 and 15, respectively. 

\subsection{Example 1: 2D channelized models from a single scenario}

\begin{figure}[htbp]
	\centering
	\includegraphics[width=0.4\textwidth]{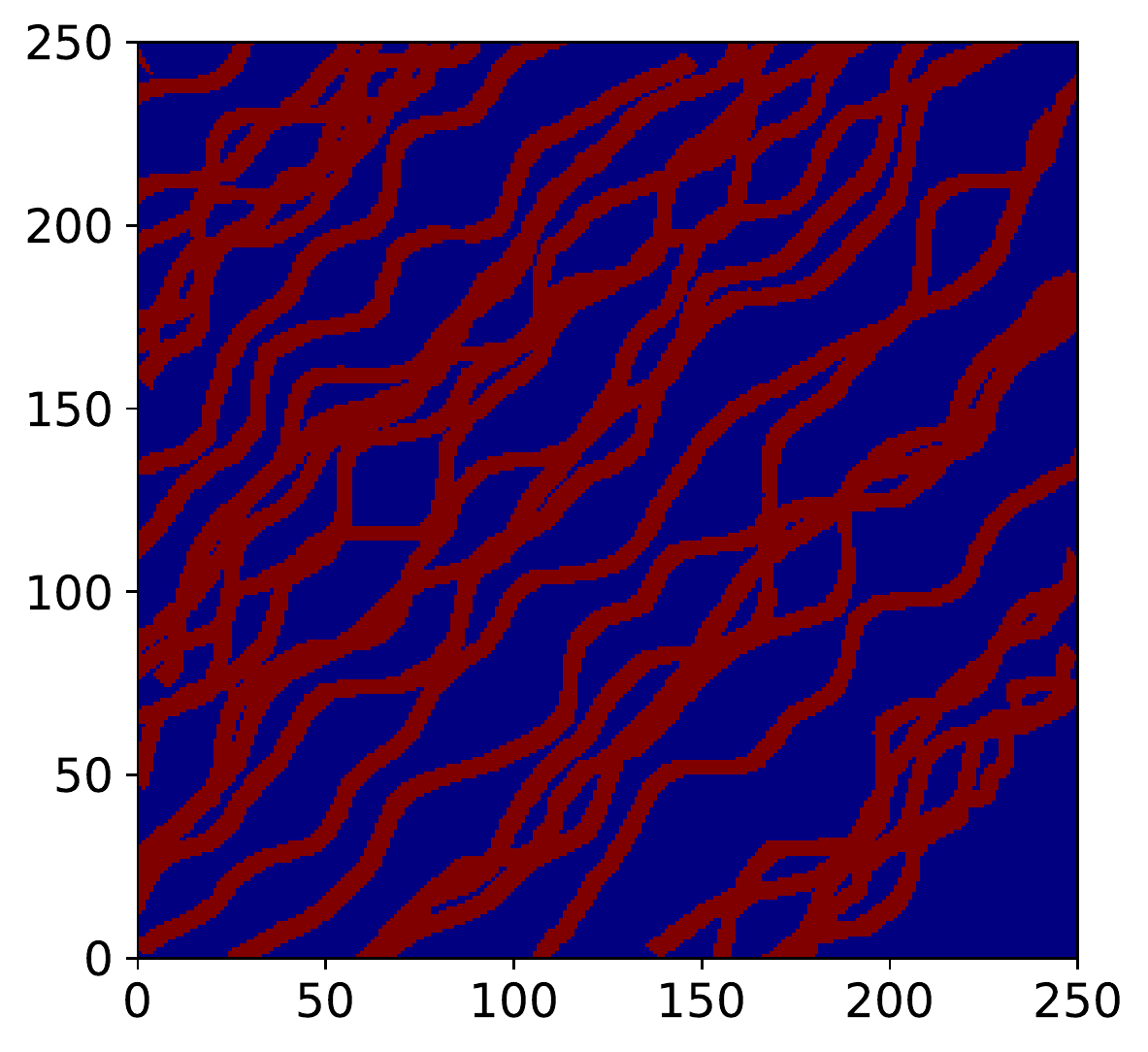}
	\caption{Training image for 2D channelized facies model. Red represents high-permeability channel sand and blue represents low-permeability mud/shale (Example~1).}
	\label{fig:ti}
\end{figure}

\begin{figure}[!htb]
    \centering
    \begin{subfigure}[b]{0.33\textwidth}
        \includegraphics[width=1\textwidth]{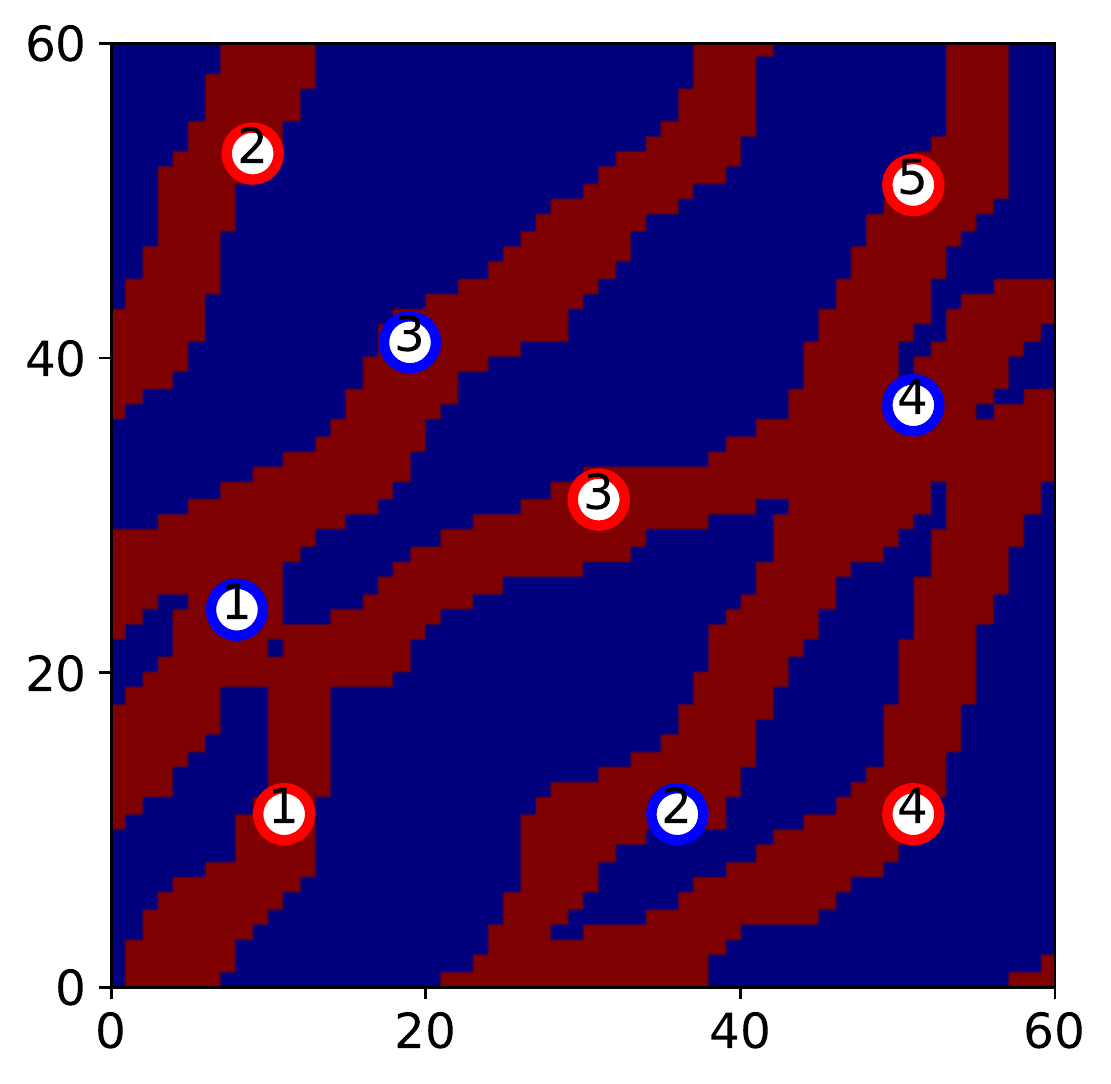}
        \caption{Realization A}
    \end{subfigure}%
    ~
    \begin{subfigure}[b]{0.33\textwidth}
        \includegraphics[width=1\textwidth]{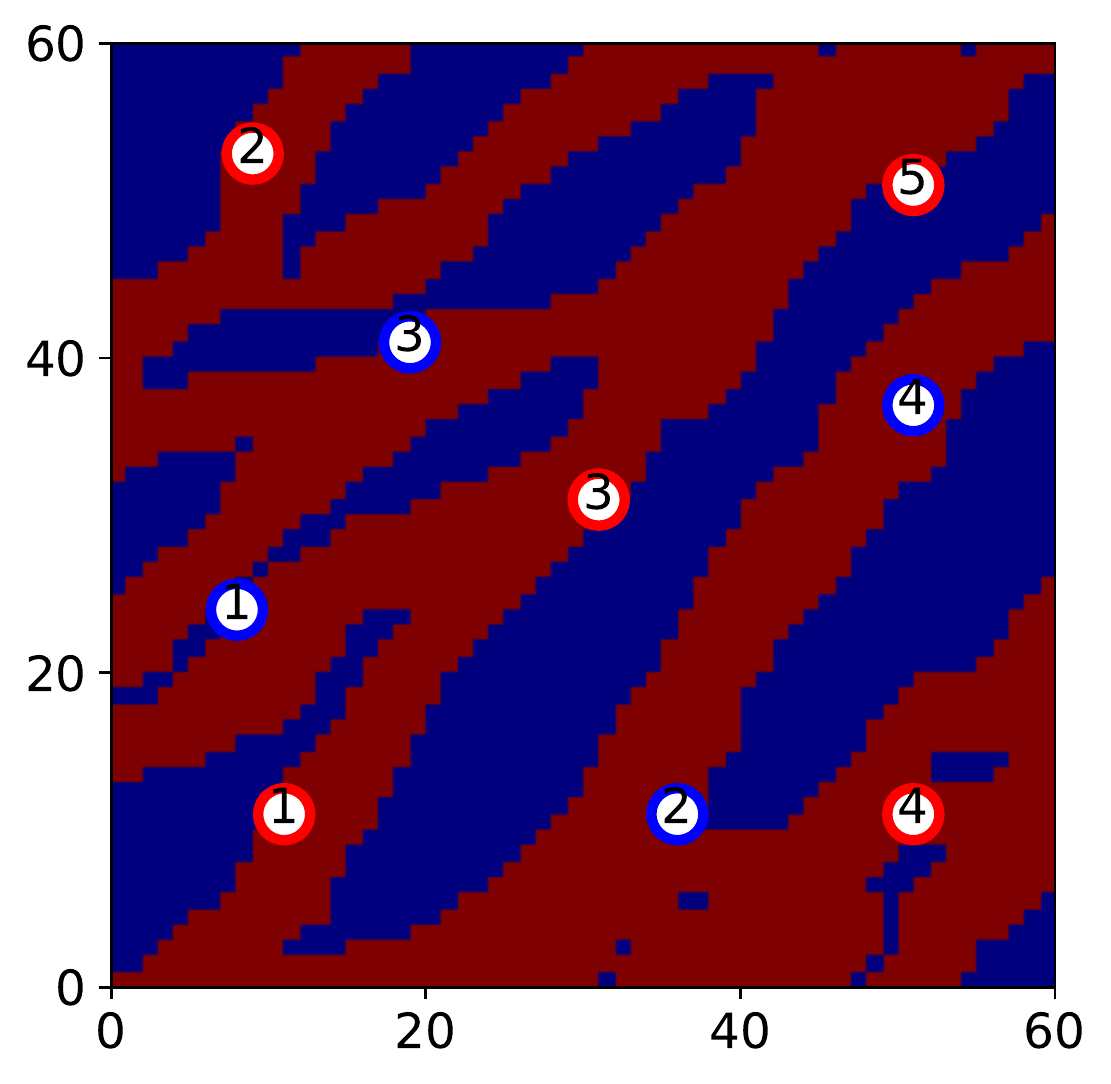}
        \caption{Realization B}
    \end{subfigure}%
    ~
    \begin{subfigure}[b]{0.33\textwidth}
        \includegraphics[width=1\textwidth]{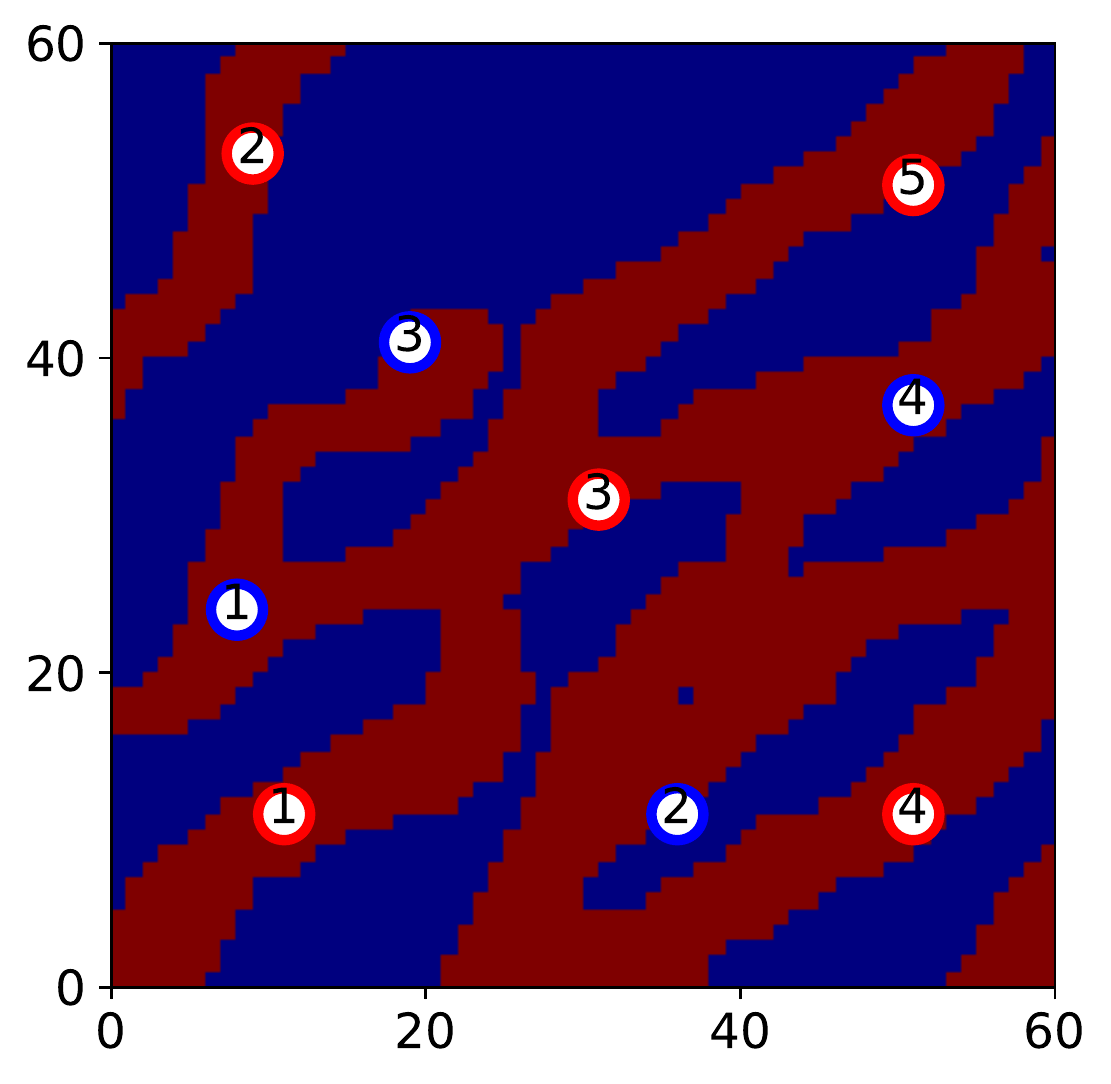}
        \caption{Realization C}
    \end{subfigure}%

    \caption{Three channelized realizations conditioned to facies type at the well locations. Red circles indicate producers and blue circles denote injectors (Example~1).}
    \label{fig:random_realz}
\end{figure}

In this example, we consider 2D binary geological models. The geological features of the channelized system are defined by the training image (from \cite{liu2019deep}) shown in Fig.~\ref{fig:ti}. The training image, defined on a $250 \times 250$ grid, extends over a region much larger than the realizations generated from it. Using this training image, along with conditioning to facies type (sand or mud) at the well locations, we generate 1000 conditional realizations using the SNESIM geostatistical algorithm \cite{strebelle2002conditional}. The models contain $60 \times 60$ cells ($N_b = 3600$, where $N_b$ is the total number of cells), with $\Delta x = \Delta y = 38$~m and $\Delta z = 9$~m. Three realizations of the channelized system are shown in Fig.~\ref{fig:random_realz}. The locations of five producers and four injectors, all located in channel sand (shown in red), are also shown. The mud permeability is specified as 40~md, while the sand permeability is set to 1700~md. Note that the realizations resemble one another in terms of geological style, though the channel locations, and thus the connectivity between wells (via high-permeability channels), differ. 

%

\subsubsection{Control policy training}
\label{sec:pol_training_2d}

We divide the 1000 geological models into 45 clusters using the approach described in Section~\ref{sec:cont_pol_opt}. The control policy is trained with 952 realizations, excluding the centroid (representative realization) of each cluster and the three randomly selected realizations shown in Fig.~\ref{fig:random_realz}. The excluded realizations will be used for evaluating the performance of the control policy after training.

At each training iteration, six realizations are sampled from each cluster and then simulated, which results in $6 \times 45=270$ flow simulations per iteration. These simulations are run in a distributed fashion with 135 processors. The training is terminated after 500 iterations. We thus perform 135,000 total simulations, which require an elapsed time equivalent to that for 1000 sequential simulations. We observe that 500 iterations is sufficient to obtain a policy that performs well on the test set while reducing the risk of overfitting to the training models.

\begin{figure}[t]
	\centering
	\includegraphics[width=0.7\textwidth]{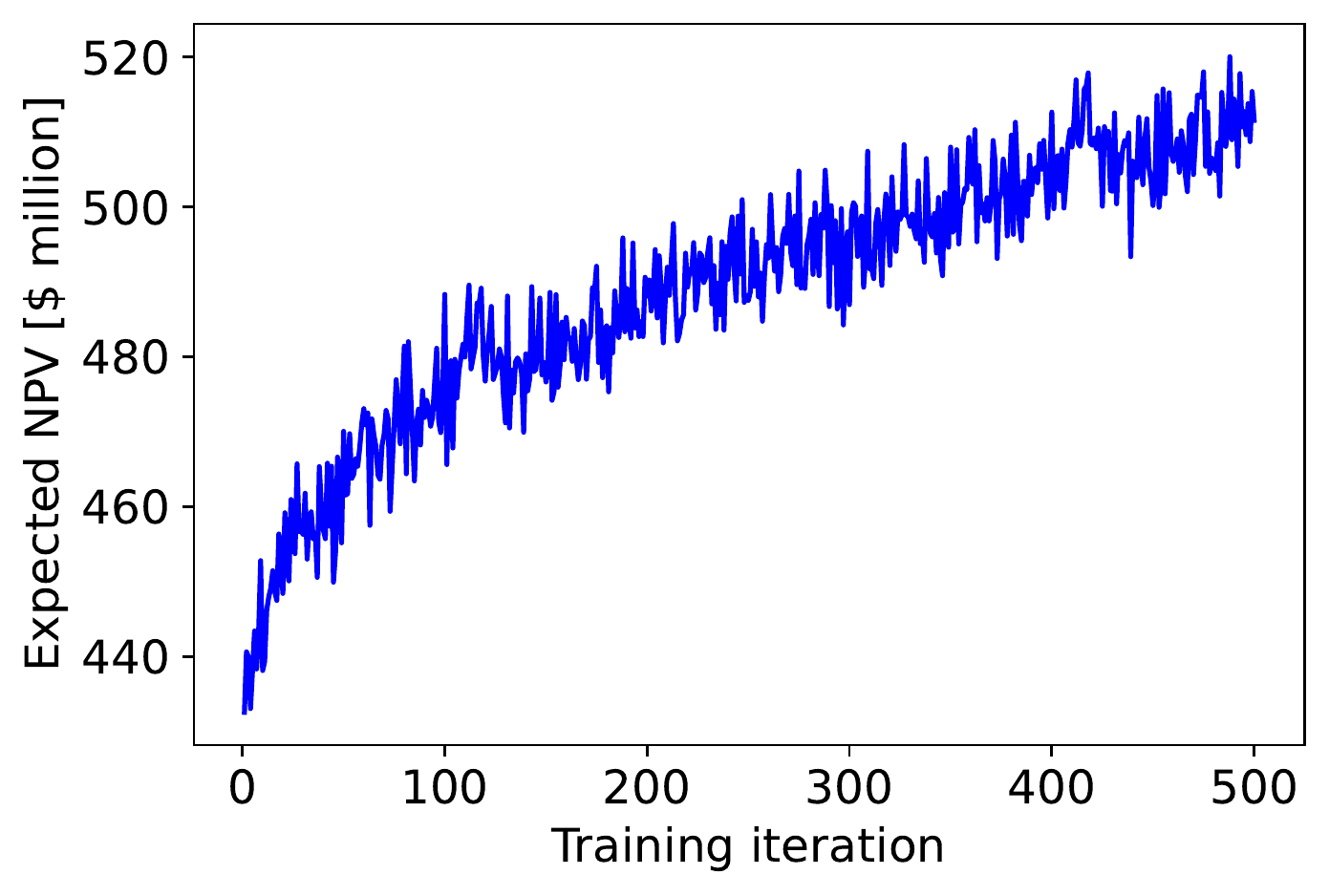}
	\caption{Evolution of expected NPV (Eq.~\ref{eq:exp_return}) computed with the sampled geological models and sampled actions in each training iteration (Example~1).}
	\label{fig:training_npv_chan}
\end{figure}

Figure~\ref{fig:training_npv_chan} displays the evolution of the expected NPV, given in Eq.~\ref{eq:exp_return}, computed with the geological models sampled at the given iteration and the well settings defined by the most recent policy. It is evident that the expected NPV, in general, increases as the training progresses. The fluctuations are due to the sampling of the geological models and the sampling of actions (from the action distribution) during training. The expected NPV of the randomly initialized policy (\$432~million) increases by 20.4\%, to \$520~million, after about 485 iterations. 

\begin{figure}[htbp]
	\centering
	\includegraphics[width=0.7\textwidth]{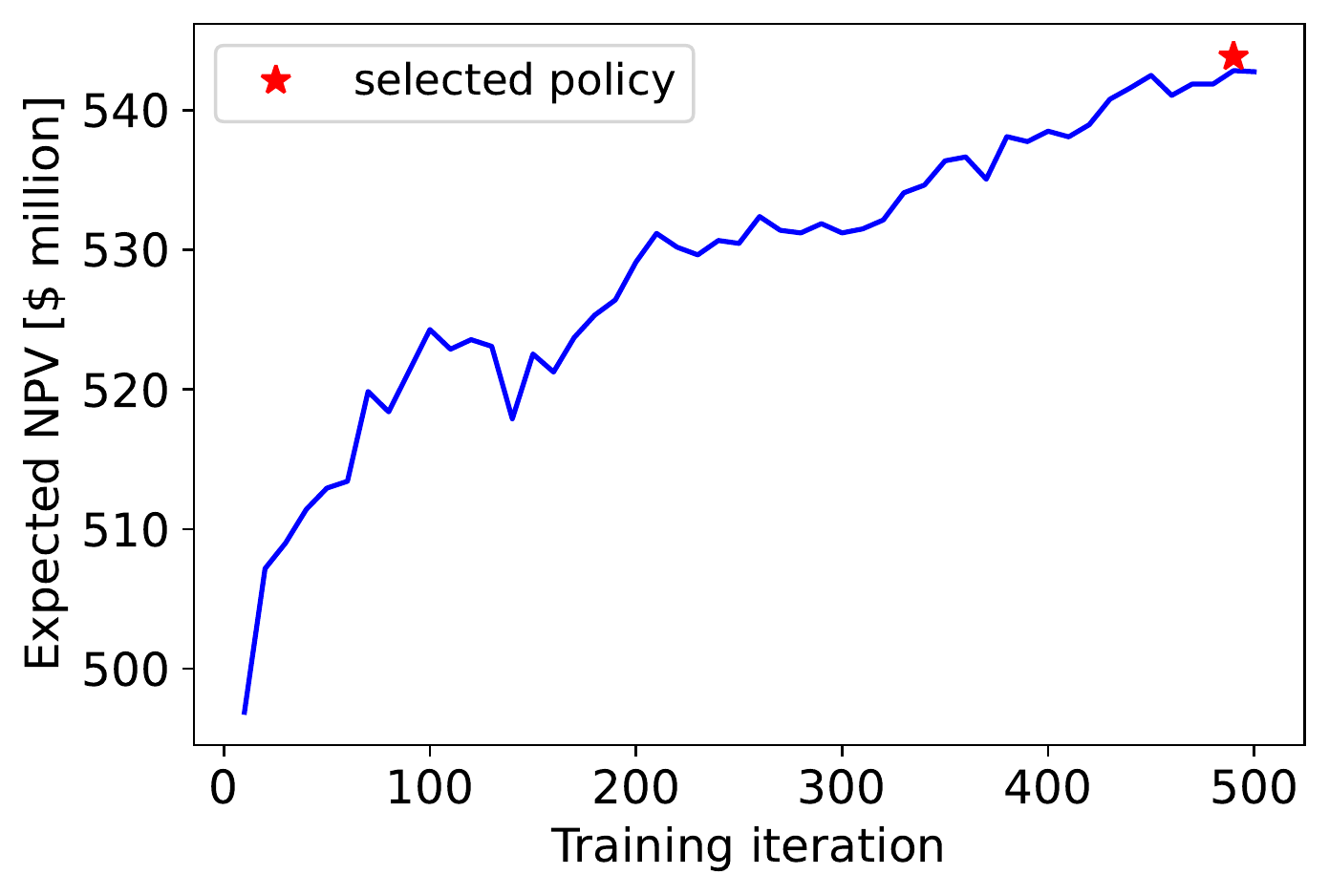}
	\caption{Evolution of expected NPV for the 45 test-case geological models (Example~1).}
	\label{fig:test_npv_chan}
\end{figure}

We next evaluate the performance of the trained policies with the 45 test-case geological models. These correspond to the centroids of the clusters into which the 1000 geological models are assigned. The evolution of expected NPV for these 45 models, obtained by applying the well settings from the most recent policy after every ten iterations, is shown in Fig.~\ref{fig:test_npv_chan}. Importantly, the test curve generally mimics the training performance in Fig.~\ref{fig:training_npv_chan}. 
A deviation from this pattern is observed if we use many more than 500 training iterations. This occurs because the policy is overfitting to the training models. We select the optimal control policy to be that with the highest expected NPV, shown as the red star in Fig.~\ref{fig:test_npv_chan}.


\subsubsection{Comparison of control policy to prior optimization}

In this and the following two subsections, we compare the performance of the DRL-based policy (described in Section~\ref{sec:pol_training_2d}) to three different benchmarks -- robust optimization over prior models, deterministic optimization, and traditional CLRM. In the first approach, considered here, well controls are optimized by applying Eq.~\ref{eq:opt_uncertainty} with prior geological models, i.e., without any model update step. Thus this (robust) optimization does not use any production data, either implicitly (as in the DRL policy) or explicitly (as in CLRM). 

The robust optimization is accomplished using a derivative-free particle swarm optimization –- mesh adaptive direct search (PSO-MADS) hybrid algorithm \cite{isebor2014a}. PSO is a population-based stochastic search method that allows for global exploration of the search space. 
MADS is a pattern-search algorithm that involves local search (polling), in random directions, around the best solution found thus far in the optimization. The hybrid PSO-MADS algorithm has been shown to benefit from the global exploration accomplished by PSO in combination with the local search provided by MADS. Please see \cite{isebor2014a} for a detailed description.

The optimization is performed using the 45 representative prior geological realizations (cluster centroids). The number of PSO particles is set to 50, which results in 2250 simulations (50 particles $\times$ 45 realizations) at each PSO iteration. In the MADS version used here, the number of polling points is twice the number of decision variables. This results in 5670 simulations (2 $\times$ 63 decision variables $\times$ 45 realizations) at each MADS iteration. This optimization is run with 250 processors.

\begin{figure}[htbp]
	\centering
	\includegraphics[width=0.6\textwidth]{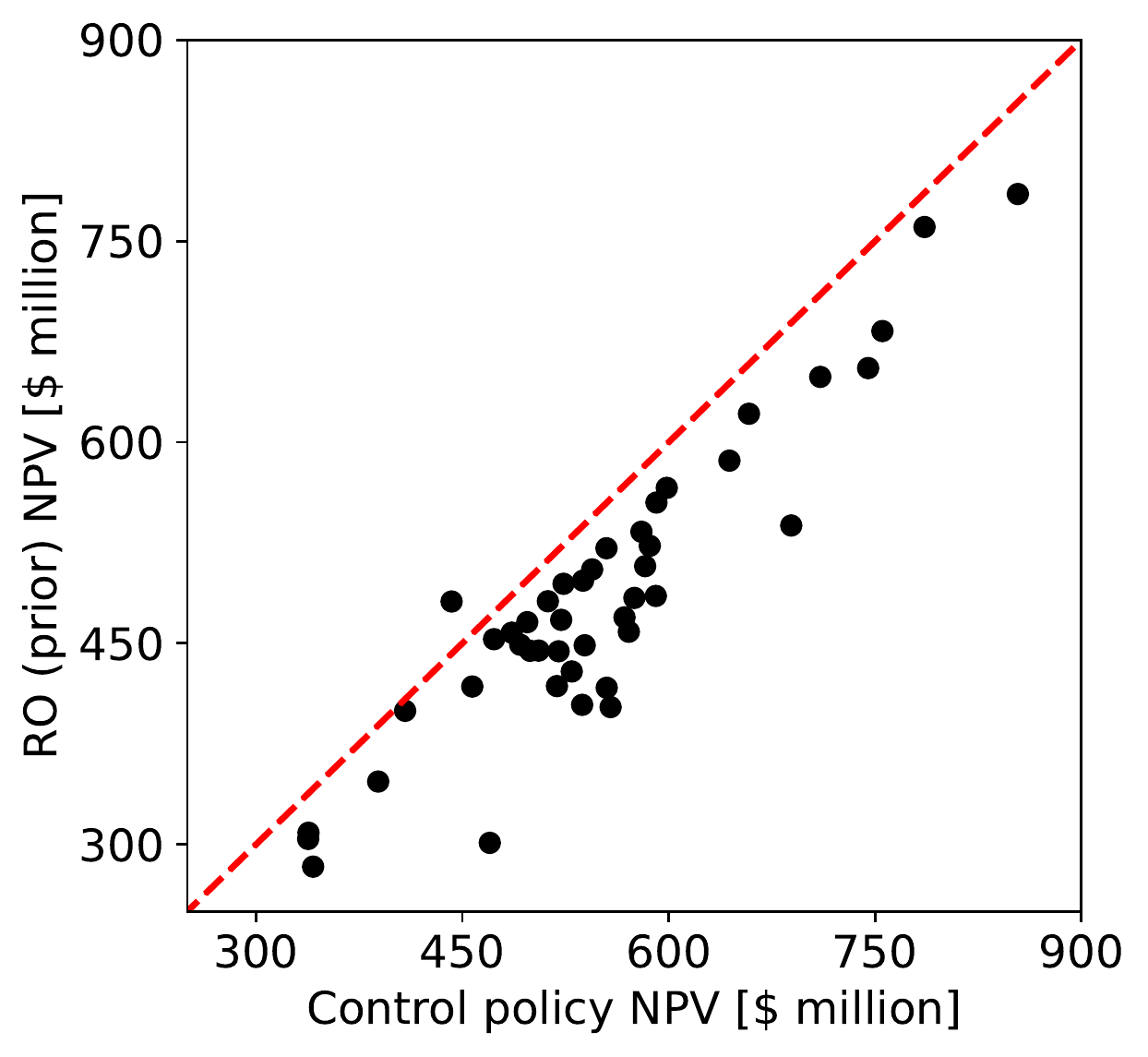}
	\caption{Comparison of solutions from robust (prior) optimization to those using the DRL-based control policy (Example~1).}
	\label{fig:drl_vs_ro_chan}
\end{figure}

The single set of well controls obtained from the robust optimization are applied to each of the 45 geological models. The resulting NPVs are compared to those obtained with the DRL-based control policy. Figure~\ref{fig:drl_vs_ro_chan} presents a cross-plot showing this comparison. It is evident from the figure that, for 44 of the 45 models, the DRL-based control policy outperforms robust optimization over prior models. The control policy leads to an average improvement over all geological models, relative to prior optimization, of \$64.3~million, or 14.7\%. The improvement from the DRL-based approach is due to the ability of the method to (implicitly) tailor the controls to the observed data on a realization-by-realization basis.

\begin{figure}[!htb]
    \centering
    \begin{subfigure}[b]{0.5\textwidth}
        \includegraphics[width=1\textwidth]{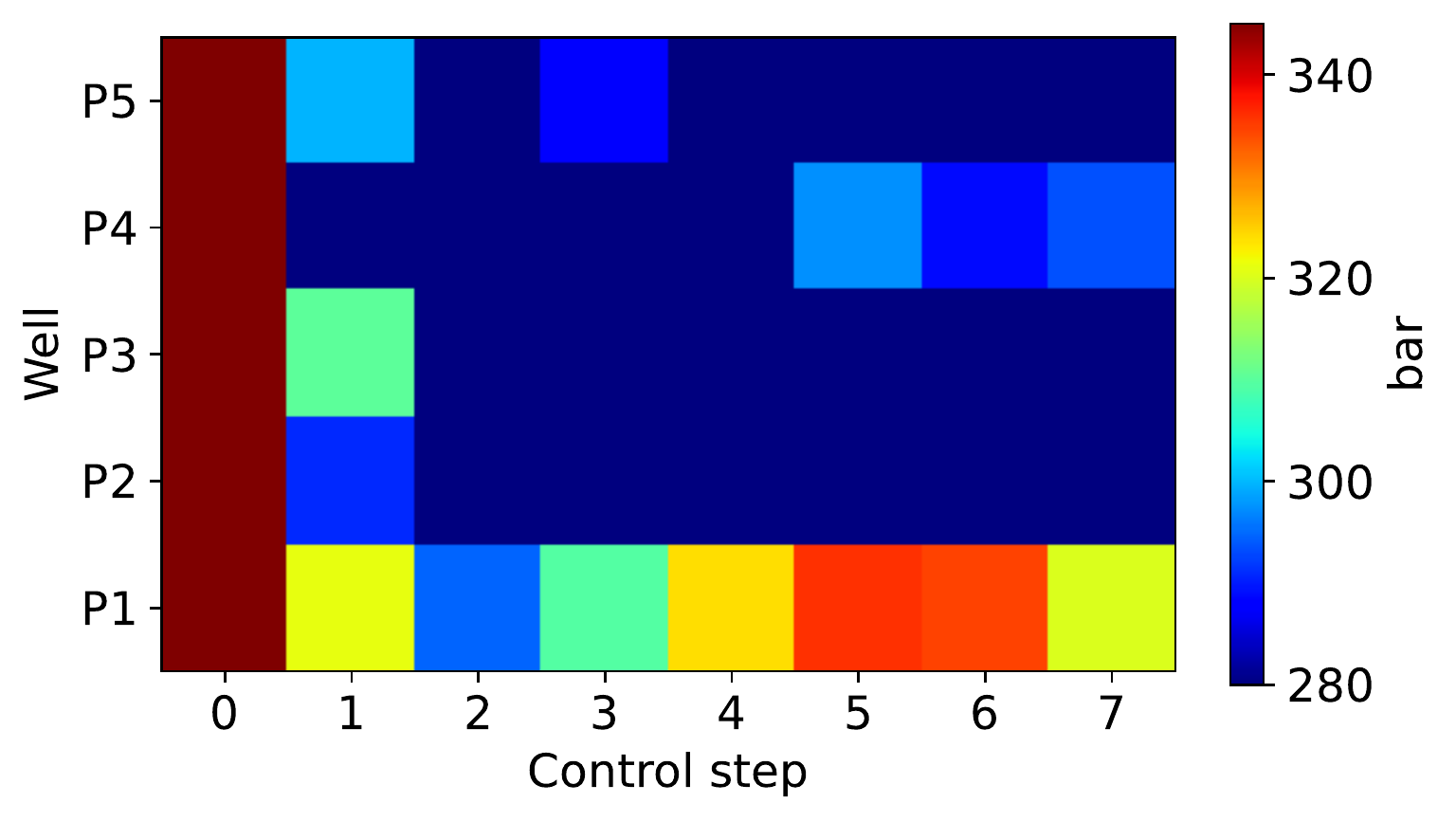}
        \caption{Control policy (P$_{25}$ model)}
    \end{subfigure}%
    ~
    \begin{subfigure}[b]{0.5\textwidth}
        \includegraphics[width=1\textwidth]{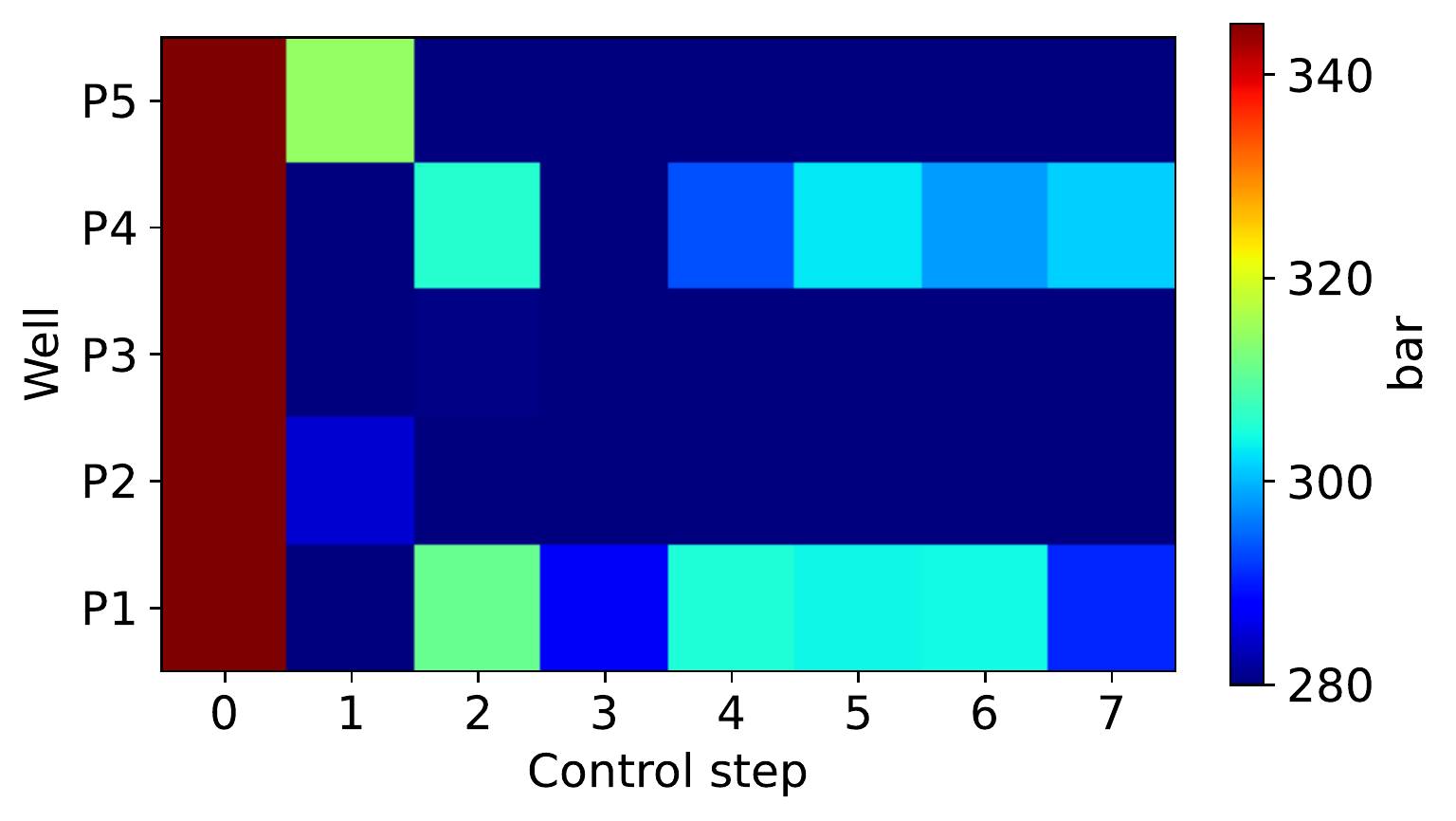}
        \caption{Control policy (P$_{75}$ model)}
    \end{subfigure}%
    
    \begin{subfigure}[b]{0.5\textwidth}
        \includegraphics[width=1\textwidth]{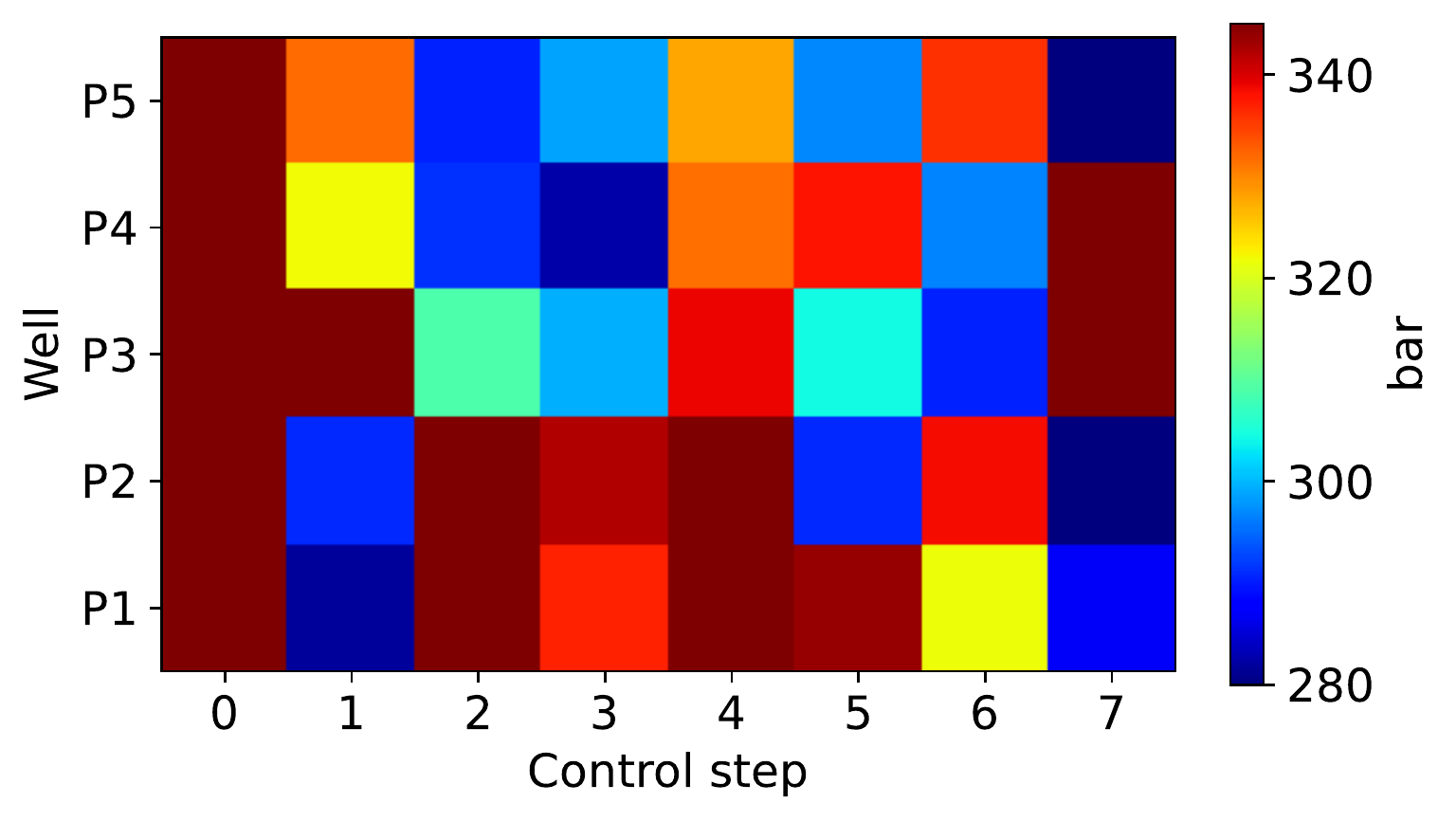}
        \caption{RO (prior)}
    \end{subfigure}%

    \caption{Well settings proposed by the control policy for the P$_{25}$ and P$_{75}$ models determined by ranking the NPVs achieved through robust (prior) optimization (Example~1).}
    \label{fig:prod_bhps_ex_1}
\end{figure}

After applying the well settings obtained from the robust optimization, we rank the 45 geological models based on their NPVs. The production well settings for the 25th and 75th percentile (denoted as P$_{25}$ and P$_{75}$) models from this ranking, obtained through use of the control policy, are shown in Fig.~\ref{fig:prod_bhps_ex_1}(a) and (b). It is evident that the well settings obtained from the control policy are different for the P$_{25}$ and P$_{75}$ models. The well settings obtained from the robust optimization are shown in Fig.~\ref{fig:prod_bhps_ex_1}(c). Interestingly, for this case robust optimization provides BHPs that show more overall variation in time.

\subsubsection{Comparison of control policy to deterministic optimization}

We now compare results from the control policy against those from deterministic optimization for each of the 45 test-case models. This entails optimization of each `true' model individually, assuming the permeability field is known. Although this (deterministic) result cannot be achieved in practice because the geology is always unknown, this allows us to compare the control policy with the theoretically best solution achievable. Due to the high computational cost associated with optimizing each model using PSO-MADS, we use the gradient-based optimization algorithm SNOPT \cite{gill2005snopt} for these runs. Although very efficient, gradient-based approaches may converge to a relatively poor local optimum. For this reason, we run each case three times starting from different initial guesses. In the results below, we show the solution with the maximum NPV from the three runs. 

\begin{figure}[htbp]
	\centering
	\includegraphics[width=0.6\textwidth]{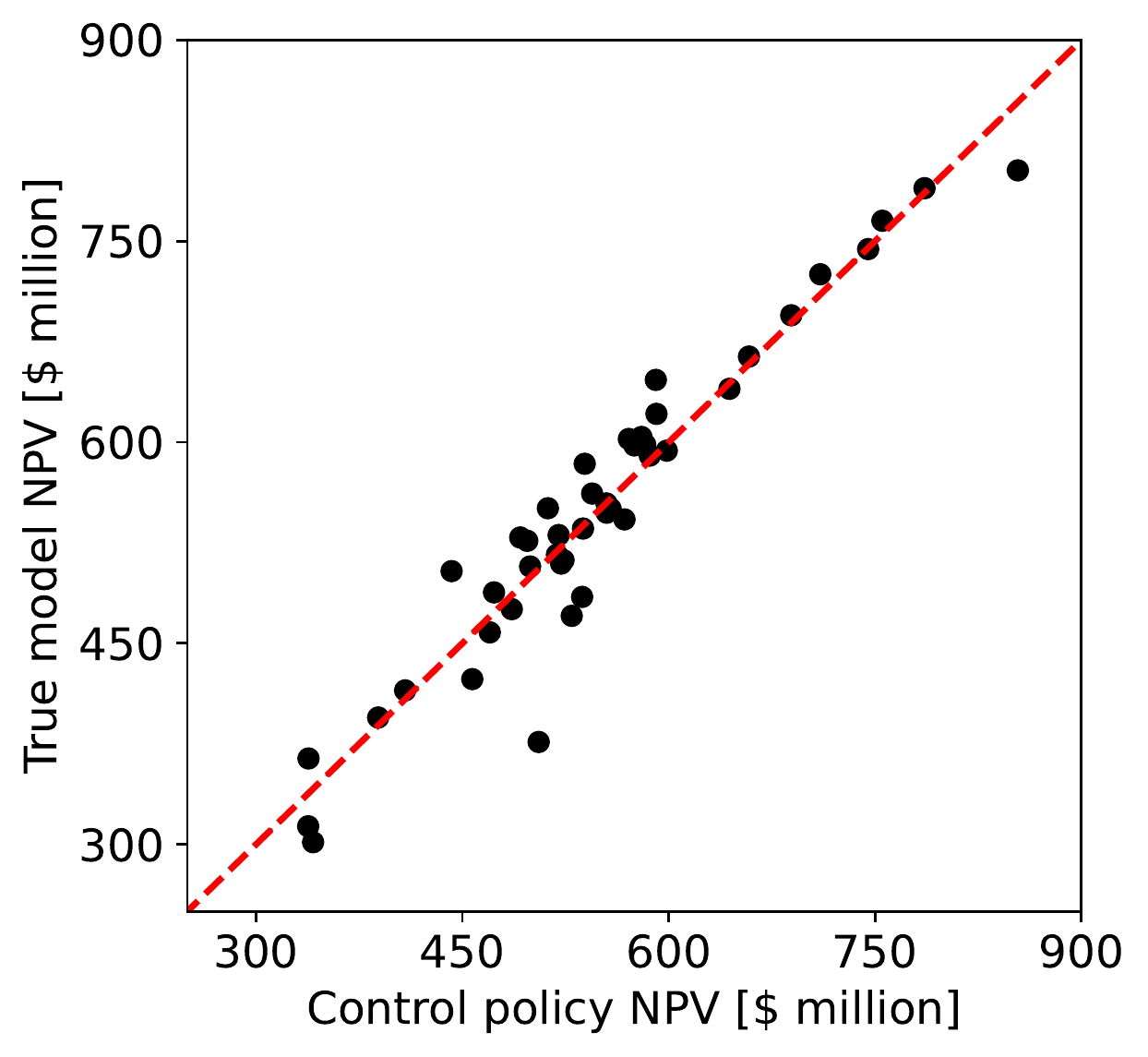}
	\caption{Comparison of solutions from deterministic optimization (performed separately for each `true' model) to those using the DRL-based control policy (Example~1).}
	\label{fig:drl_vs_true_chan}
\end{figure}

Figure~\ref{fig:drl_vs_true_chan} displays a comparison of the deterministic `true' model NPVs to NPVs from the DRL-based control policy. It is evident from Fig.~\ref{fig:drl_vs_true_chan} that the control policy solutions are comparable to those from deterministic optimization. If we were consistently obtaining the global optimum with SNOPT, these NPVs would be at least as high as those from the control policy. This is not achieved in many of the cases, and we see that the control policy approach outperforms deterministic optimization in nearly half (47\%) of the cases. The key observation here, however, is that the control policy approach provides results comparable to those from deterministic optimization. This is significant, because deterministic optimization is not achievable in practice, while the control policy approach uses data that are available in practical settings.

\begin{figure}[!htb]
    \centering
    \begin{subfigure}[b]{0.45\textwidth}
        \includegraphics[width=1\textwidth]{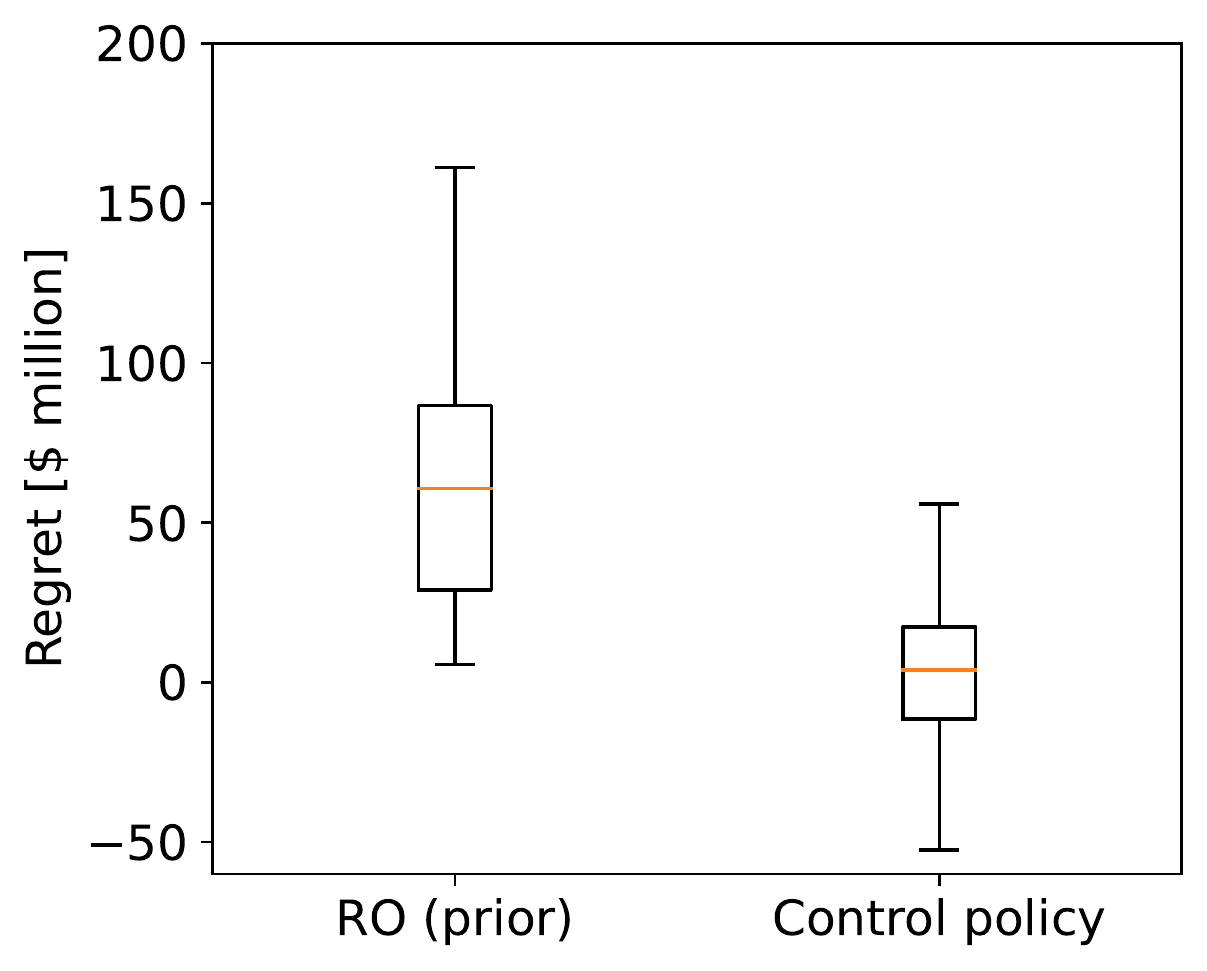}
        \caption{Box plots of regret}
    \end{subfigure}%
    \hspace{2\baselineskip}
    \begin{subfigure}[b]{0.45\textwidth}
        \includegraphics[width=1\textwidth]{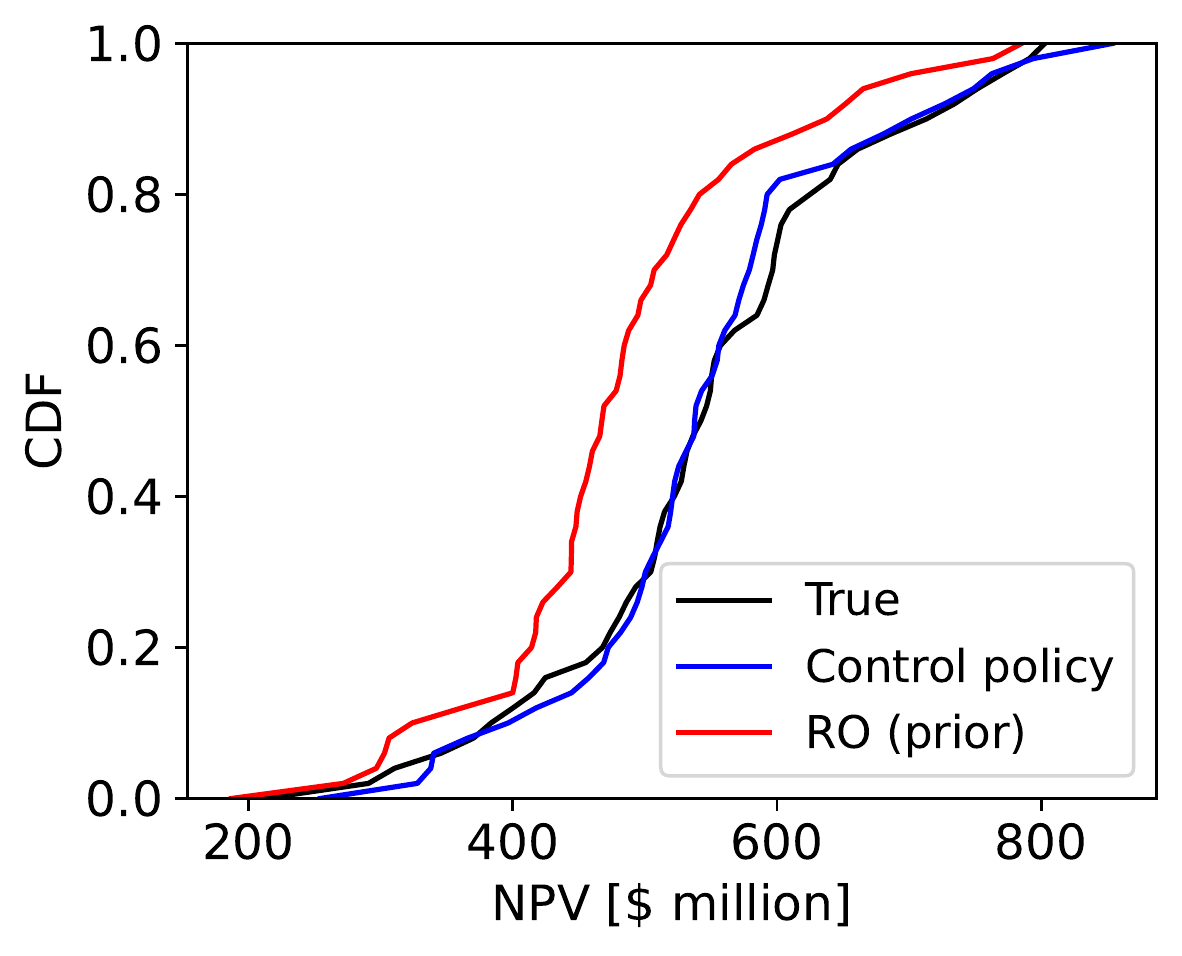}
        \caption{CDFs of the optimum NPVs}
    \end{subfigure}%

    \caption{Comparison of results from the three approaches for the 45 test-case geological models (Example~1).}
    \label{fig:bench_mark_chan}
\end{figure}

We next compute the `regret' using the prior optimization and control policy approaches for the 45 geological models. Regret is computed by subtracting the NPV obtained from that of deterministic optimization (lower regret values are better). Results are presented in terms of box plots in Fig.~\ref{fig:bench_mark_chan}(a). The box plots show the minimum, maximum, P$_{25}$, P$_{50}$ and P$_{75}$ regrets for the control policy and prior optimization procedures. The maximum regret using prior optimization is \$161.2~million, while that of the control policy is \$61.7~million. The use of the control policy also leads to a much lower median regret (\$3.8~million) than prior optimization (\$60.7~million). In Fig.~\ref{fig:bench_mark_chan}(b), we present CDFs of NPV for the 45 geological models obtained from the three procedures. The close correspondence between control-policy results and deterministic optimization results is evident, as is the fact that both approaches outperform prior optimization. 

\subsubsection{Comparison of control policy to traditional CLRM}

We now compare the performance of the DRL-based control policy with the traditional CLRM approach. We perform this comparison for three randomly selected realizations, which act as the `true' models. These models are shown in Fig.~\ref{fig:random_realz}. As noted earlier, these `true' models are not included in the set of the geological realizations used in training the control policy. The observed data $\textbf{d}^{*}$ for each `true' model is given by
\begin{equation}
    \textbf{d}^{*} = \textbf{d}_{true} + \boldsymbol{\epsilon},
    \label{eq:obs_data}
\end{equation}
where $\textbf{d}_{true}$ is the data obtained from simulating the true model and $\boldsymbol{\epsilon}$ is the measurement error (with standard deviations given in Section~\ref{sec:drl}).

The optimizations in traditional CLRM are performed using PSO-MADS. The data assimilation is accomplished using the randomized maximum likelihood (RML) method, with the geological models parameterized using an optimization-based principal component analysis (O-PCA) approach \cite{vo2014new}. O-PCA provides a differentiable representation of the permeability field, which enables the use of a gradient-based method (SNOPT) for the minimizations required by RML. We now briefly describe this history matching procedure.

O-PCA provides a low-dimensional parameterization of the geological models, thus reducing the number of variables that must be determined during history matching. The method essentially provides a post-processing of a standard PCA representation to better characterize models described by non-Gaussian spatial statistics. In O-PCA, we first construct a PCA representation from a set of $N$ prior realizations. Here we use 997 realizations, conditioned to hard data and generated using SNESIM. Each realization is expressed as a vector and inserted (as a column, after centering) into a data matrix $X_c \in \mathbb{R}^{N_b \times N}$ 
\begin{equation}
    X_c = [\textbf{m}_1 - \Bar{\textbf{m}}, \textbf{m}_2 - \Bar{\textbf{m}}, \ldots, \textbf{m}_N - \Bar{\textbf{m}}],
    \label{eq:centered_matrix}
\end{equation}
where $\Bar{\textbf{m}}$ is the mean of the $N$ realizations. A singular value decomposition (SVD) of $X_c$, truncated at $n$ singular values/vectors, is then performed. This allows us to write $X_c \approx U_n \Sigma_n V^T_n$. In PCA, new realizations can be generated through application of $\textbf{m}_{pca} = U_n \Sigma_n \boldsymbol{\xi} + \Bar{\textbf{m}}$, where $\boldsymbol{\xi}$ is the low-dimensional variable. If the goal is to generate new (random) realizations, we sample $\boldsymbol{\xi}$ from $\mathcal{N}(\textbf{0}, \textbf{1})$ (of dimension $n$); if the goal is history matching, $\boldsymbol{\xi}$ is determined such that simulation with the resulting permeability field minimizes a data mismatch.

For the binary case considered here, the elements of \textbf{m} are either 0 (corresponding to mud) or 1 (channel sand). With O-PCA, new realizations are constructed by solving the separable minimization problem given by
\begin{equation}
    \textbf{m}_{opca} = \argmin{\textbf{u}}{||U_n \Sigma_n \boldsymbol{\xi} + \Bar{\textbf{m}} - \textbf{u}||_2^2 \ + \ \gamma \textbf{u}^T(\textbf{1} - \textbf{u})}.
    \label{eq:o_pca_real}
\end{equation}
This representation can again be used to generate random realizations or for history matching. The regularization term $\textbf{u}^T(\textbf{1} - \textbf{u})$ (here $\textbf{1}$ is a unit vector of dimension $N_b$), weighted by $\gamma$, shifts the values of $\textbf{m}_{opca}$ toward 0 or 1. This maintains a reasonable degree of consistency with the original SNESIM realizations $\textbf{m}_{i}$, $i=1, \dots, N$. Importantly, however, the O-PCA representation retains values between 0 and 1, thus enabling Eq.~\ref{eq:o_pca_real} to be differentiated. 

Posterior realizations, defined in terms of $\Bxirml \in \mathbb{R}^n$, are generated through the application of RML. This entails solving the minimization problem 
\begin{equation}
\Bxirml = \argmin{\Bxi} \  \big(\Bd_{\Bxi}-\Bdobs^*\big)^T\Cdinv\big(\Bd_{\Bxi}-\Bdobs^*\big) + \big(\Bxi-\Bxiuc\big)^T\big(\Bxi-\Bxiuc\big),
\label{eq:rml}
\end{equation}
where $\Bd_{\Bxi}$ is the observed data from the flow simulation of the model obtained from the projection of $\Bxi$ to the full model space (using Eq.~\ref{eq:o_pca_real}), $C_{\text{d}}$ is the covariance matrix of the data measurement error, and $\Bxiuc$ is sampled from $\mathcal{N}(\textbf{0}, \textbf{1})$ (of dimension $n$).

The optimization in Eq.~\ref{eq:rml} is performed using the SNOPT algorithm. Details of the O-PCA RML approach for generating posterior models can be found in \cite{vo2015data}. We set $n$ to the value required to explain 85\% of the variance of the random geological models in $X_c$ (which corresponds to $n = 225$). At each data assimilation step, we generate 30 posterior models by solving Eq.~\ref{eq:rml} a total of 30 times, each time with a different prior realization $\Bxiuc \sim \mathcal{N}(\textbf{0}, \textbf{1})$.

\begin{figure}[t]
    \centering
    \begin{subfigure}[b]{0.45\textwidth}
        \includegraphics[width=1\textwidth]{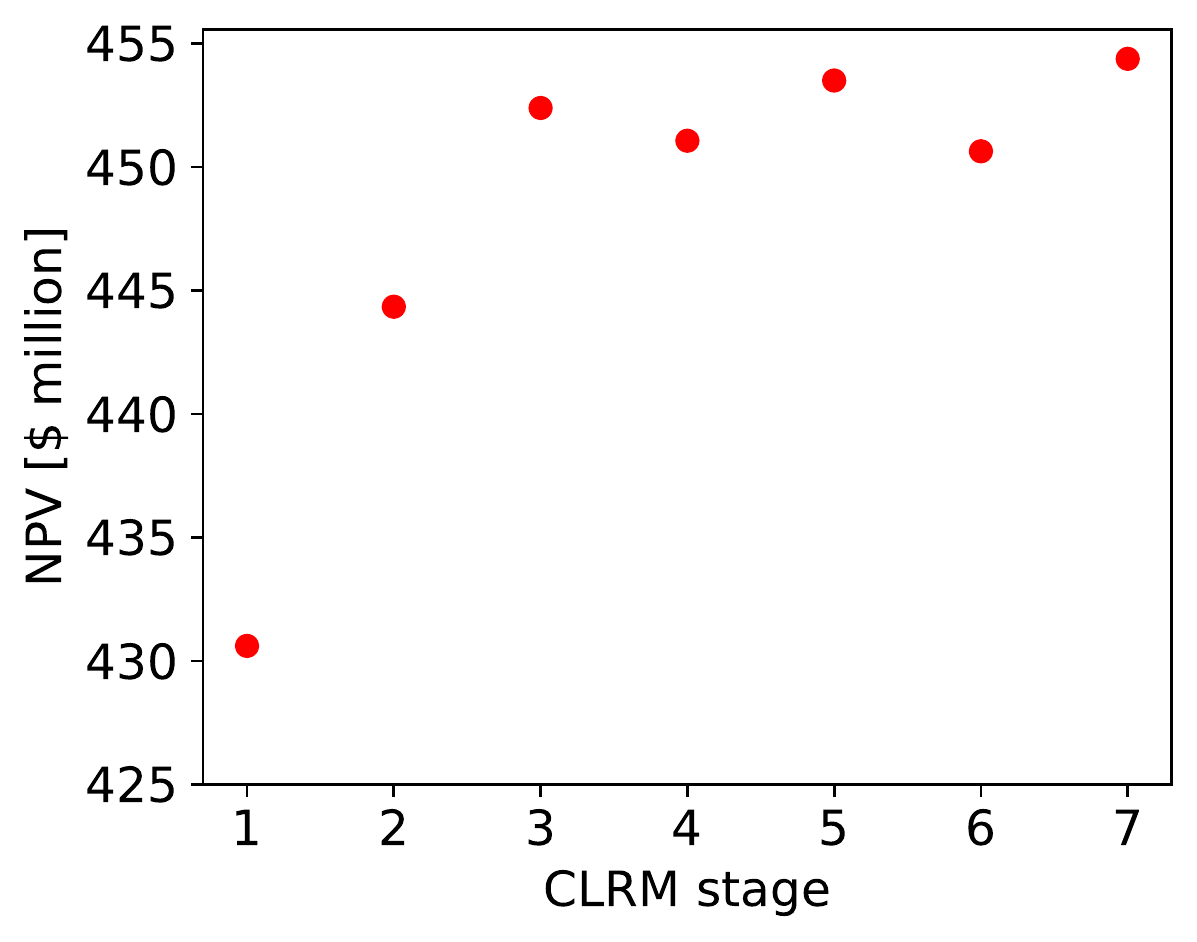}
        \caption{True model A}
    \end{subfigure}%
    \hspace{2\baselineskip}
    \begin{subfigure}[b]{0.45\textwidth}
        \includegraphics[width=1\textwidth]{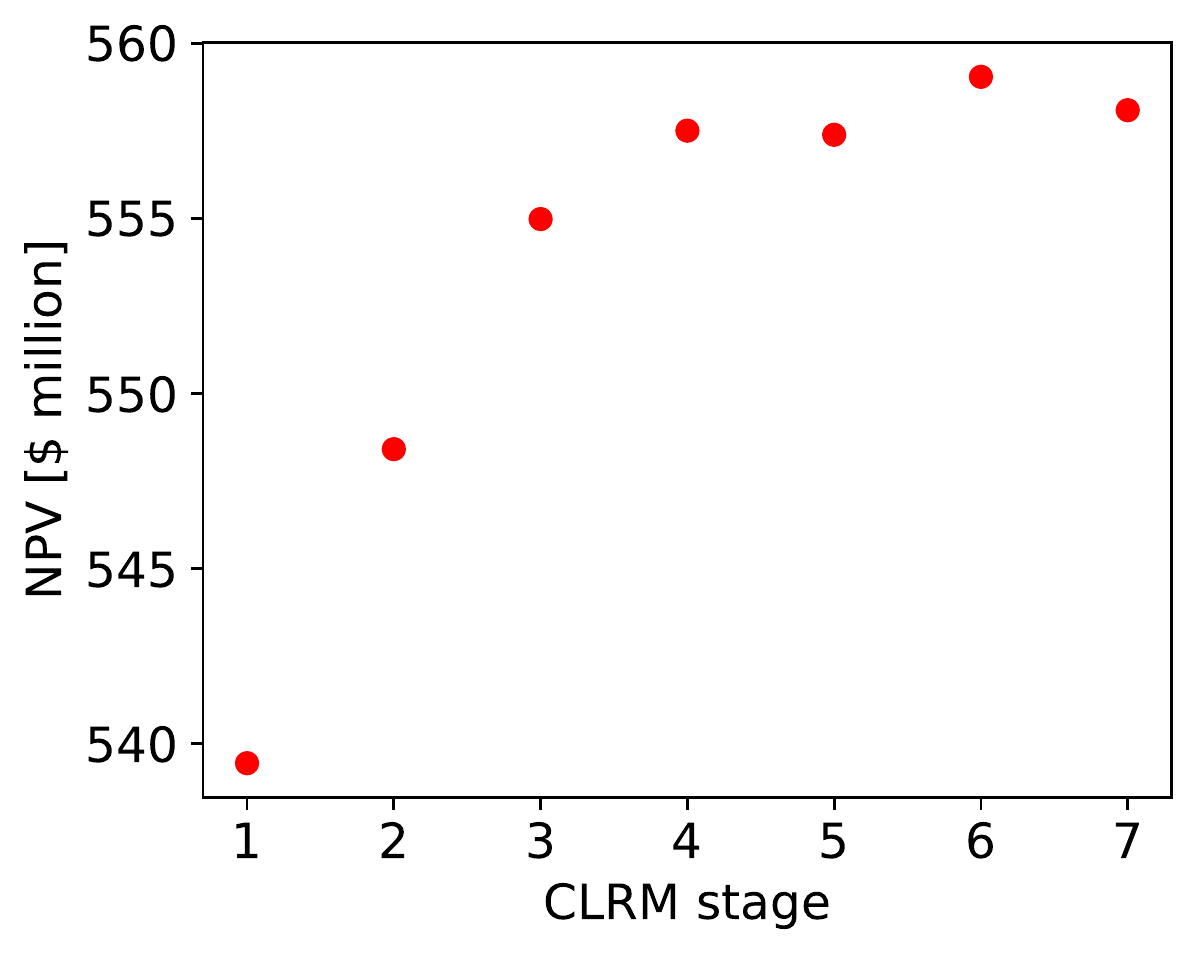}
        \caption{True model B}
    \end{subfigure}%
    
    \begin{subfigure}[b]{0.45\textwidth}
        \includegraphics[width=1\textwidth]{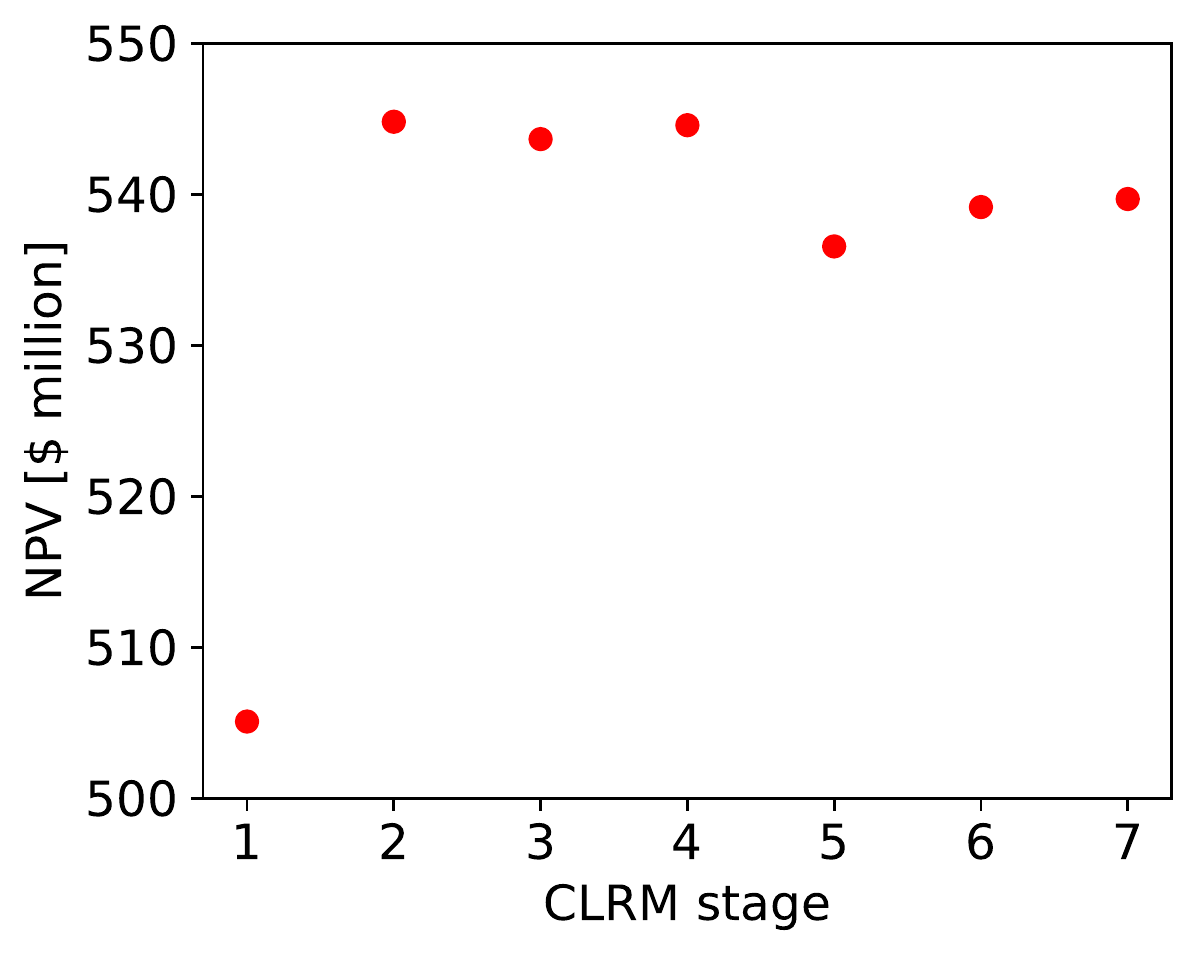}
        \caption{True model C}
    \end{subfigure}%

    \caption{Evolution of NPV for the three true models using traditional CLRM (Example~1).}
    \label{fig:clrm_true_npv}
\end{figure}

Due to the stochastic nature of PSO-MADS and RML, we perform the traditional CLRM three times for each true model. We present results for the best solution from the three runs. The well settings determined after the data assimilation and optimization in each CLRM stage are applied to the true models. The resulting NPVs are shown in Fig.~\ref{fig:clrm_true_npv}. The NPV at each stage is an estimate based on the current (uncertain) model realizations. These shift from stage to stage because, as new data are collected and assimilated, the models change, which results in new well settings at all future stages. It is common to observe significant improvement relative to the first stage, since the geological uncertainty is reduced substantially, relative to the prior, even with limited data. Non-monotonicity in NPV is also common, as the evolving geological models correspond to different optimal well settings and to shifts in NPV.


%

\begin{figure}[!htb]
    \centering
    \begin{subfigure}[b]{0.45\textwidth}
        \includegraphics[width=1\textwidth]{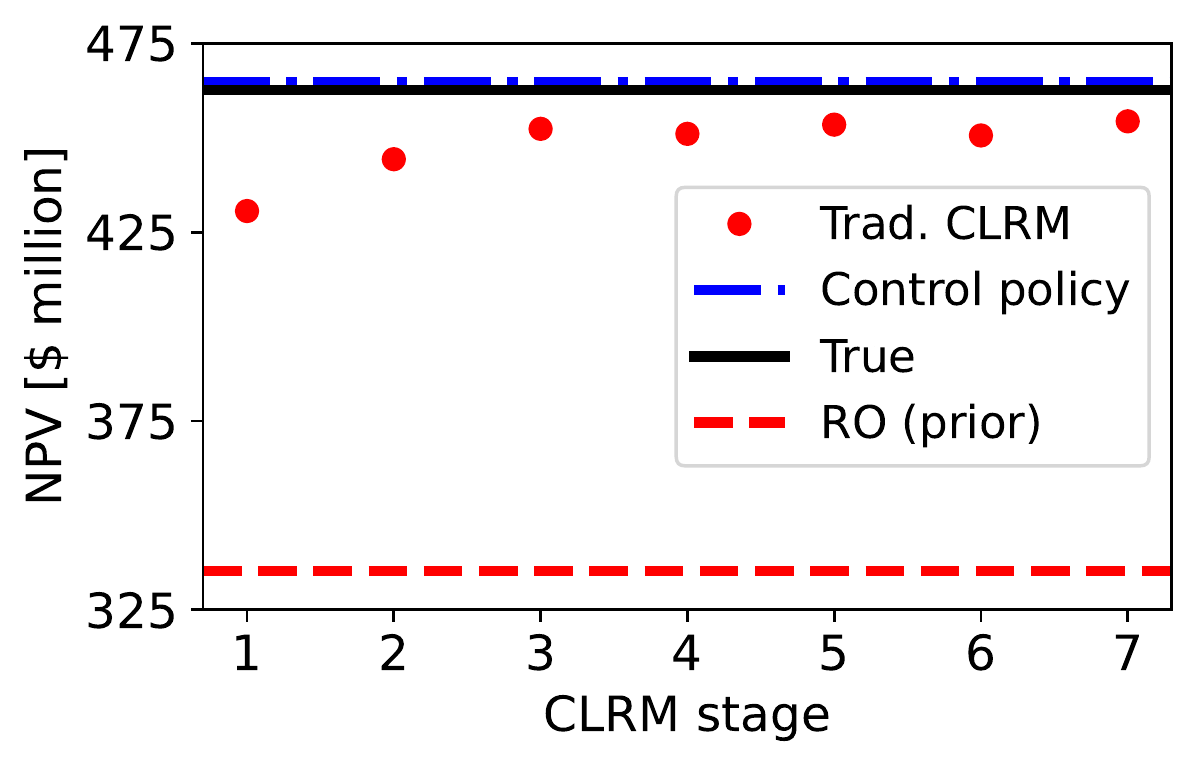}
        \caption{True model A}
    \end{subfigure}%
    \hspace{2\baselineskip}
    \begin{subfigure}[b]{0.45\textwidth}
        \includegraphics[width=1\textwidth]{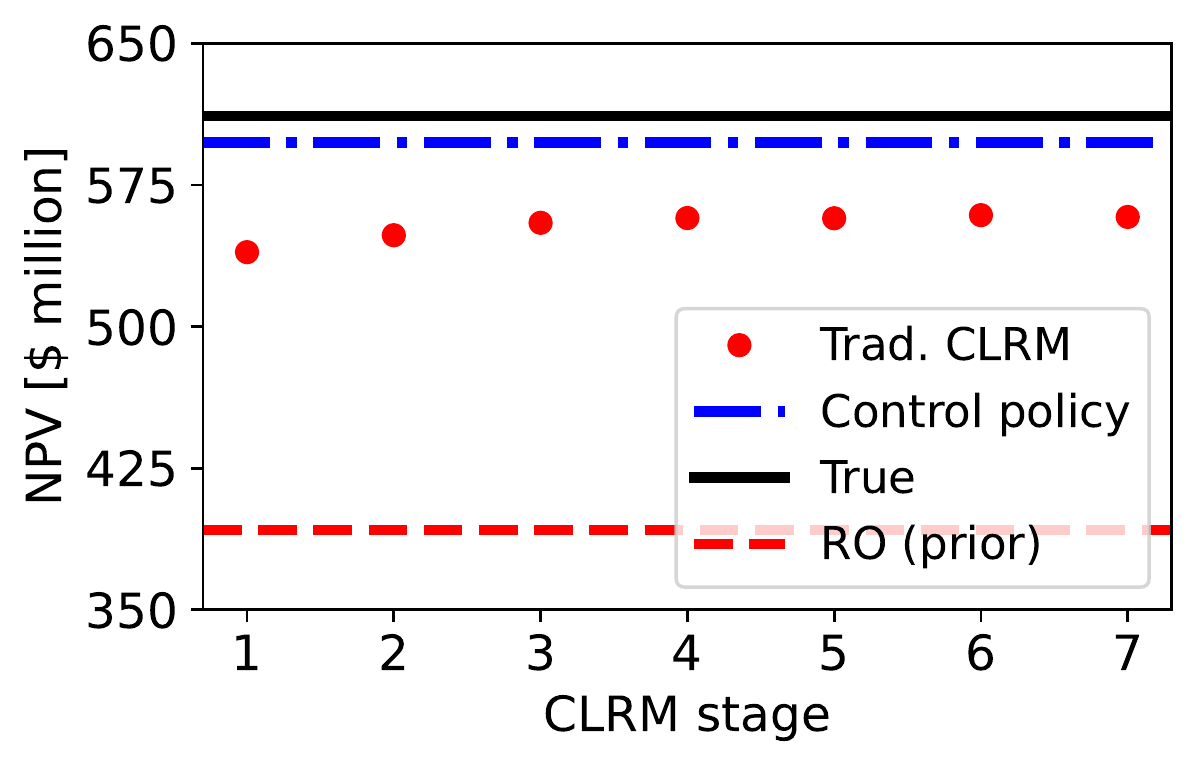}
        \caption{True model B}
    \end{subfigure}%
    
    \begin{subfigure}[b]{0.45\textwidth}
        \includegraphics[width=1\textwidth]{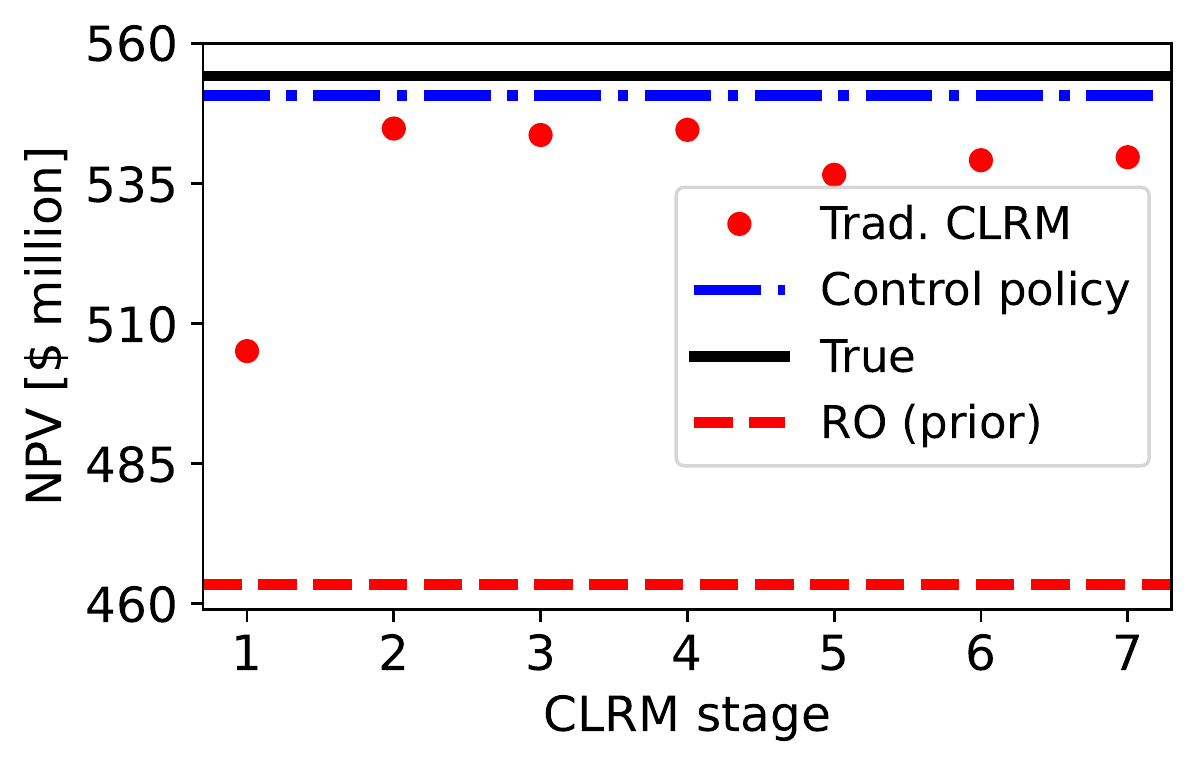}
        \caption{True model C}
    \end{subfigure}%

    \caption{NPV for the three true models using prior robust optimization, all stages of traditional CLRM, control policy, and true (deterministic) optimization  (Example~1).}
    \label{fig:control_policy_vs_clrm}
\end{figure}

Figure~\ref{fig:control_policy_vs_clrm} and Table~\ref{tab:npv_comp} compare traditional CLRM, the DRL-based control policy, and optimization over prior models, for the three true models considered for CLRM. The use of the traditional CLRM results in NPV increases (over the NPV from robust optimization with prior models) of 35.5\%, 42.6\%, and 16.6\%, for True models~A, B, and C, while the use of the control policy leads to NPV increases of 38.8\%, 52.5\%, and 19.0\% for the three true models. The control policy solutions result in an average regret, for the three models, of \$5~million, while traditional CLRM leads to an average regret of \$25.3~million. These results demonstrate the superior performance of the control policy over traditional CLRM for these cases.

\begin{table}[!htb]
    \centering
    \caption{NPV (in million USD) for prior robust optimization, final stage of traditional CLRM, control policy, and true (deterministic) optimization (Example~1).}
     \begin{tabular}{ccccc}
		\hline
True model  & RO (prior) &  Trad.~CLRM  &  Control policy   &   True  \\
        \hline
        A &   335  &    454     &   465    &    463  \\
        B &   392  &    559     &   598    &    612  \\
        C &   463  &    540     &   551    &    554  \\
          \hline
          \end{tabular}
    \label{tab:npv_comp}
\end{table}

\begin{figure}[!htb]
    \centering
    \begin{subfigure}[b]{0.45\textwidth}
        \includegraphics[width=1\textwidth]{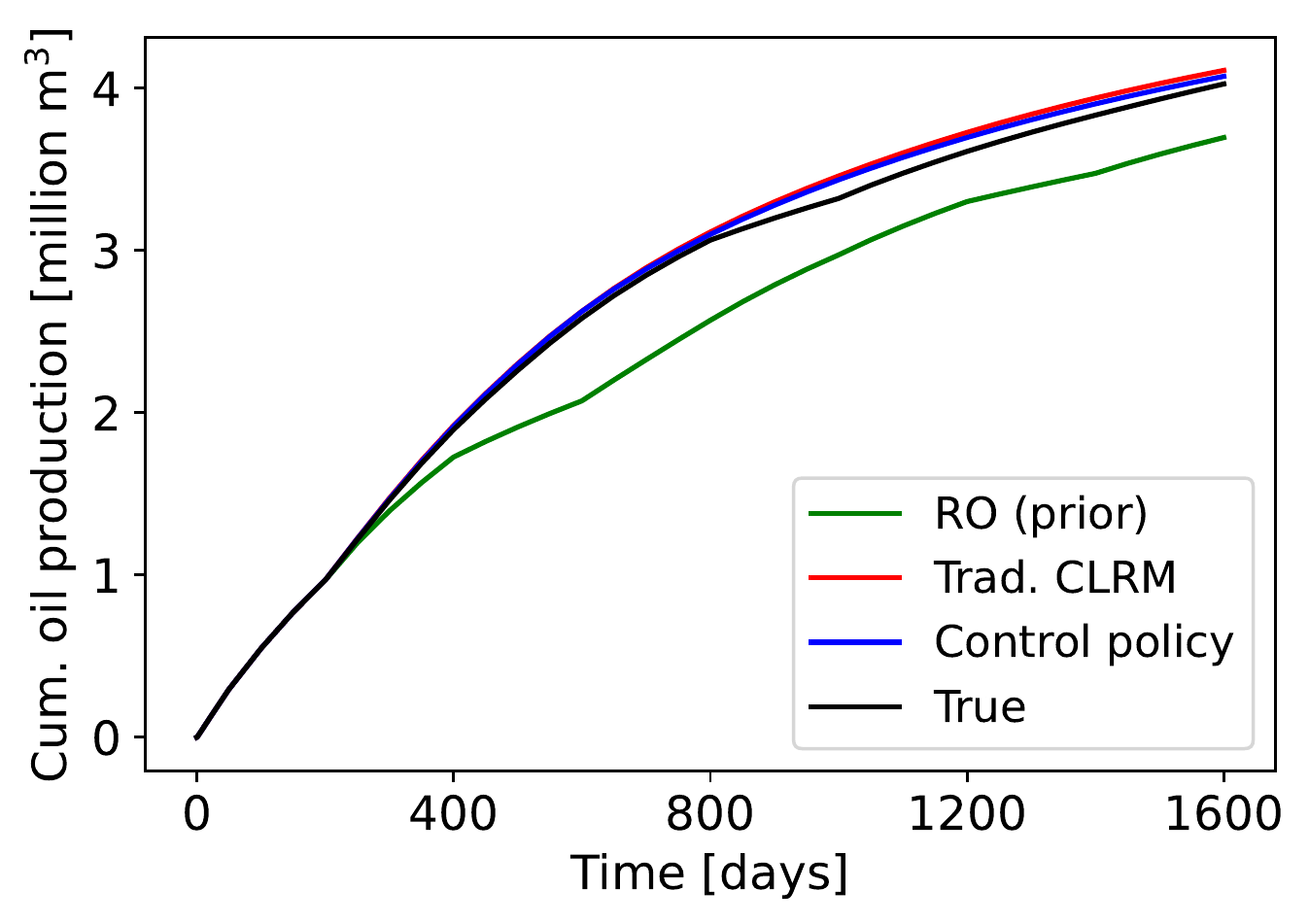}
        \caption{Cumulative oil production}
    \end{subfigure}%
    \hspace{2\baselineskip}
    \begin{subfigure}[b]{0.45\textwidth}
        \includegraphics[width=1\textwidth]{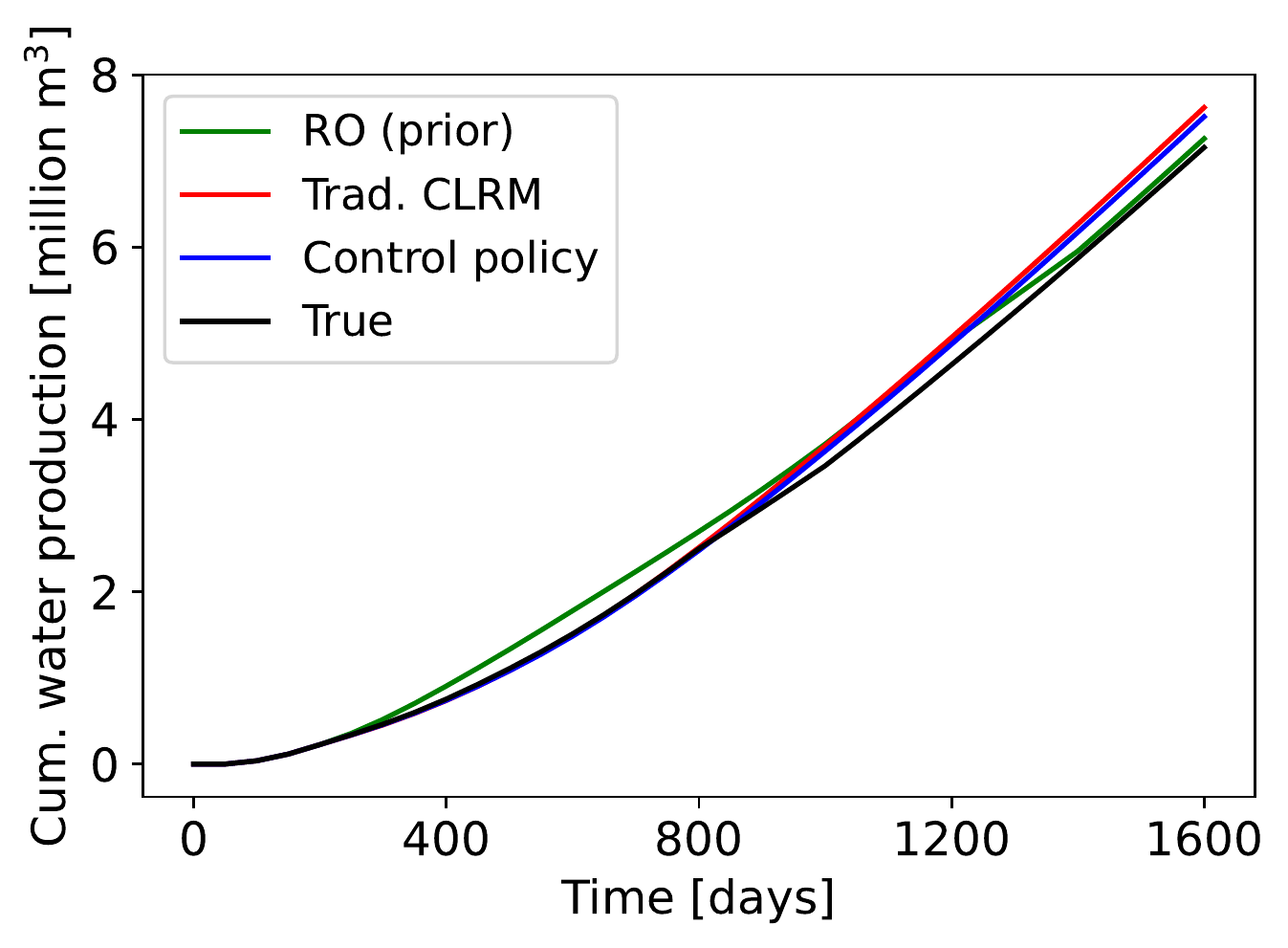}
        \caption{Cumulative water production}
    \end{subfigure}%
    
    \begin{subfigure}[b]{0.45\textwidth}
        \includegraphics[width=1\textwidth]{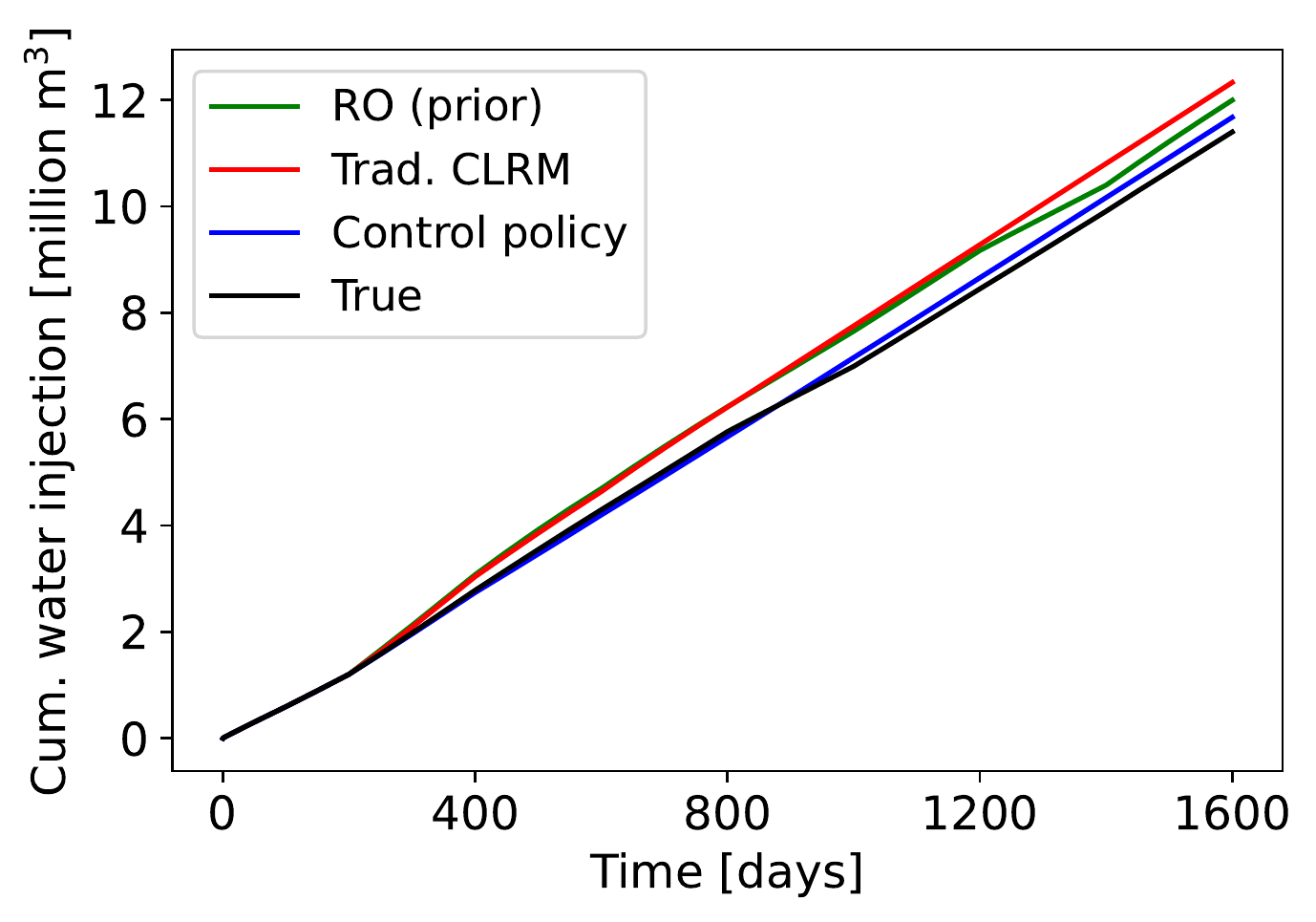}
        \caption{Cumulative water injection}
    \end{subfigure}%

    \caption{Cumulative oil and water production and water injection, as determined from robust optimization over prior models, traditional CLRM, DRL-based control policy, and deterministic optimization on the true model (Example~1, True model~A).}
    \label{fig:cum_prod_inj_2d}
\end{figure}

Finally, in Fig.~\ref{fig:cum_prod_inj_2d}, we present the field-wide cumulative oil and water production and cumulative water injection for True model~A with each of the three approaches. The traditional CLRM and control policy solutions result in comparable cumulative oil production. This oil production clearly exceeds that achieved by optimizing over prior models. Small differences between the various solutions are evident in cumulative water production and injection, with the traditional CLRM solution corresponding to more water injected and produced.

\subsection{Example 2: 3D bimodal models from multiple scenarios}

In the previous example, the geological models were generated from a single geological scenario. In this example, we consider geological models from five different 3D geological scenarios. The channel geometries (shape, size) and orientation vary from one scenario to another. Realizations are then drawn from each scenario, with detailed channel locations varying between realizations.

The five geological scenarios are defined by the parameters given in Table~\ref{tab:scenarios}. Here amplitude, wavelength, width, thickness and orientation define the average geometry of the sand channels. The spatial correlations between properties within the channels and mud are defined by the variogram range (given in terms of number of grid blocks).

A training image defined on a $250 \times 250 \times 20$ grid is constructed for each scenario. A total of 500 binary channelized realizations, conditioned to facies type at the well locations, are generated for each scenario. The realizations are defined on grids containing $40 \times 40 \times 5$ cells, with $\Delta x = \Delta y = 38$~m and $\Delta z = 3.6$~m. 

\begin{table}[!htb]
    \centering
    \caption{Parameter values defining the 3D channelized geological scenarios (Example~2).  
    }
     \begin{tabular}{ccccccc}
		\hline
        Scenario  &  Amplitude   &   Wavelength  & Width & Thickness&  Orientation  & Variogram range \\
        
        {}  &  (m)   &   (m)  & (m) & (m)   &  (degrees)   & (blocks) \\
        
        \hline
        1 &   122    &    914  &  91   & 27   &    45   &   20  \\
        2 &   152    &    762  &  122  & 37   &    0    &   30  \\
        3 &   122    &    838  &  61   & 47   &    45   &   40  \\
        4 &   152    &    762  &  122  & 27   &   -45   &   20  \\
        5 &   182    &    838  &  61   & 37   &    0    &   30  \\
          \hline
          \end{tabular}
    \label{tab:scenarios}
\end{table}

The models used in this example are bimodal, meaning there is property variation within each facies, as opposed to the binary models (constant properties within facies) considered in Example~1. A cookie-cutter approach \cite{2007PhDCastro} is used to generate the bimodal realizations. This entails the (additional) generation of separate sand and mud permeability fields for each binary channelized realization. This is accomplished by generating full ($40 \times 40 \times 5$) permeability fields for each facies, and then assigning permeability for each grid block based on the facies type. A spherical variogram (with range given in Table~\ref{tab:scenarios}) is used to generate the sand and mud permeability fields. These permeability fields are conditioned to permeability values at the well locations. The ratio of vertical to horizontal permeability is set to 0.1.

\begin{figure}[t]
    \centering
    \begin{subfigure}[b]{0.34\textwidth}
        \includegraphics[width=1\textwidth]{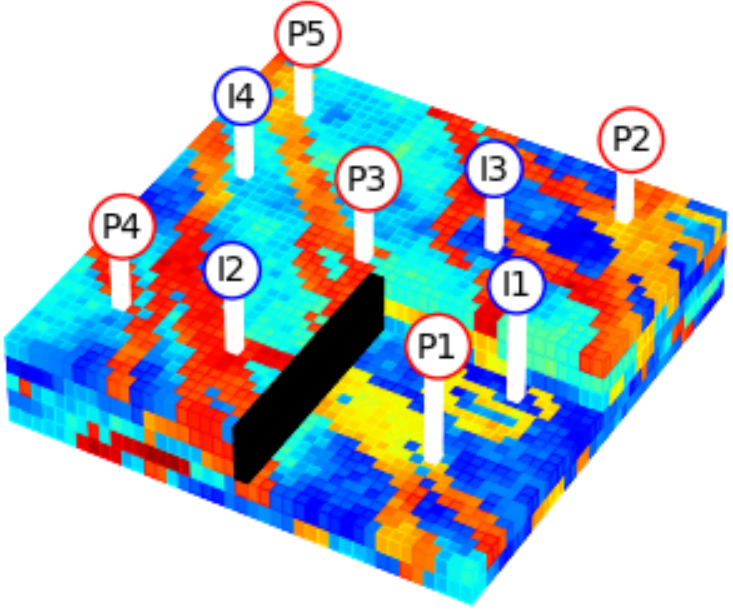}
        \caption{Scenario 1}
    \end{subfigure}%
    ~
    \begin{subfigure}[b]{0.34\textwidth}
        \includegraphics[width=1\textwidth]{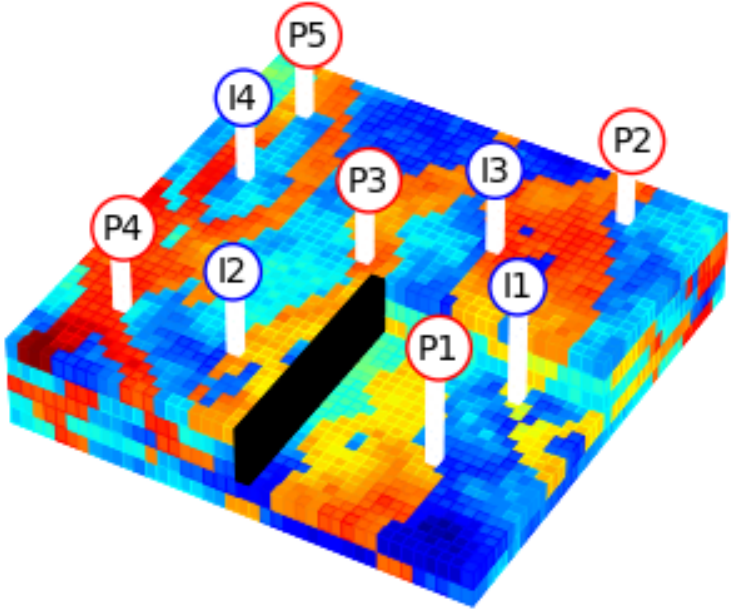}
        \caption{Scenario 2}
    \end{subfigure}%
    ~
    \begin{subfigure}[b]{0.34\textwidth}
        \includegraphics[width=1\textwidth]{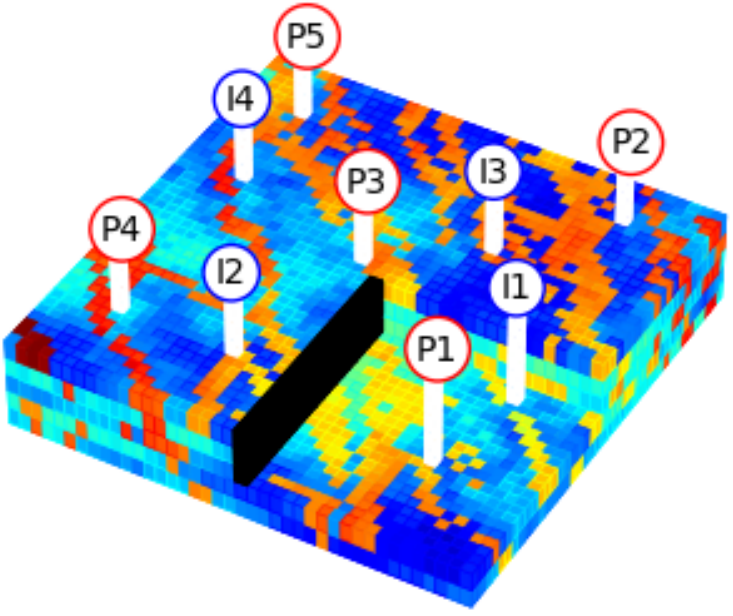}
        \caption{Scenario 3}
    \end{subfigure}%
    
     \begin{subfigure}[b]{0.34\textwidth}
        \includegraphics[width=1\textwidth]{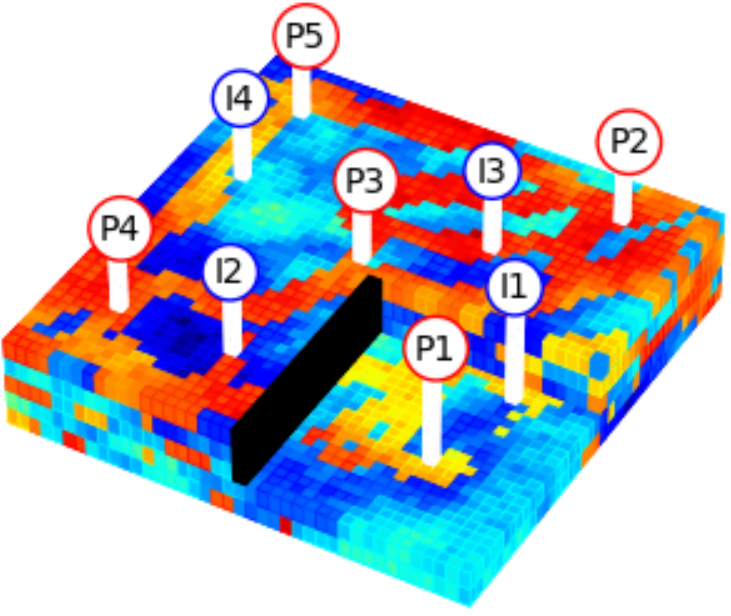}
        \caption{Scenario 4}
    \end{subfigure}%
    ~
    \begin{subfigure}[b]{0.34\textwidth}
        \includegraphics[width=1\textwidth]{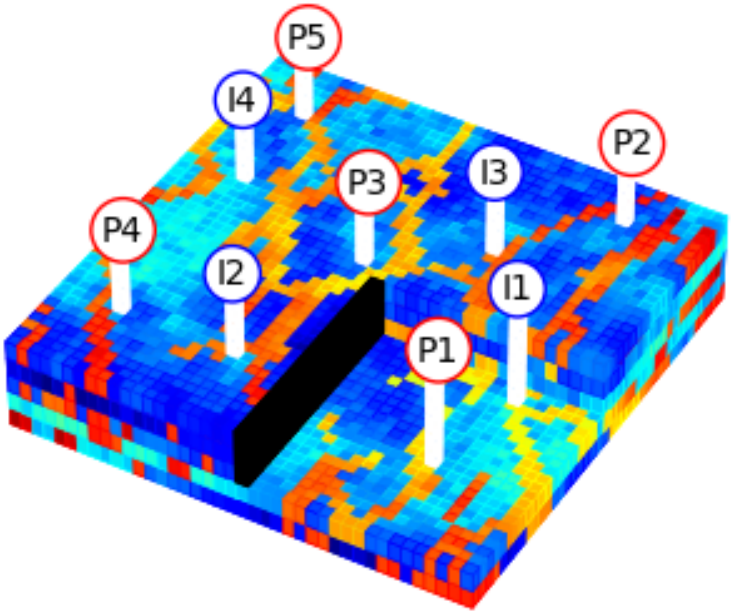}
        \caption{Scenario 5}
    \end{subfigure}%

    \caption{3D channelized bimodal realizations for each scenario (Example~2).}
    \label{fig:scen_3d}
\end{figure}

\begin{figure}[htbp]
	\centering
	\includegraphics[width=0.7\textwidth]{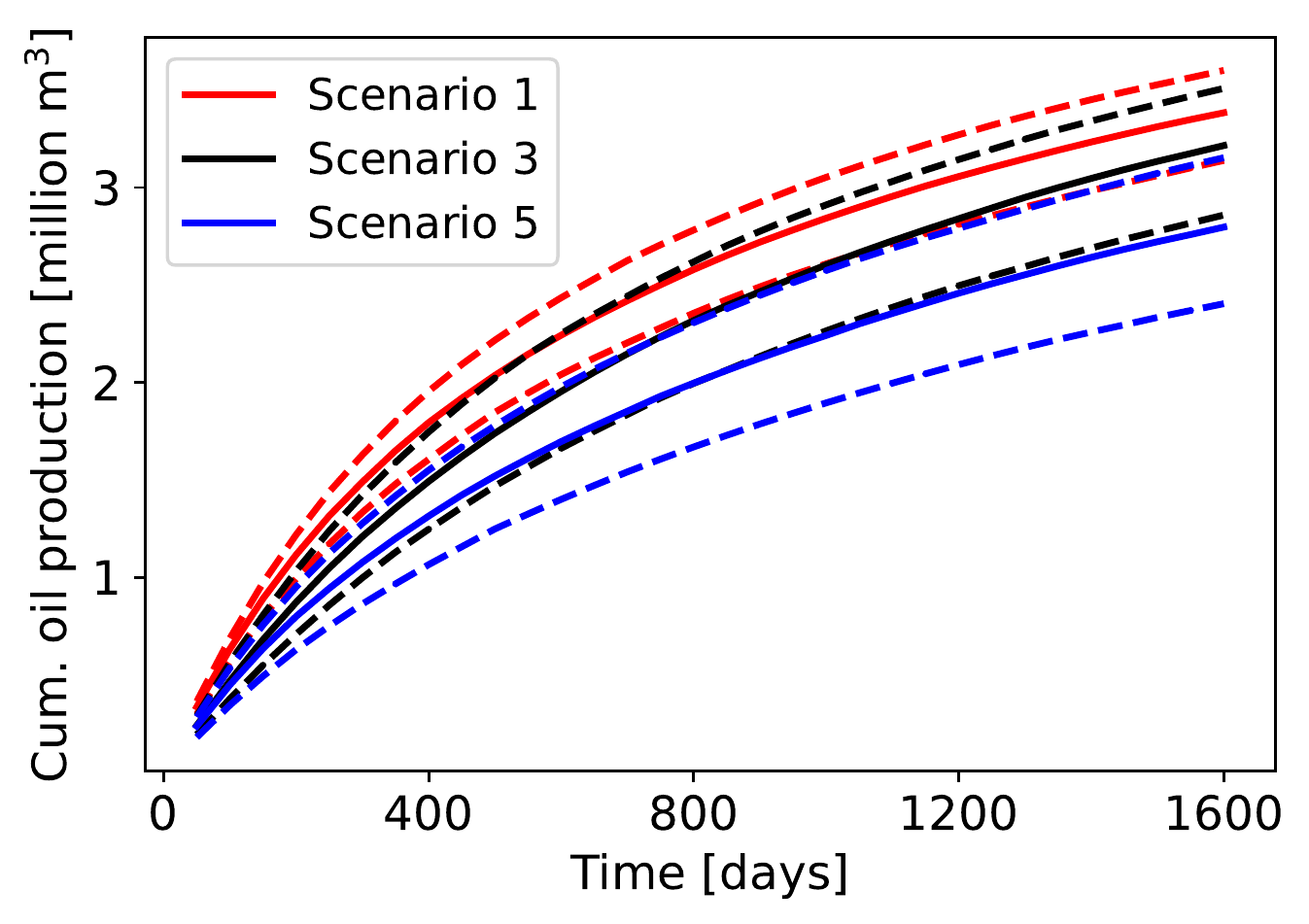}
	\caption{Field-wide cumulative oil production statistics for Scenarios~1, 3 and 5. The dashed curves correspond to the P$_{10}$ and P$_{90}$ results over the 500 realizations in each scenario, while the solid curves represent P$_{50}$ results (Example~2).}
	\label{fig:cum_oil_bimodal}
\end{figure}

Figure~\ref{fig:scen_3d} shows one realization for each scenario. Differences in channel orientation and geometry between the scenarios are evident, as is the variation of permeability within facies. The wells penetrate all layers, i.e., they extend through the entire model in the $z$-direction. The general flow behavior also differs between scenarios. This is illustrated in Fig.~\ref{fig:cum_oil_bimodal}, where we show P$_{10}$, P$_{50}$ and P$_{90}$ results for field-wide cumulative oil production for Scenarios~1, 3 and 5. 


\subsubsection{Control policy training}
\label{sec:cont_pol_opt_3d}

The 2500 geological models are divided into 40 clusters for the training of the control policy. The centroids of the 40 clusters, which are representative of the full set, are excluded from the training. At each training iteration, 12 realizations are sampled from each cluster, resulting in 480 flow simulations per iteration. The simulations are performed with 240 processors. We terminate the training after 500 iterations, resulting in 240,000 total simulation runs.

\begin{figure}[htbp]
	\centering
	\includegraphics[width=0.7\textwidth]{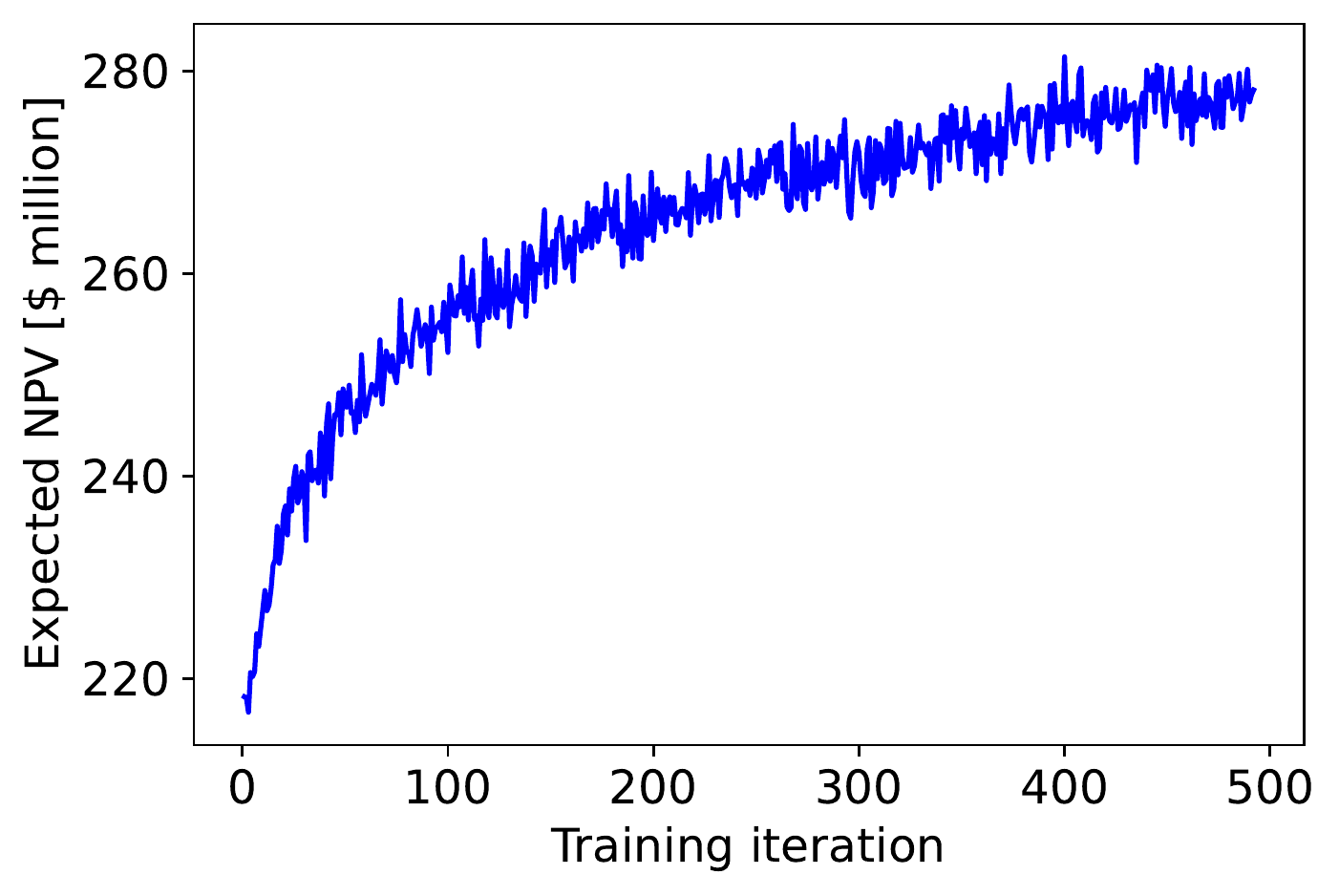}
	\caption{Evolution of expected NPV (Eq.~\ref{eq:exp_return}) computed with the sampled geological models and sampled actions in each training iteration (Example~2).}
	\label{fig:training_npv_3d}
\end{figure}

\begin{figure}[htbp]
	\centering
	\includegraphics[width=0.7\textwidth]{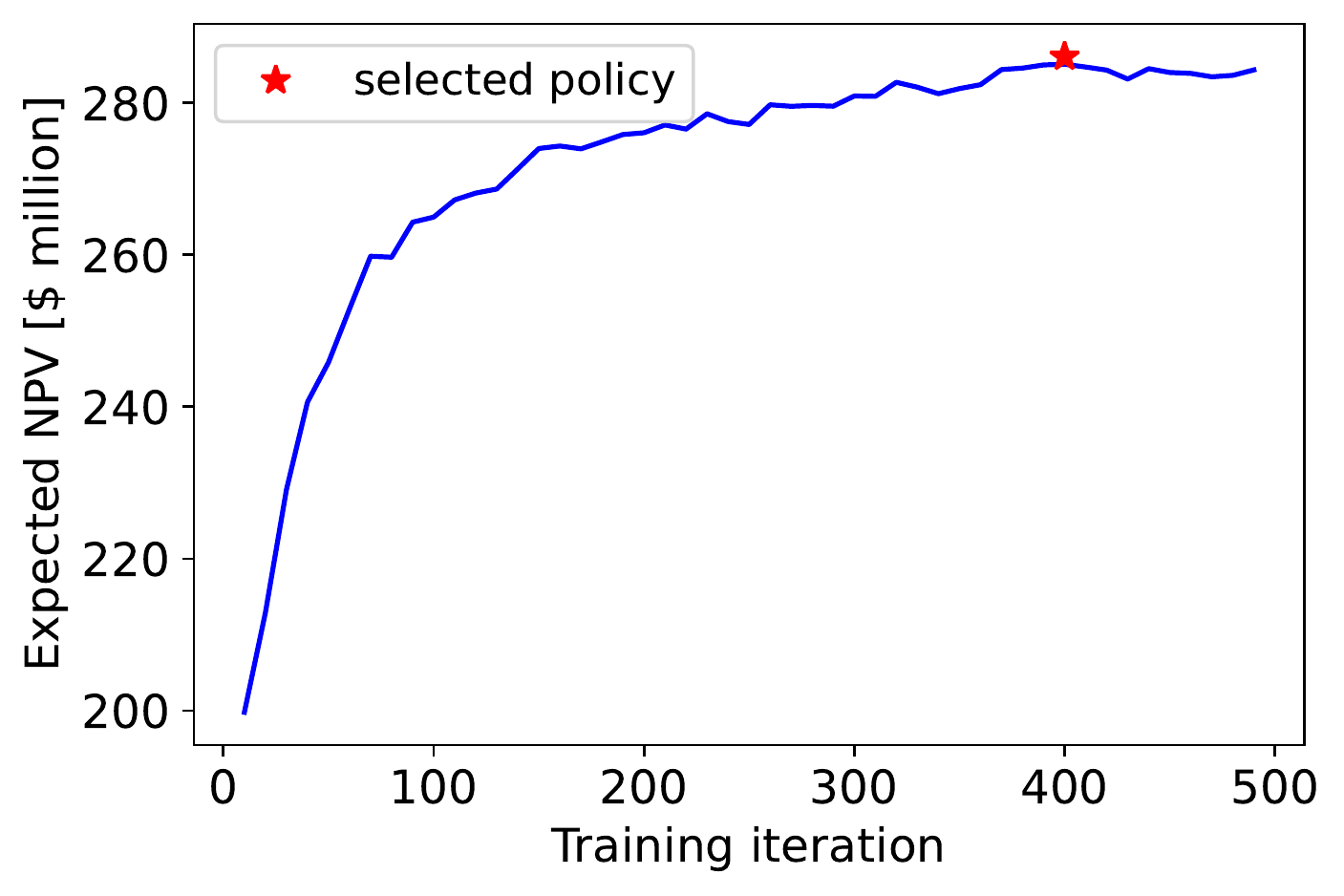}
	\caption{Evolution of the expected NPV for the 40 test-case geological models (Example~2).}
	\label{fig:test_npv_3d}
\end{figure}

The evolution of the expected NPV during training is shown in Fig.~\ref{fig:training_npv_3d}. The expected NPV of the random initial policy (\$218.2 million) increases by 29\% after about 400 iterations. The 40 representative realizations are used for evaluating the updated control policies after every ten iterations. The expected NPV of the control policies over the 40 test-case geological models are shown in Fig.~\ref{fig:test_npv_3d}. As in the previous example, the optimal control policy is chosen as the policy with the highest expected NPV (shown as the red star in Fig.~\ref{fig:test_npv_3d}).


\subsubsection{Comparison of control policy to prior optimization}

As in Example~1, we compare the control policy to robust optimization with PSO-MADS. The optimization is performed using the 40 test-case geological models excluded from the control policy training. With 50 particles, we perform a total of 2000 simulations at each PSO iteration, while 5040 simulations are required for a MADS iteration. We use 250 processors for the robust optimization.

\begin{figure}[htbp]
	\centering
	\includegraphics[width=0.6\textwidth]{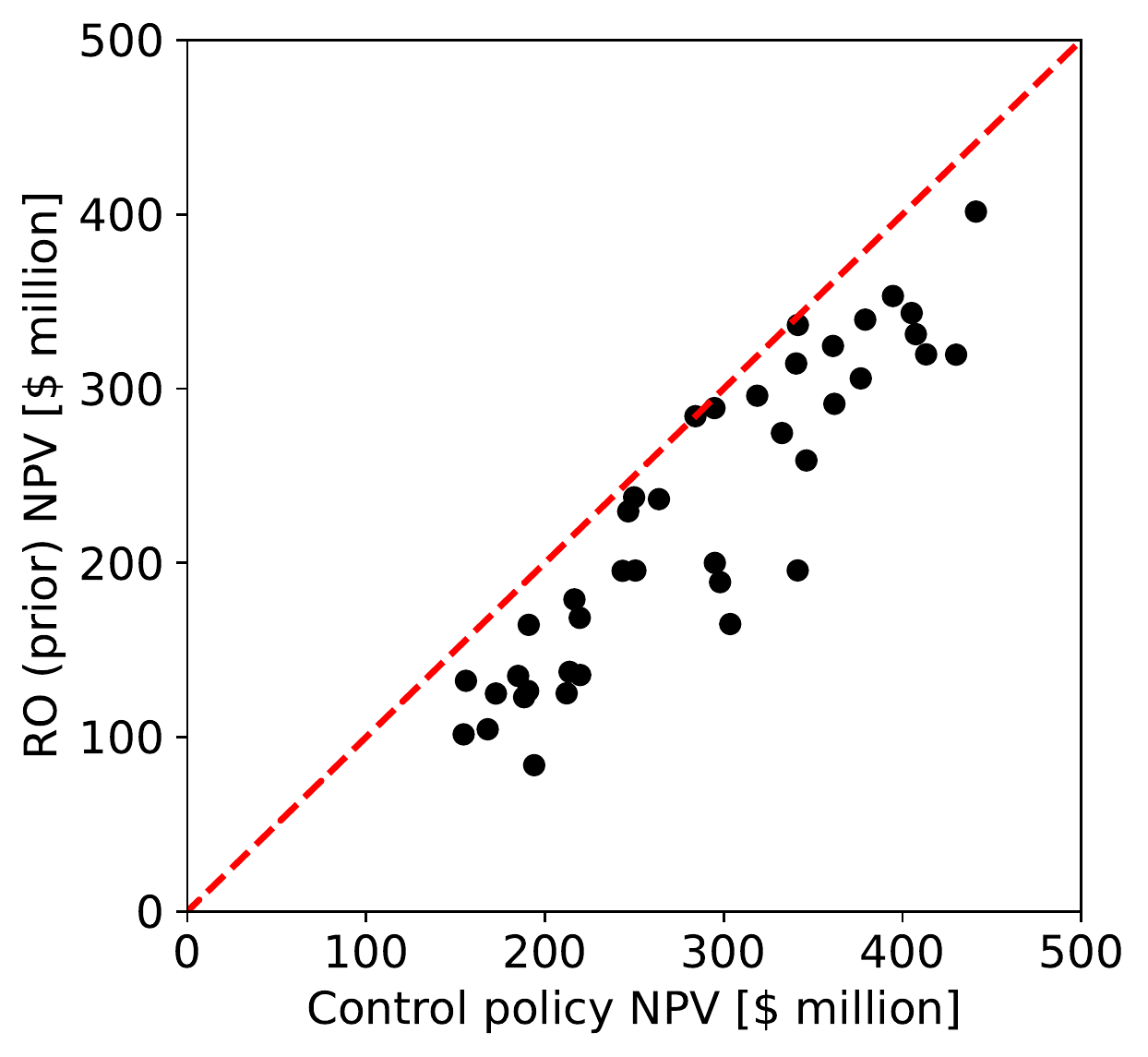}
	\caption{Comparison of solutions from robust (prior) optimization to those using the DRL-based control policy (Example~2).}
	\label{fig:drl_vs_ro_3d}
\end{figure}

The well settings obtained from the robust optimization are then applied to each of the 40 geological models. Figure~\ref{fig:drl_vs_ro_3d} displays a cross plot comparing the NPVs obtained from this prior optimization to those of the DRL-based control policy. The control policy approach clearly outperforms prior optimization, consistent with the results in Fig.~\ref{fig:drl_vs_ro_chan} for Example~1. The control policy provides an average improvement of 32.7\% (\$58.3 million) relative to prior optimization.

\begin{figure}[!htb]
    \centering
    \begin{subfigure}[b]{0.5\textwidth}
        \includegraphics[width=1\textwidth]{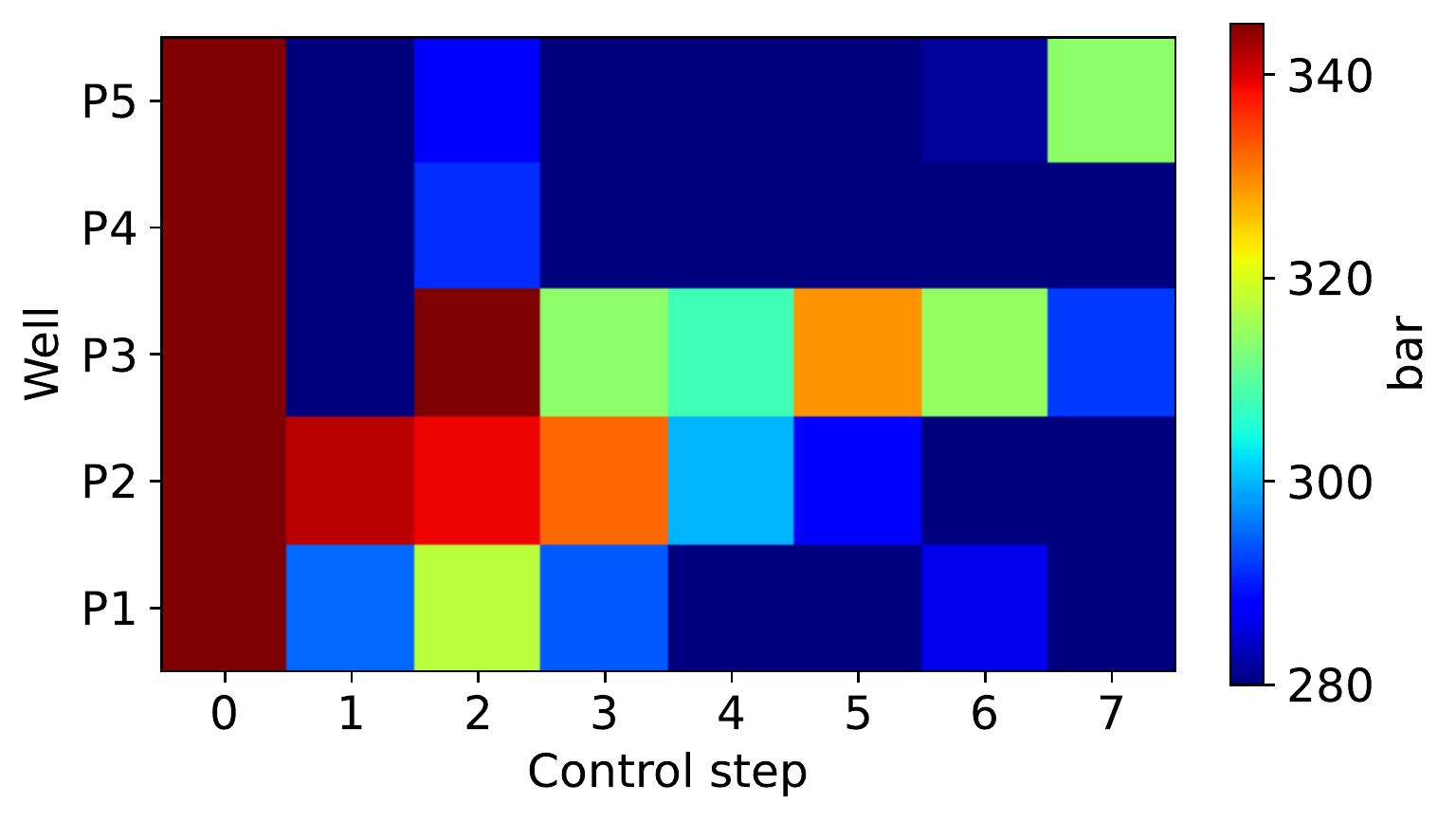}
        \caption{Control policy (P$_{10}$ model)}
    \end{subfigure}%
    ~
    \begin{subfigure}[b]{0.5\textwidth}
        \includegraphics[width=1\textwidth]{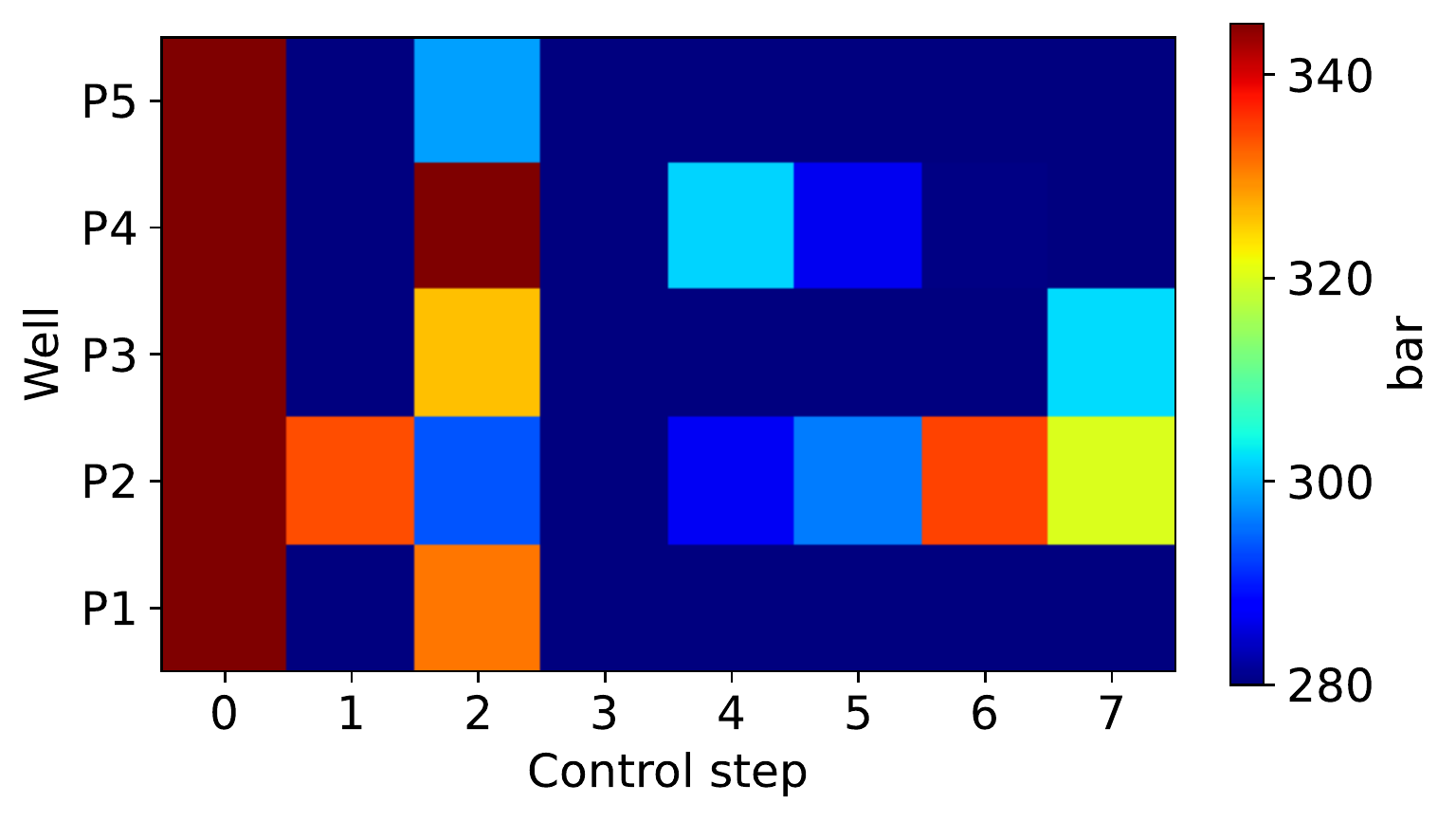}
        \caption{Control policy (P$_{90}$ model)}
    \end{subfigure}%
    
    \begin{subfigure}[b]{0.5\textwidth}
        \includegraphics[width=1\textwidth]{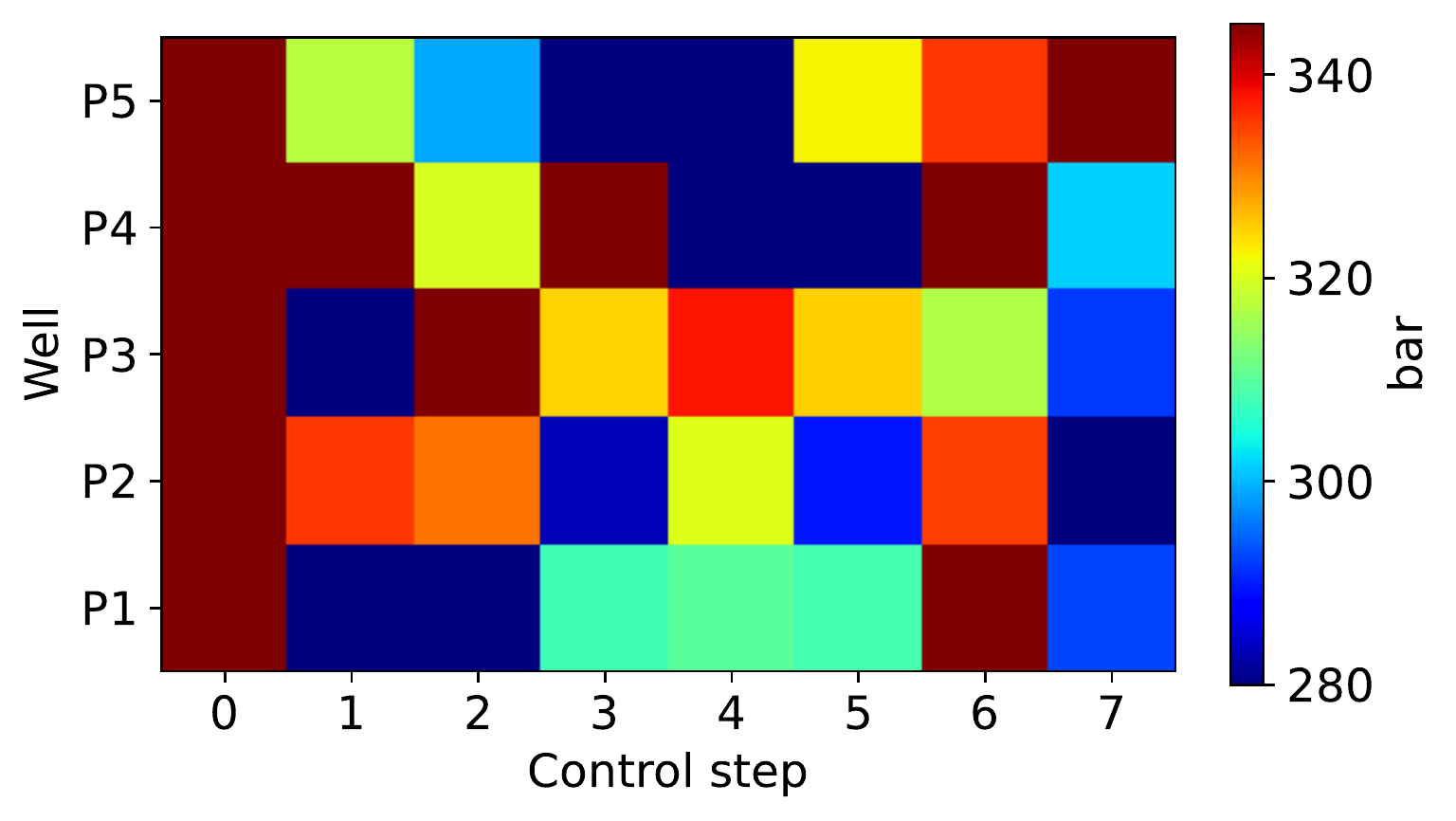}
        \caption{RO (prior)}
    \end{subfigure}%

    \caption{Well settings proposed by the control policy for the P$_{10}$ and P$_{90}$ models determined by ranking the NPVs achieved through robust (prior) optimization (Example~2).}
    \label{fig:prod_bhps_ex_2}
\end{figure}

As in Example~1, we rank the geological models based on their NPVs from robust (prior) optimization. Figure~\ref{fig:prod_bhps_ex_2}(a) and (b) shows the well settings for the producers, obtained through use of the control policy, for the P$_{10}$ and P$_{90}$ models. The well settings obtained from the robust (prior) optimization are shown in Fig.~\ref{fig:prod_bhps_ex_2}(c). The well settings obtained from the control policy differ between the two models, and they differ significantly from the robust (prior) optimization result. This again demonstrates that the policy adjusts the controls for each particular realization.

\subsubsection{Comparison of control policy to deterministic optimization}

\begin{figure}[htbp]
	\centering
	\includegraphics[width=0.6\textwidth]{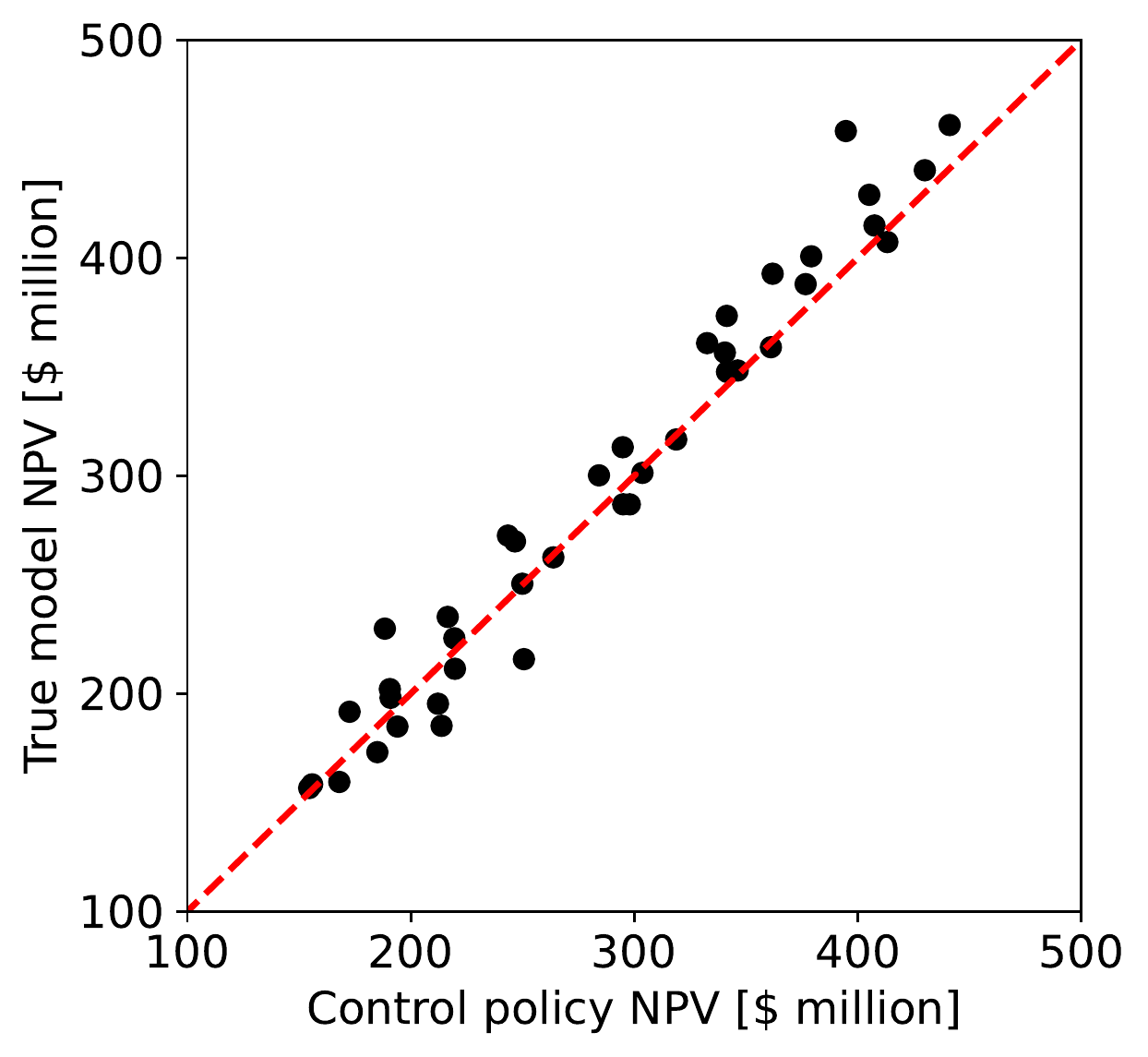}
	\caption{Comparison of solutions from deterministic optimization (performed separately for each `true' model) to those using the DRL-based control policy (Example~2).}
	\label{fig:drl_vs_true_3d}
\end{figure}

We now optimize the 40 test-case geological models individually using SNOPT. The cross-plot of the true NPV (from deterministic optimization) for each model versus that obtained using the DRL-based control policy is shown in Fig.~\ref{fig:drl_vs_true_3d}. Although the true NPV is higher than that of the control policy in 26 of the models, there is generally close agreement between the NPVs from both approaches. As mentioned earlier, this true NPV cannot be expected in practice, though it provides a useful (ideal) benchmark for the solutions obtained from the control policy approach.

\begin{figure}[!htb]
    \centering
    \begin{subfigure}[b]{0.45\textwidth}
        \includegraphics[width=1\textwidth]{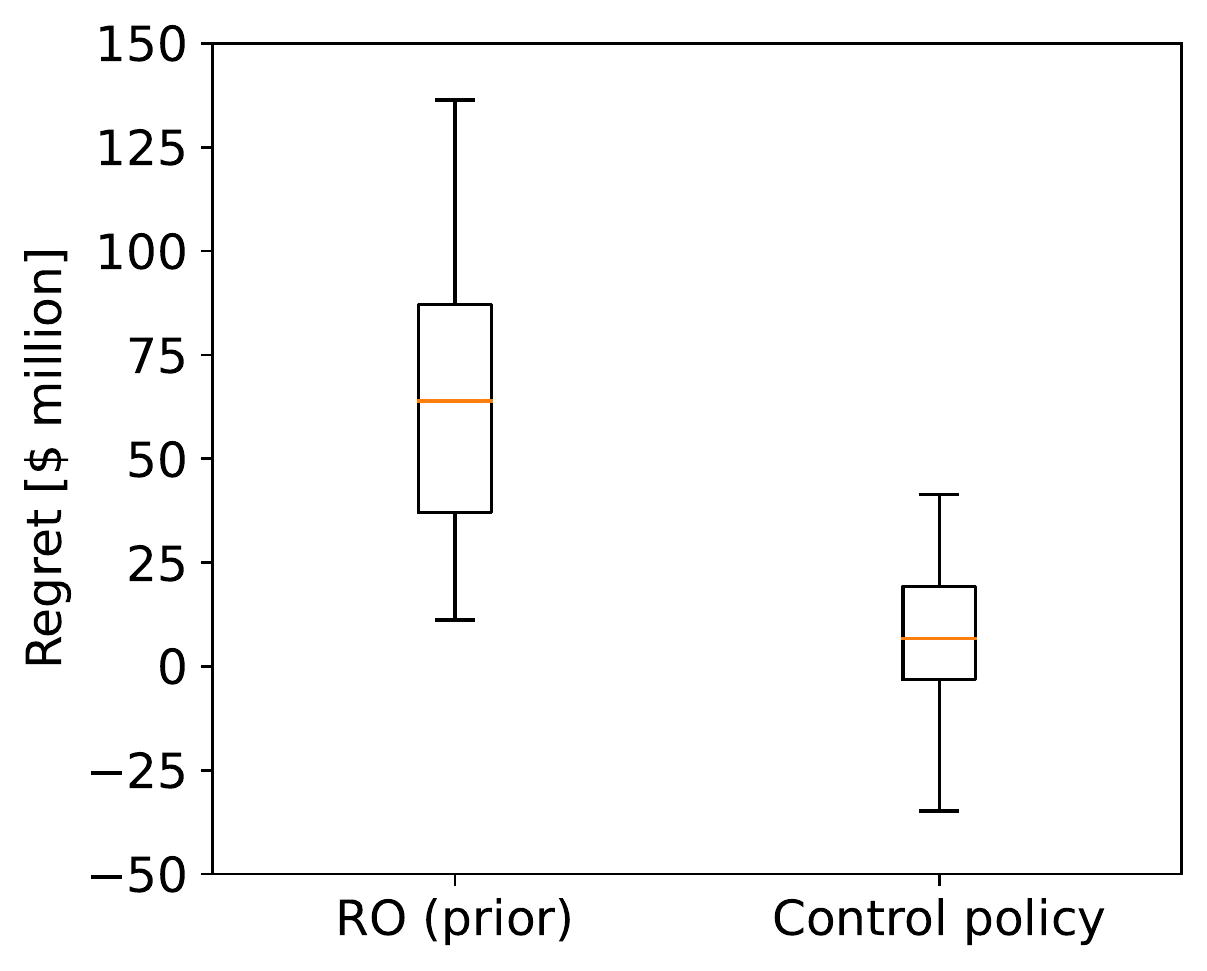}
        \caption{Box plots of regret}
    \end{subfigure}%
    \hspace{2\baselineskip}
    \begin{subfigure}[b]{0.45\textwidth}
        \includegraphics[width=1\textwidth]{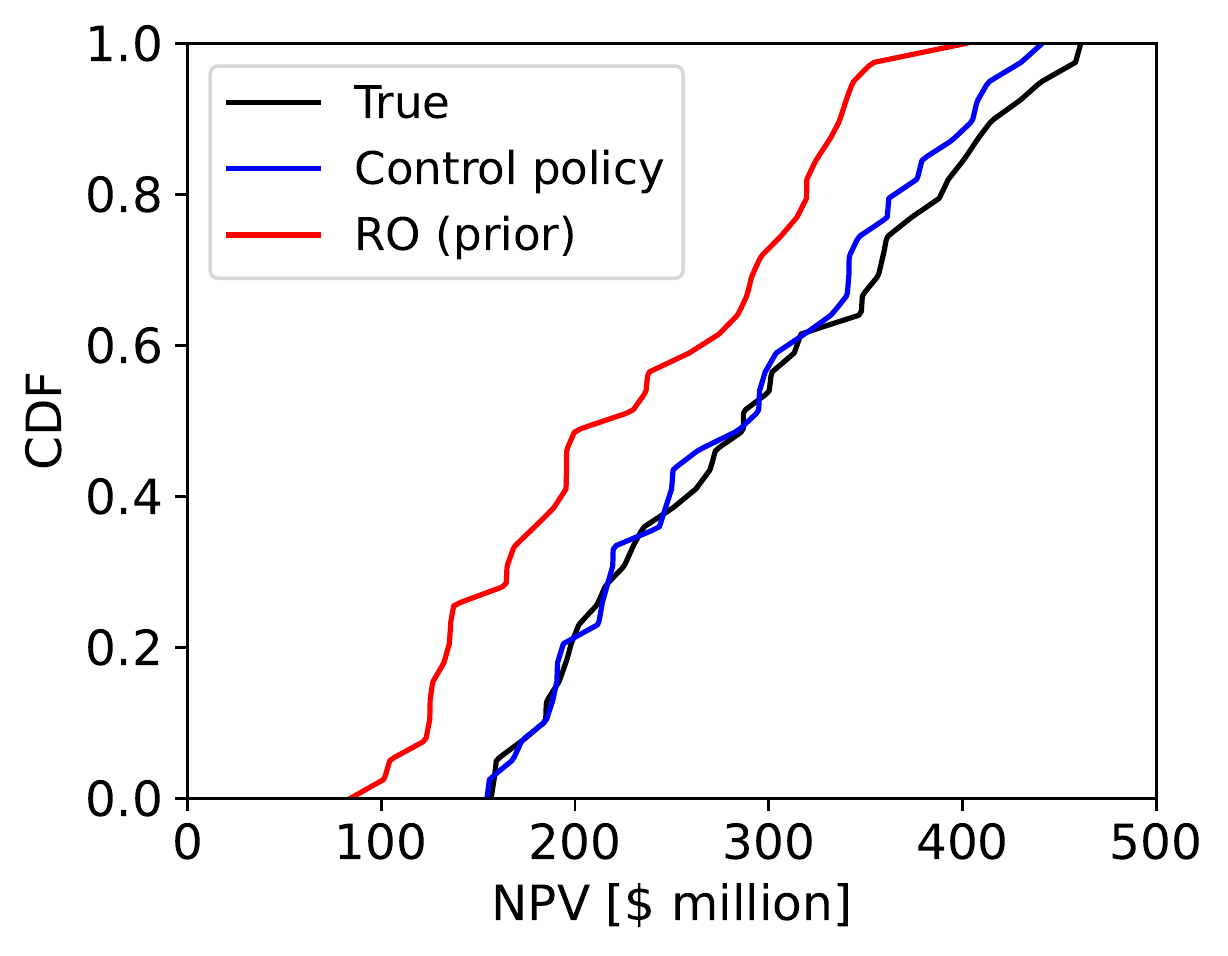}
        \caption{CDFs of the optimum NPVs}
    \end{subfigure}%

    \caption{Comparison of results from the three approaches for the 40 test-case geological models (Example~2).}
    \label{fig:bench_mark_3d}
\end{figure}

The regret for the robust (prior) optimization and control policy are shown in Fig.~\ref{fig:bench_mark_3d}(a). The use of prior robust optimization leads to a maximum regret of \$177 million compared to \$64 million for the control policy. The P$_{50}$ regret for prior optimization and the control policy are \$63.9 million and \$6.7 million, respectively. Figure~\ref{fig:bench_mark_3d} displays CDFs of the NPVs using the three approaches. We see close agreement between the CDFs from deterministic optimization and the control policy approach, with both procedures outperforming prior optimization. These results are consistent with those obtained in Example~1 (Fig.~\ref{fig:bench_mark_chan}) and further demonstrate the advantages of the DRL-based control policy methodology.

\begin{figure}[!htb]
    \centering
    \begin{subfigure}[b]{0.45\textwidth}
        \includegraphics[width=1\textwidth]{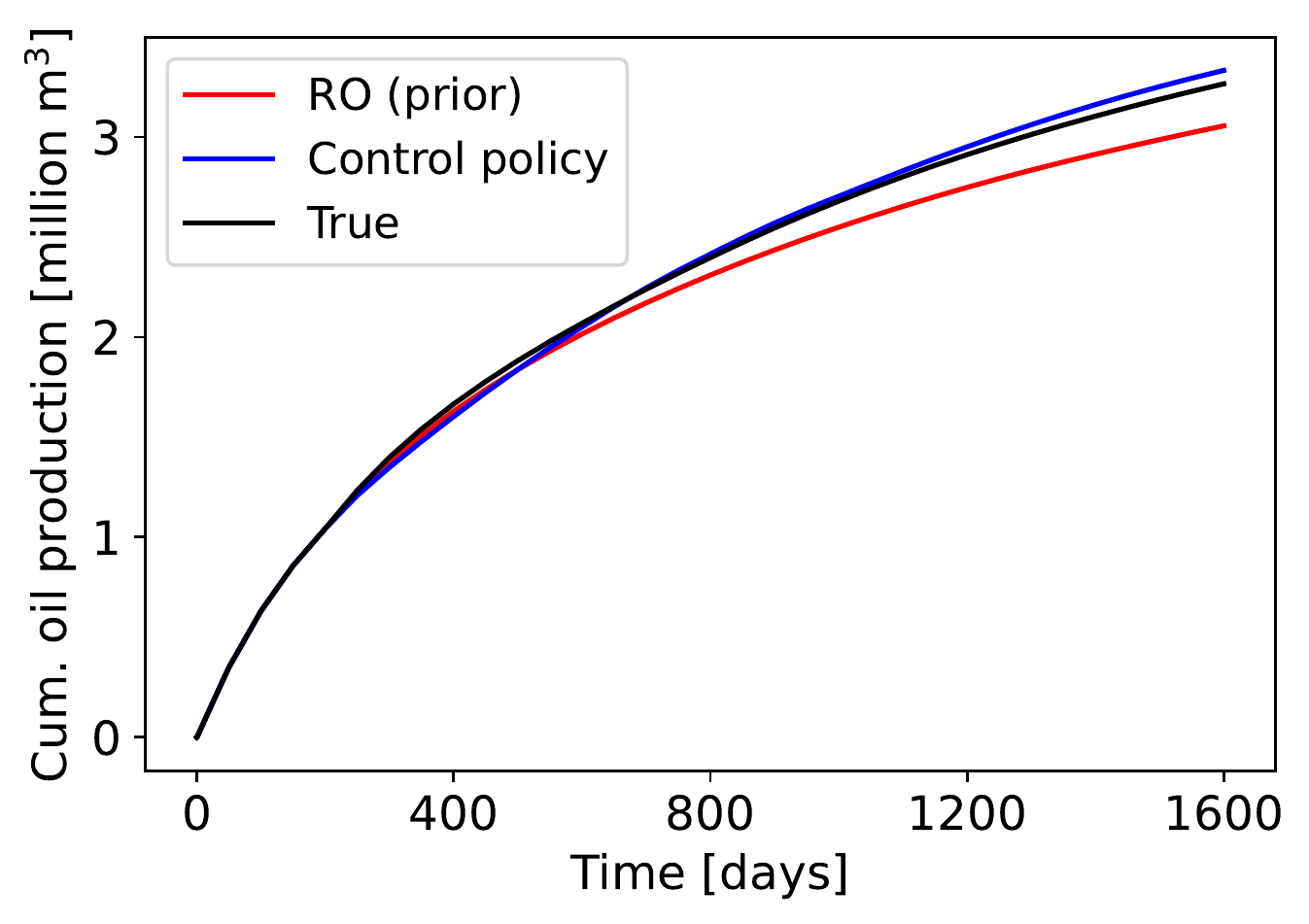}
        \caption{Cumulative oil production}
    \end{subfigure}%
    \hspace{2\baselineskip}
    \begin{subfigure}[b]{0.45\textwidth}
        \includegraphics[width=1\textwidth]{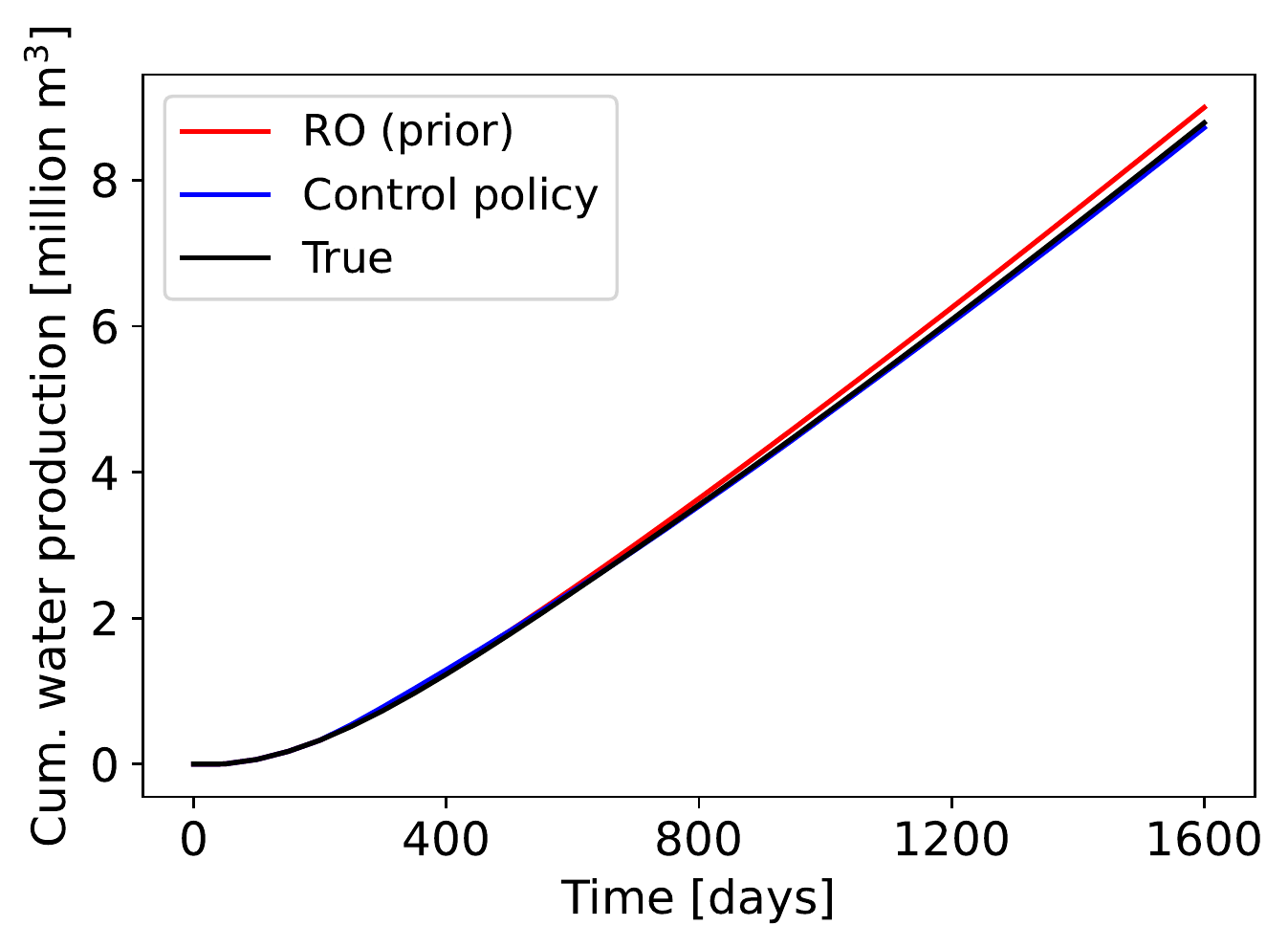}
        \caption{Cumulative water production}
    \end{subfigure}%
    
    \begin{subfigure}[b]{0.45\textwidth}
        \includegraphics[width=1\textwidth]{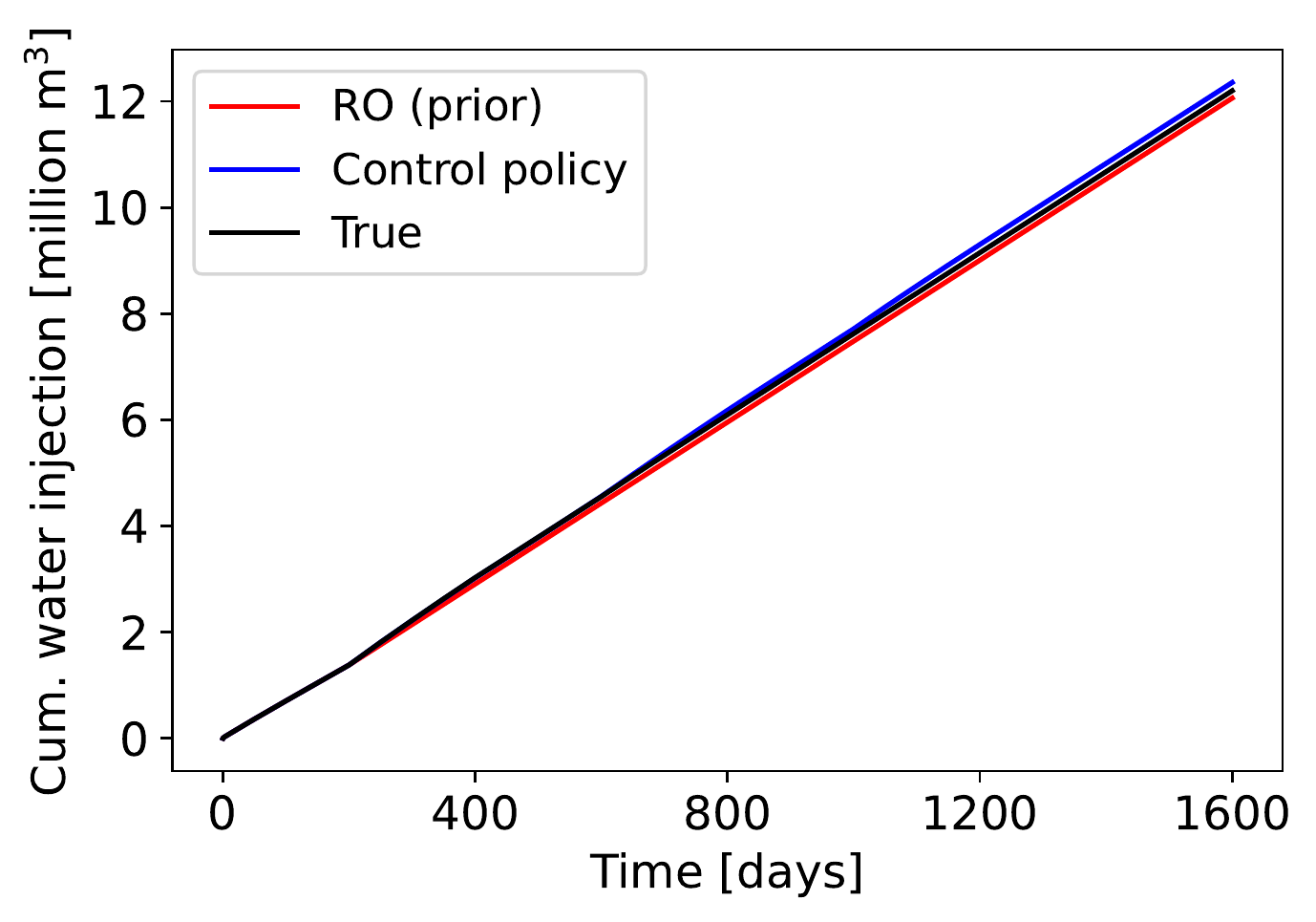}
        \caption{Cumulative water injection}
    \end{subfigure}%

    \caption{Cumulative oil and water production and water injection, as determined from robust optimization over prior models, DRL-based control policy, and deterministic optimization, for the model with P$_{50}$ prior optimization regret (Example~2).}
    \label{fig:cum_prod_inj_3d}
\end{figure}

Field-wide cumulative oil and water production and water injection, for the geological model with the P$_{50}$ regret from prior optimization, are shown in Fig.~\ref{fig:cum_prod_inj_3d}. As is evident from Fig.~\ref{fig:cum_prod_inj_3d}(a), the use of the control policy results in higher cumulative oil production compared to robust (prior) optimization and deterministic optimization. The solutions from the three procedures display comparable cumulative water production and injection profiles (Fig.~\ref{fig:cum_prod_inj_3d}(b) and (c)).

We do not apply traditional CLRM for this example (as we did in Example~1) as this would require the treatment of multiple geological scenarios, which substantially complicates the history matching procedure. More specifically, to treat multiple scenarios we would need to implement a two-stage history matching approach, where we first determine the likely geological scenario or scenarios, and then construct realizations within those scenarios that match observed data. Thus, in this case we only compare the DRL-based control policy to robust prior optimization and to deterministic (model-by-model) optimization.

\subsection{Computational cost of the various approaches}

Because the various methods use different codes written in different languages, we base our assessment of computational demands on the number of flow simulation runs required by each approach. The number of runs required for traditional CLRM (denoted $C_{tc}$) can be expressed as
\begin{equation}
    C_{tc} = a N_r N_c(N_h  + N_o N_{ps}),
    \label{eq:comp_cost_tc}
\end{equation}
where $N_{r}$ is the number of realizations used in optimization and history matching, $N_c$ is the number of CLRM stages, $N_h$ is the number of history matching iterations per CLRM stage, $N_o$ is the number of optimization iterations, and $N_{ps}$ denotes the average number of runs per PSO-MADS iteration. Because restarts can be used (due to the fact that previous control steps are not optimized), the coefficient $a$ differs from unity, and is well approximated as $a \approx 0.5$. Note that the assessment in Eq.~\ref{eq:comp_cost_tc} neglects the (relatively small) cost of the backward pass required to calculate the adjoint gradients used by SNOPT in the history matching step.

In the training of the control policy, with $N_s$ geological models simulated per iteration and a total of $N_{ic}$ iterations, the number of simulations ($C_{cp}$) is simply
\begin{equation}
    C_{cp} = N_{ic} N_s.
    \label{eq:comp_cost_cp}
\end{equation}
We ignore the cost of training the control policy with stochastic gradient descent as this is insignificant compared to the time for $N_{ic} N_s$ simulation runs. This is because the neural network is moderately sized (as is common in DRL) and only a few epochs are performed per iteration. 

The simulations in the different approaches can be performed in a distributed fashion with multiple processors. Prior robust optimization, and one control step in traditional CLRM, require $N_{ps}N_{r}$ processors to achieve full parallelization. For control policy training, $N_s$ processors are required for full parallelization. The computational costs of the different approaches, in both serial and full-parallelization modes, are reported in Table~\ref{tab:comp_compl}.

\begin{table}[!htb]
    \centering
    \caption{Computational cost, in terms of number of simulations, for robust (prior) optimization, traditional CLRM, and control policy, in both serial and full-parallelization modes.}
     \begin{tabular}{p{5cm}p{4cm}p{3cm}}
		\hline
Method  & \multicolumn{2}{c}{Computational cost \ \ \ \ \ \ \ \ \ \ \ \ \ \ } \\
         {} & Serial & Parallel \\
        \hline
        RO (prior)       &   $N_o N_r N_{ps}$  &    $N_o$ \\
        Traditional CLRM &   $0.5N_r N_c(N_h  + N_o N_{ps})$  &    $0.5N_c(N_h  + N_o)$     \\
        Control policy   &   $N_{ic} N_s$  &    $N_{ic}$     \\
          \hline
          \end{tabular}
    \label{tab:comp_compl}
\end{table}

For Example~1, the values of the variables in Table~\ref{tab:comp_compl} are as follows: $N_o = 106$, $N_{r} = 30$, $N_{ps} = 50$, $N_c = 7$, $N_{h} = 30$, $N_{ic} = 500$ and $N_s = 270$. Thus we have a computational cost of approximately 159,000 flow simulations for prior robust optimization, 560,000 for the traditional CLRM, and 135,000 for the control policy approach. It is important to note that the costs associated with prior robust optimization and traditional CLRM could be reduced considerably if adjoint-gradient methods are used for optimization. However, the implementation of these methods requires access to the simulator source code, and they may not perform well with certain types of nonlinear constraints.

Under full parallelization, using the values for Example~1, prior robust optimization requires an elapsed time equivalent to 106 simulations, traditional CLRM requires 476 equivalent simulations, and the control policy approach requires 500 equivalent simulations. Thus the control policy and CLRM procedures are comparable in this setting. However, traditional CLRM and robust (prior) optimization require more than 1500 processors (considering PSO and MADS iterations) to achieve full parallelization, while control policy training requires only 270 processors for full parallelization. 

For Example~2, we have $N_{r} = 40$ and $N_s = 480$. The values of the other variables in Table~\ref{tab:comp_compl} for prior robust optimization and the control policy approach are the same as in Example~1. This results in a computational cost of approximately 212,000 simulations for prior robust optimization, and 240,000 for the control policy approach. 

We note finally that the computational cost for the control policy approach is entirely from the preprocessing (training) step. Once trained, the control policy can immediately provide (optimal) well settings without any time lag. This is in contrast to traditional CLRM, which requires about 160,000 additional simulation runs (using the values from Example~1) at each control step. 
 
\section{Concluding remarks}
\label{sec:concl}

In this work, we introduced a general nonintrusive control policy framework based on deep reinforcement learning for the closed-loop management of subsurface flow operations. The CLRM problem is formulated as a partially observable Markov decision process, where decisions are made based on quantities available from well data. The control policy, which is represented by a temporal convolution and gated transformer blocks, is trained using a proximal policy optimization algorithm. This entails the solution of a single optimization problem involving a set of prior geological models. This is in contrast to traditional CLRM workflows, where the repeated application of data assimilation and robust optimization steps is required. At each policy training iteration, representative samples of the geological models are simulated and the parameters of the control policy are updated using gradient descent. At each decision stage of the online reservoir management process, the trained control policy instantaneously maps observed data to optimal production and injection well settings. 

The new framework was applied to 2D and 3D example cases. In the 2D case, binary channelized geological models, corresponding to realizations drawn from a single geological scenario, were considered. The training of the control policy required 135,000 total flow simulations, which is equivalent to 500 sequential simulations in a fully parallelized setting. This represents only 24\% of the simulations required for traditional CLRM (using the algorithms and parameter values considered in this study). The DRL-based approach was shown to provide solutions close to those from deterministic optimization of individual geological realizations. This is a significant finding, as deterministic optimization is not possible in practice because geological uncertainty is always present. Our results clearly demonstrate the advantages of the control policy approach relative to both robust optimization over prior geological models and to the traditional CLRM approach. Specifically, the control policy approach led to an average improvement of 14.7\% in NPV relative to robust (prior) optimization, and to an average increase of 3.8\% compared to traditional CLRM.

The second example involved 3D bimodal geological models drawn from five different geological scenarios. The use of multiple scenarios complicates the history matching steps in traditional CLRM, but this does not introduce additional complications for the DRL-based control policy. For this case, the control policy framework was compared to robust optimization over prior geological models and to deterministic optimization of the individual geological models (traditional CLRM was not considered due to the complications associated with multiple scenarios). Consistent with the first example, the use of the control policy again provided better solutions than prior robust optimization; here we achieved an average improvement of 32.7\% in NPV. The average regret (loss) in NPV for the control policy approach relative to deterministic optimization was only 1.9\%, indicating very comparable performance. The results for the 2D and 3D cases clearly demonstrate the efficacy of the control policy procedure for problems of the type addressed by traditional CLRM workflows.

There are a number of directions for future work in this area. The computations required for training could be accelerated through use of deep learning \cite{kim2022recurrent} or flow network \cite{ren2019implementation} surrogate models, and the use of such treatments should be investigated. The incorporation of practical constraints, including limits on the shifts in well settings from control step to control step, should be incorporated. The workflow could then be tested on real field problems. The DRL-based framework is quite general, and our approach should be applicable in other areas where closed-loop modeling is used. Within the context of subsurface flow, for example, our procedures could be generalized to treat aquifer management, CO$_2$ storage, and geothermal production operations.

\section*{Acknowledgements}
We thank the Stanford Graduate Fellowship program and the Stanford Smart Fields Consortium for financial support. We are grateful to Yong Do Kim for providing the traditional CLRM code. We acknowledge the Stanford Center for Computational Earth \& Environmental Science for providing the computational resources used in this study. 

\bibliography{drl}

\end{document}